\documentclass[pdflatex]{sn-jnl}

\usepackage[numbers,sort&compress]{natbib}

\usepackage{multirow}%
\usepackage{amsmath,amssymb,amsfonts}%
\usepackage{amsthm}%
\usepackage{mathrsfs}%
\usepackage[title]{appendix}%
\usepackage{xcolor}%
\usepackage{textcomp}%
\usepackage{manyfoot}%
\usepackage{booktabs}%
\usepackage{enumitem}
\usepackage{comment}
\usepackage{algorithm}%
\usepackage{algorithmicx}%
\usepackage{algpseudocode}%
\usepackage{listings}%
\usepackage{subcaption}
\usepackage{tabularx}
\usepackage{amssymb}  
\usepackage{graphicx}
\usepackage{subcaption}
\usepackage{float}      
\usepackage{placeins}
\usepackage{booktabs,tabularx,threeparttable,array}
\newcolumntype{T}[1]{>{\ttfamily\raggedright\arraybackslash}p{#1}}
\newcolumntype{L}[1]{>{\raggedright\arraybackslash}p{#1}}
\newcolumntype{C}[1]{>{\centering\arraybackslash}p{#1}}
\newcolumntype{Y}{>{\raggedright\arraybackslash}X}

\usepackage{caption}
\makeatletter
\renewcommand{\fnum@figure}{\figurename~\thefigure}
\makeatother
\renewcommand{\figurename}{Figure} 
\newcommand{\SItext}{Supplementary Information}
\newcommand{\SIFig}[1]{Figure~S#1~(\SItext)}

\newtheoremstyle{nobold}
  {\topsep}   {\topsep}
  {\itshape}           
  {}                   
  {\itshape}           
  {.}                  
  {.5em}               
  {}                   

\theoremstyle{nobold}
\newtheorem{theorem}{Theorem}
\newtheorem*{theorem*}{Theorem}
\newtheorem{proposition}[theorem]{Proposition}%
\newtheorem{lemma}[theorem]{Lemma}%
\newtheorem*{lemma*}{Lemma}
\newtheorem*{corollary*}{Corollary}
\newtheorem{corollary}[theorem]{Corollary}
\newtheorem{remark}[theorem]{Remark}
\newtheorem{assumption}{Assumption}

\theoremstyle{thmstyletwo}%

\theoremstyle{thmstylethree}%

\begin{document}

\title[]{Article Title}

\title{A Multiparty Homomorphic Encryption Approach to Confidential Federated Kaplan–Meier Survival Analysis}
\author[1]{\fnm{Narasimha Raghavan} \sur{Veeraragavan}}\email{Narasimha.Raghavan.Veeraragavan@fhi.no}
\author[2]{\fnm{Svetlana} \sur{Boudko}}\email{svetlana@nr.no}
\author[1,3]{\fnm{Jan Franz} \sur{Nygård}}\email{Jan.Franz.Nygard@fhi.no}

\affil[1]{\orgdiv{Department of Registry Informatics, Cancer Registry of Norway}, \orgname{Norwegian Institute of Public Health}, \orgaddress{ \city{Oslo},  \country{Norway}}}
\affil[2]{ \orgname{Norwegian Computing Center}, \orgaddress{ \city{Oslo},  \country{Norway}}}
\affil[3]{\orgdiv{Department of Physics and Technology}, \orgname{The Arctic University of Norway}, \orgaddress{\city{Tromsø},  \country{Norway}}}


\abstract{
The proliferation of real-world health data enables multi-institutional survival studies, yet privacy constraints preclude centralizing sensitive records. We present a privacy-preserving federated Kaplan--Meier framework based on threshold CKKS (Cheon-Kim-Kim-Song) homomorphic encryption that supports approximate floating-point computation and encrypted aggregation of per-time-point counts while exposing only public outputs. Sites compute aligned at-risk and event tallies on a shared time grid and encrypt compact vectors; a coordinator aggregates ciphertexts; and a decryptor committee produces partial shares fused per block to recover aggregated plaintexts without releasing per-time-point tables. We prove correctness, stability, and slot-optimal vector packing, and derive scaling laws showing that communication grows linearly with the number of sites and predictably with the number of time points. Empirically, using synthetic breast-cancer data (N=60,000) distributed across 500 sites, encrypted federated curves match the pooled oracle to numerical precision. In contrast, plaintext protocols permit trivial reconstruction by subtraction; our threshold-gated design precludes this attack under the stated threat model, enabling high-fidelity survival estimation with predictable overhead and substantially reduced privacy risk.
}

\keywords{Federated Analytics, Multiparty Homomorphic Encryption, Kaplan-Meier, survival Analysis}

\maketitle

\section{Introduction}
Survival analysis is central to clinical and epidemiological research, yet privacy and governance constraints hinder centralizing sensitive records. Federated analytics (FA) alleviates this by keeping data in place while coordinating computation across institutions. We focus on the Kaplan--Meier (KM) estimator and adopt multiparty homomorphic encryption (HE) to protect intermediate statistics throughout the pipeline. A concise comparison of representative leveled fully homomorphic encryption (FHE) schemes appears in Table~\ref{tab:fhe-comparison}; we adopt CKKS with threshold decryption for its native support of approximate real arithmetic and efficient single-instruction multiple-data (SIMD) packing.
 
\begin{table}[htbp]
\centering
\footnotesize
\setlength{\tabcolsep}{3pt}
\renewcommand{\arraystretch}{1.2}
\begin{tabularx}{\linewidth}{@{}%
  >{\raggedright\arraybackslash}p{0.16\linewidth}%
  >{\raggedright\arraybackslash}p{0.14\linewidth}%
  >{\centering\arraybackslash}p{0.12\linewidth}%
  >{\raggedright\arraybackslash}p{0.13\linewidth}%
  >{\raggedright\arraybackslash}p{0.11\linewidth}%
  >{\centering\arraybackslash}p{0.09\linewidth}%
  >{\raggedright\arraybackslash}X@{}}
\toprule
\textbf{Scheme} & \textbf{Arithmetic} & \textbf{Precision} & \textbf{Bootstr.} & \textbf{Efficiency} & \textbf{Threshold} & \textbf{Use cases} \\
\midrule
BFV~\cite{cryptoeprint:2012/144,32009BZ}
& Integer (mod $q$) & Exact & Optional (leveled) & Moderate & Yes & Private queries; encrypted databases \\
BGV~\cite{BrakerskiGV12}
& Integer (mod $q$) & Exact & Optional & Efficient with batching & Yes & Complex integer analytics; encrypted ML \\
CKKS~\cite{ckks}
& Real/complex & Approximate & Optional & High (SIMD packing) & Yes & Floating-point analytics; encrypted ML \\
FHEW~\cite{10.1007/978-3-662-46800-5_24}
& Boolean gates & Exact & Required & Fast per gate & Limited & Lightweight logic gates \\
TFHE~\cite{cryptoeprint:2018/421}
& Boolean gates & Exact & Required (per gate) & Less efficient overall & Limited & Encrypted control flow; logic circuits \\
CGGI~\cite{10.1007/978-3-662-53887-6_1}
& Boolean gates & Exact & Required & Fast per gate & Limited & LUTs; conditional logic \\
\bottomrule
\end{tabularx}
\caption{Comparison of representative FHE schemes used in privacy-preserving analytics. ``Threshold'' indicates widely available threshold decryption in mainstream libraries.}
\label{tab:fhe-comparison}
\end{table}
 
\begin{table}[htbp]
\centering
\footnotesize
\setlength{\tabcolsep}{2.6pt}
\renewcommand{\arraystretch}{1.18}
\begin{tabularx}{\linewidth}{@{}%
  >{\raggedright\arraybackslash}p{0.22\linewidth}%
  >{\raggedright\arraybackslash}p{0.18\linewidth}%
  >{\centering\arraybackslash}p{0.14\linewidth}%
  >{\raggedright\arraybackslash}p{0.18\linewidth}%
  >{\centering\arraybackslash}p{0.12\linewidth}@{}}
\toprule
\textbf{Work} & \textbf{Primitive(s)} & \textbf{Arith.} & \textbf{Reveals $(n_t,d_t)$?} & \textbf{Thresh.} \\
\midrule
Froelicher et al.~\cite{Froelicher2021} & BGV/BFV (MHE) & Integer & Often yes (intermediate tables) & Yes \\
Geva et al.~\cite{geva2023collaborative} & CKKS (HE) & Floating (approx.) & Varies by release & Optional \\
Rahimian et al.~\cite{rahimian2024private} & DP (+FA) & Real via DP & No (noisy outputs only) & N/A \\
Veeraragavan et al.\ (2025)~\cite{veeraragavan2025} & DP (node-level) & Real via DP & No (private curves only) & N/A \\
\textbf{This work} & \textbf{CKKS + threshold} & \textbf{Floating (approx.)} & \textbf{No (outputs gated to $\hat S_{\mathrm{HE}}(t)$)} & \textbf{Yes} \\
\bottomrule
\end{tabularx}
\caption{Representative federated KM systems across three families: integer-HE (BGV/BFV), approximate-HE (CKKS), and differential privacy (DP). Our framework differs by (i) threshold CKKS with \emph{output gating} to the public survival curve $\hat S_{\mathrm{HE}}(t)$ (no per-time-point tables), (ii) KM-specific estimator-level guarantees, and (iii) closed-form \emph{communication} and \emph{computational} scaling laws.}
\label{tab:related-km-systems}
\end{table}

Existing federated KM systems have advanced practice, but four gaps remain that we address.
 
\textbf{A key privacy gap (Gap 1)} lies in subtraction-based reconstruction in plaintext two-round KM, which has rarely been quantified for overlapping, multi-site settings. We formalize and empirically demonstrate its practicality and then eliminate the channel by never releasing per-time-point tables.
 
\textbf{An estimator-level theory gap (Gap 2)} persists beyond cryptographic security proofs and microbenchmarks, with limited analysis of how approximate HE perturbs the \emph{KM estimator}. To address this, we provide (i) correctness equalities for both plain and HE-exact settings, (ii) a CKKS perturbation bound with a uniform-convergence corollary, and (iii) an identifiability result showing that publishing survival alone reveals hazard ratios but not per-time totals or any per-site split.
 
\textbf{A scaling-laws gap (Gap 3)} exists in the form of \emph{communication} and \emph{computational} scaling laws specific to KM. These include the effects of packing choices, the stepwise dependence on the size of the time grid, and the linear dependence on the number of sites and decryptors, none of which are stated in a way that enables cost prediction \emph{a priori}. We address this by deriving simple laws that align with observed behavior.
 
\textbf{A packing design gap (Gap 4)} remains despite the widespread use of CKKS packing, as we are not aware of a formal \emph{optimality} statement for KM's two streams \((d_t,n_t)\) under add-only aggregation. We prove a slot-count lower bound and show that interleaved co-packing is optimal under this model.
 
Federated Kaplan--Meier (KM) systems to date fall into three main lines: (i) \emph{integer} multiparty/threshold HE (e.g., BFV/BGV) exemplified by FAMHE~\cite{Froelicher2021}, which guarantees exact arithmetic but requires fixed-point encodings for real-valued workflows; (ii) \emph{approximate} HE with CKKS~\cite{geva2023collaborative}, which enables floating-point analytics but, to our knowledge, does not center on estimator-level guarantees for KM nor enforce output gating of per-time-point tables; and (iii) \emph{differential privacy} (DP) approaches~\cite{rahimian2024private,veeraragavan2025}, which are lightweight yet can degrade curve fidelity in sparse-event regimes. Secure multiparty computation (MPC) solutions~\cite{VONMALTITZ2021} offer strong cryptographic assurances but typically entail heavier interaction. 
 
Our work is distinct in four respects (Table~\ref{tab:related-km-systems}). 
First, we couple \textbf{threshold CKKS} with \textbf{output gating} and return only the public survival outputs $\hat S_{\mathrm{HE}}(t)$ (and optional bands), never the plaintext $(n_t,d_t)$ tables, thereby closing the subtraction-based reconstruction channel we quantify. When the combiner is operated by the decryptor committee (off-server), the coordinator observes only ciphertexts throughout; if the combiner is co-located with the coordinator, the coordinator may transiently view aggregated $(n_t,d_t)$, but the system policy \emph{does not release} these tables to sites and only publishes $\hat S_{\mathrm{HE}}(t)$ (and optional bands).
 
Second, we provide \textbf{estimator-level theory} specific to KM: plain/HE-exact equalities, a CKKS \emph{perturbation bound} with a uniform-convergence corollary, and an \emph{identifiability} result showing that publishing $\hat S_{\mathrm{HE}}(t)$ fixes hazard ratios but not counts. 
Third, we prove a \textbf{packing optimality} lemma for interleaving $(n_t,d_t)$ under add-only aggregation and derive \textbf{KM-specific communication and computational laws} that predict costs and explain the observed interleaving speedups. 
 
Fourth, we evaluate up to \textbf{$K{=}500$} sites on a large synthetic cohort (and up to 100 sites on NCCTG lung cancer in Section~S9 of the \SItext; see \SIFig{1}--\SIFig{3}), showing numerical indistinguishability from the pooled oracle under both packings and an empirically eliminated reconstruction risk under threshold decryption.
 
Compared with prior CKKS-based systems, our framework combines threshold CKKS, output gating, estimator-level guarantees for Kaplan--Meier analysis, and explicit communication and computational scaling laws.
 
Our contributions are:
 
1. A complete multiparty CKKS framework for federated KM with threshold decryption and output gating that returns only public outputs (the KM curve and optional bands).
 
2. We ensure estimator-level guarantees: (i) plain federated KM equals the pooled oracle; (ii) exact HE reproduces the oracle; (iii) an explicit CKKS perturbation bound with uniform convergence as noise vanishes; and (iv) an identifiability proposition (hazards determined, counts not).
 
3. A packing optimality lemma for interleaving \((n_t,d_t)\) and a \emph{computational scaling law} that separates server additions, committee share generation, fusion, and the final KM pass.
 
4. We consider closed-form \emph{communication scaling laws} for the two KM rounds and rules of thumb: encrypted-round bandwidth scales linearly with the number of sites and increases in predictable \emph{steps} with the number of time points; decryption-share traffic scales linearly with the decryptor committee size.
 
5. An empirical study on a large synthetic cohort (up to 500 sites) and an NCCTG lung dataset (see Section~S9 of the \SItext; \SIFig{1}--\SIFig{3}) showing numerical indistinguishability from the pooled estimator under both packings and quantifying the reconstruction risk in plaintext vs.\ its removal under threshold HE.
 
We employ a two-round scheme: \textbf{R1} discovers the global time grid by unifying local survival times (plaintext); \textbf{R2} uploads encrypted per-time-point counts for aggregation and threshold decryption. We consider two packing modes: \textbf{interleaved}, which co-packs \((n_t,d_t)\) pairs to minimize ciphertext count, and \textbf{separate}, which packs the two streams independently. Interleaving reduces bandwidth and decryption-share traffic whenever it fits more \((n_t,d_t)\) pairs into the same slot budget; exact formulas and scaling laws appear in Sections~\ref{sec:methods} and~\ref{sec:theory-results}. We evaluate deliberately non-IID (independent and identically distributed) client partitions via Dirichlet label splits ($\alpha=0.2$) and report all main metrics under both regimes, with figures cross-referenced in Section~\ref{sec:exp_results}.
 
Security and efficiency of HE schemes are well studied, but estimator-level KM results under CKKS are, to our knowledge, not formalized in prior federated work. We are not aware of (i) a KM-specific CKKS perturbation bound with a uniform-convergence corollary, (ii) an identifiability statement for publishing $\hat S_{\mathrm{HE}}(t)$ without counts in a federated setting, or (iii) a packing optimality proof tailored to KM's add-only aggregation. Prior systems typically report empirical throughput or asymptotic crypto costs; our scaling laws are end-to-end for the KM workload and align with measured trends. Packing and communication are key practical determinants of add-only Kaplan--Meier deployment because they govern ciphertext count, uplink and share traffic, and runtime at scale. These quantities also inform deployment choices such as packing layout, committee size, and expected bandwidth.
\section{Results}
\label{sec:exp_results}
 
We evaluate the federated Kaplan--Meier (KM) pipeline along three axes---privacy,
numerical fidelity, and system scalability---organized as six research questions
(RQ1--RQ6). We adopt the following notation:
$K$ (number of clients), $|T|$ (size of the event-time grid), $n$ (CKKS ring
degree; packing $B{=}n/2$ complex slots), $|Q|$ (total modulus bitlength),
$R$ (committee size), and $\theta$ (decryption threshold in a
$\theta$-of-$R$ scheme).
 
Unless stated otherwise, the pooled (centralized) KM curve serves as the
\emph{oracle} reference. For each metric, we report the mean and 95\%
confidence intervals (CIs) across repeated trials. An RQ is considered
\emph{met} when all pre-specified criteria in Section~\ref{sec:methods} are
satisfied across the swept hyperparameters.
 
The six questions address privacy (RQ1), numerical and statistical fidelity
(RQ2--RQ3), packing (RQ4), runtime (RQ5), and communication (RQ6), followed
by analytical results.
 
\begin{figure}[htbp]
  \centering
  \includegraphics[width=\linewidth]{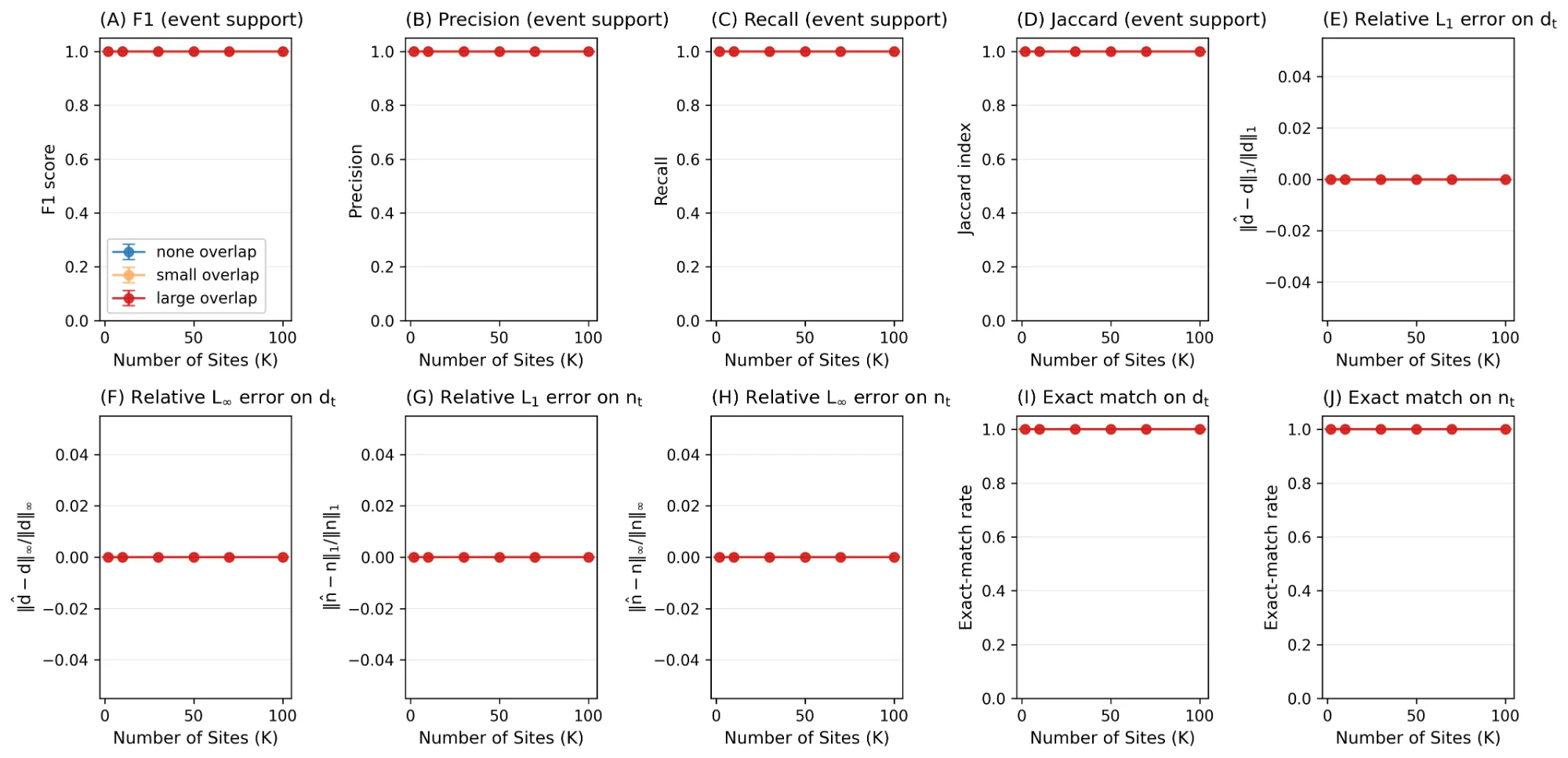}
  \caption{Reconstruction metrics vs.\ number of sites $K$ for IID data partitioning on the Synthetic Breast dataset. Ten metrics are shown: (A) F1 score for event support, (B) precision, (C) recall, (D) Jaccard index, (E) relative $L_1$ error on $d_t$, (F) relative $L_\infty$ error on $d_t$, (G) relative $L_1$ error on $n_t$, (H) relative $L_\infty$ error on $n_t$, (I) exact-match rate on $d_t$, and (J) exact-match rate on $n_t$. Curves show means with 95\% confidence intervals. Three overlap conditions are compared: none (blue), small (orange), and large (red). All conditions yielded identical results, demonstrating exact reconstruction regardless of overlap.}
  \label{fig:recon_grid_iid}
\end{figure}
 
\begin{figure}[htbp]
  \centering
  \includegraphics[width=\linewidth]{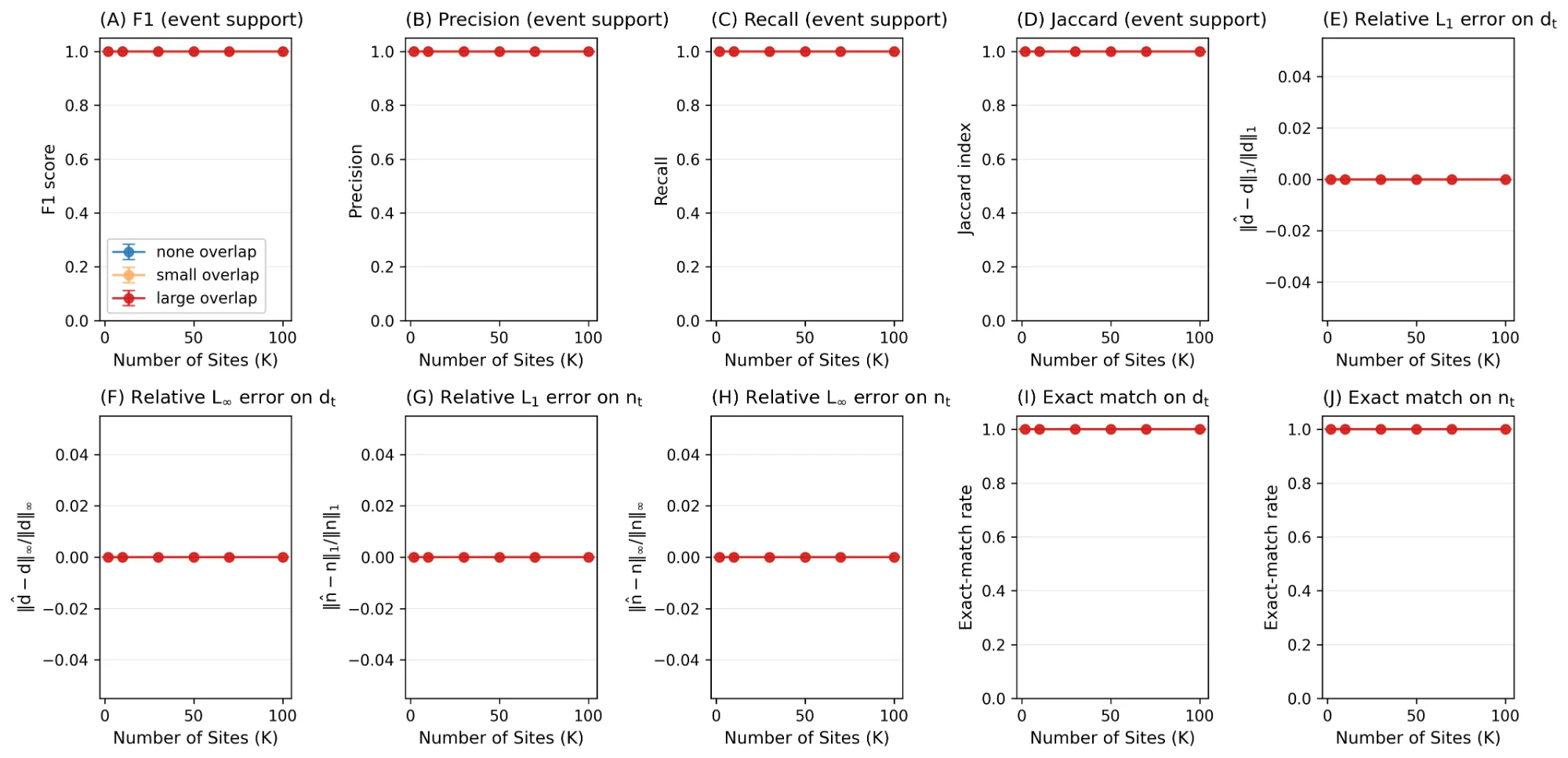}
  \caption{Reconstruction metrics vs.\ number of sites $K$ for label non-IID data partitioning (Dirichlet $\alpha=0.2$) on the Synthetic Breast dataset. The layout is identical to Figure~\ref{fig:recon_grid_iid} and shows results under label-based non-IID partitioning. All overlap conditions yielded identical results, demonstrating exact reconstruction even under heterogeneous data distributions.}
  \label{fig:recon_grid_label}
\end{figure}
 
\begin{figure}[htbp]
  \centering
  \includegraphics[width=\linewidth]{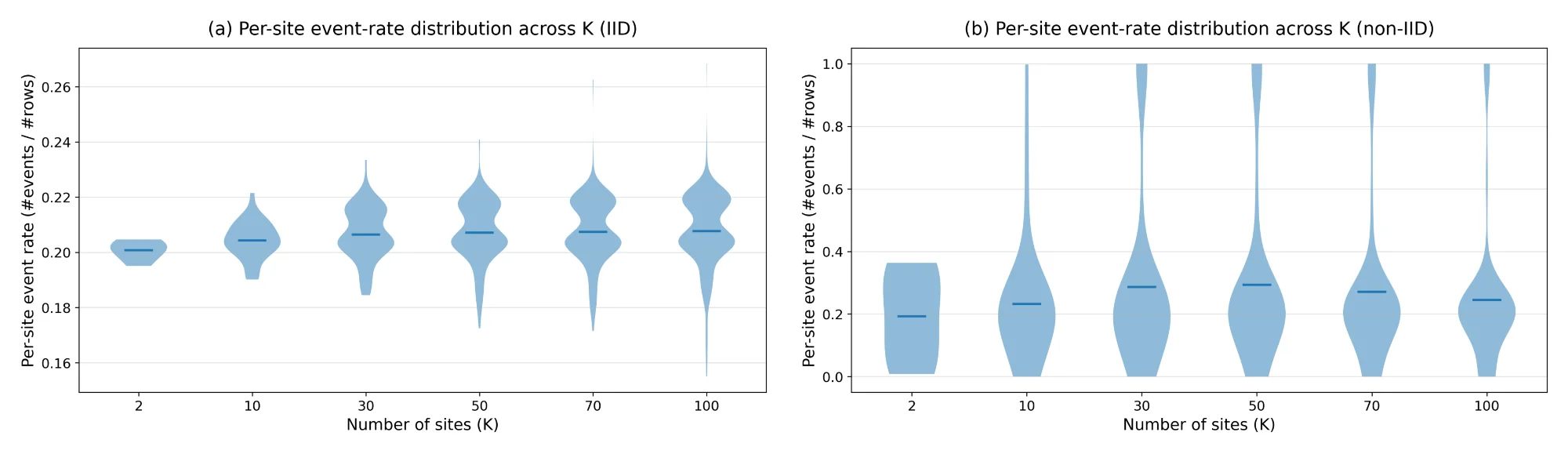}
  \caption{Per-site event-rate distributions for the Synthetic Breast dataset. Violin plots show the distribution of event rates (\#events/\#rows) across sites under (a) IID partitioning and (b) label non-IID partitioning. Wider, skewed violins under non-IID indicate heterogeneous site contributions but do not prevent exact residualization when plaintext per-time-point counts are disclosed.}
  \label{fig:recon_eventrates}
\end{figure}
 
\emph{Can a site practically reconstruct other sites' contributions when per-time-point counts are exchanged in plaintext, and does our CKKS+threshold design remove that attack surface?} (\textbf{RQ1})
 
\noindent\textit{Setup:}
We simulate a federated KM setting with $K \in \{2,10,30,50,70,100\}$ sites under three overlap regimes (\textit{none}, \textit{small}, \textit{large}) and two partitioning styles: IID and non-IID label splits (Dirichlet allocation with $\alpha{=}0.2$ over event labels to match Section~\ref{subsubsec:datapartition}). In the plaintext baseline, as commonly implemented in two-round KM, the coordinator broadcasts aggregated per-time-point counts $(n_t^{(\mathrm{Fed})}, d_t^{(\mathrm{Fed})})$ to all sites, or sites can independently inspect them at the coordinator. An honest-but-curious site $A$ then computes the residual
\[
n_t^{(\mathrm{others})} = n_t^{(\mathrm{Fed})} - n_t^{(A)}, \qquad
d_t^{(\mathrm{others})} = d_t^{(\mathrm{Fed})} - d_t^{(A)},
\]
thereby attempting to recover the combined contributions of the remaining sites.
 
We evaluate both per-time-point event counts $d_t$ and numbers at risk $n_t$
across $K$ and overlap regimes, reporting two consolidated $2{\times}5$ grids
(mean with 95\% CIs): one for IID (Figure~\ref{fig:recon_grid_iid}) and one
for non-IID label splits (Figure~\ref{fig:recon_grid_label}). Panels (A)--(D)
summarize support identification for event times (F1, precision, recall, and
Jaccard on $\mathcal{S}=\{t : d_t>0\}$; higher is \emph{worse} for privacy).
Panels (E)--(H) report relative errors (lower is better):
\[
\frac{\lVert \hat{\mathbf{d}} - \mathbf{d}\rVert_{1}}{\lVert \mathbf{d}\rVert_{1}+\varepsilon_0},\;
\frac{\lVert \hat{\mathbf{d}} - \mathbf{d}\rVert_{\infty}}{\lVert \mathbf{d}\rVert_{\infty}+\varepsilon_0},\;
\frac{\lVert \hat{\mathbf{n}} - \mathbf{n}\rVert_{1}}{\lVert \mathbf{n}\rVert_{1}+\varepsilon_0},\;
\frac{\lVert \hat{\mathbf{n}} - \mathbf{n}\rVert_{\infty}}{\lVert \mathbf{n}\rVert_{\infty}+\varepsilon_0},
\]
with $\varepsilon_0=10^{-12}$ to avoid division by zero and norms restricted to
indices where denominators are positive. Panels (I)--(J) show exact-match rates
\[
\frac{1}{|T|}\sum_t \mathbf{1}\{\hat d_t=d_t\},
\qquad
\frac{1}{|T|}\sum_t \mathbf{1}\{\hat n_t=n_t\},
\]
where higher values are worse for privacy. To contextualize heterogeneity, we
also plot per-client event-rate distributions under IID vs.\ non-IID (two-panel
figure, Figure~\ref{fig:recon_eventrates}).
 
Across all $K$ and overlap regimes, both IID and non-IID grids exhibit
saturated support metrics (A)--(D) at~1.0, vanishing relative errors (E)--(H)
near~0, and exact-match rates (I)--(J) at~1.0. In other words, once
$(n_t^{(\mathrm{Fed})}, d_t^{(\mathrm{Fed})})$ are disclosed in plaintext,
subtracting a site's own $(n_t^{(A)}, d_t^{(A)})$ exactly recovers the
combined ``others'' $(n_t^{(\mathrm{others})}, d_t^{(\mathrm{others})})$ on the
common time grid, independent of $K$, overlap, or partition style. The flat
lines with tight, often invisible, error bars reflect that this is an algebraic
identity given exact counts and synchronized times.
 
The plaintext protocol reveals precisely the information the attacker needs. If
the federated aggregator broadcasts $(n_t^{(\mathrm{Fed})},
d_t^{(\mathrm{Fed})})$, as is typical in unencrypted two-round KM, any site
can compute the residual and thereby learn the groupwise contributions of the
remaining sites; with $K{=}2$ this equals the other site exactly. This
conclusion does \emph{not} rely on IID assumptions. Non-IID label skew or
overlap only change how counts are distributed across $t$, not the identity
$x^{(\mathrm{others})} = x^{(\mathrm{Fed})} - x^{(A)}$ for
$x \in \{n,d\}$.
 
\noindent\textit{Role of IID vs.\ non-IID and overlap:}
The event-rate figure (Figure~\ref{fig:recon_eventrates}) shows that IID splits
produce tightly clustered per-client event shares, whereas non-IID splits
produce broader, skewed distributions. These differences matter for
\emph{inference from the survival curve alone}, but they do not mitigate the
plaintext subtraction attack, which operates directly on disclosed counts.
Overlap increases redundancy in shared event times, but again does not impede
exact residualization once totals are visible.
 
\noindent\textit{Impact of CKKS+threshold:}
Our CKKS-based design removes the subtraction channel by withholding
intermediate plaintext counts from sites and publishing only $\hat S_{\mathrm{HE}}(t)$, plus
optional derived outputs such as confidence bands. Sites send encrypted packed
streams; decryption requires a threshold of parties; and when the combiner runs
off-server within the decryptor committee, the coordinator observes only
ciphertexts. If the combiner is co-located with the coordinator, the
coordinator may transiently view aggregated $(n_t,d_t)$ to evaluate
$\hat S_{\mathrm{HE}}(t)$, but per-time-point tables are not disclosed to sites. Because
the attacker lacks $(n_t^{(\mathrm{Fed})}, d_t^{(\mathrm{Fed})})$ in plaintext,
the subtraction step is unavailable, and reconstructing others' counts from
$\hat S_{\mathrm{HE}}(t)$ alone is under-determined without additional side information.
 
\noindent
When per-time-point counts are exchanged in plaintext, reconstruction of other
sites' contributions is trivially exact in our experiments, with perfect
support and exact matches across all $K$ and partition styles. Encrypting
those counts with CKKS and revealing only the final survival outputs removes
the subtraction channel and thus the demonstrated privacy risk.
 
\begin{figure}[htbp]
  \centering
  \includegraphics[width=0.9\linewidth]{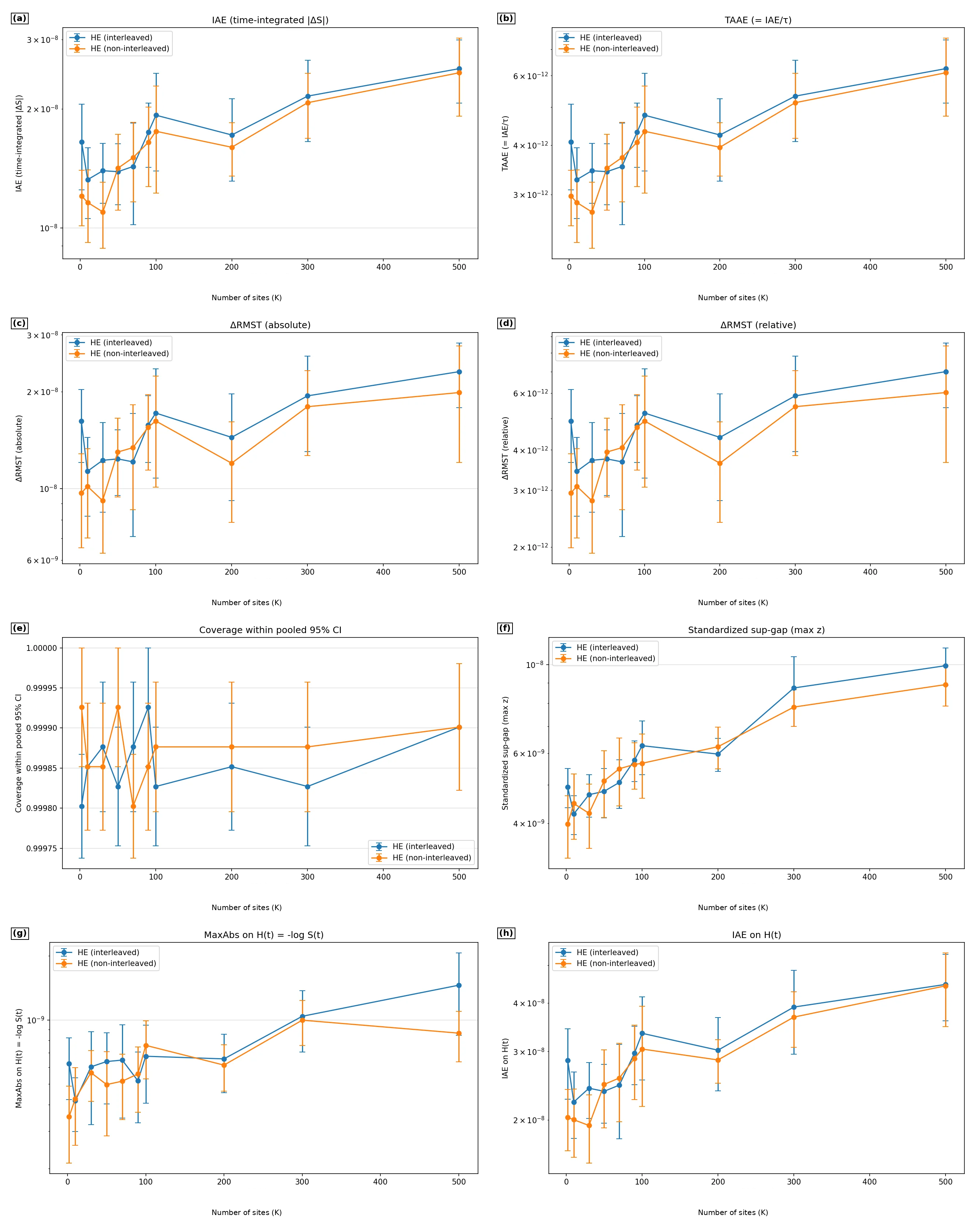}
  \caption{Numerical fidelity metrics for HE-federated Kaplan--Meier vs.\ pooled oracle across number of sites $K$ on the Synthetic Breast dataset. Eight metrics are shown: (a) integrated absolute error on survival function $\mathrm{IAE}(S)$, (b) time-averaged absolute error $\mathrm{TAAE}(S)$, (c) absolute difference in restricted mean survival time $|\Delta\!\mathrm{RMST}|$, (d) relative RMST difference $\Delta\!\mathrm{RMST}_{\mathrm{rel}}$, (e) coverage probability of 95\% confidence intervals, (f) standardized supremum gap $\mathrm{supZ}$ measuring maximum standardized deviation, (g) maximum absolute difference in cumulative hazard $\max_t|H_{\mathrm{HE}}-H_{\mathrm{oracle}}|$, and (h) integrated absolute error on cumulative hazard $\mathrm{IAE}_H$. Each panel compares interleaved (blue) and non-interleaved/separate (orange) packing strategies with means and 95\% confidence intervals. The two packing strategies are numerically indistinguishable across all metrics as $K$ increases.}
  \label{fig:nf-grid}
\end{figure}
 
\begin{figure}[htbp]
  \centering
  \includegraphics[width=\linewidth]{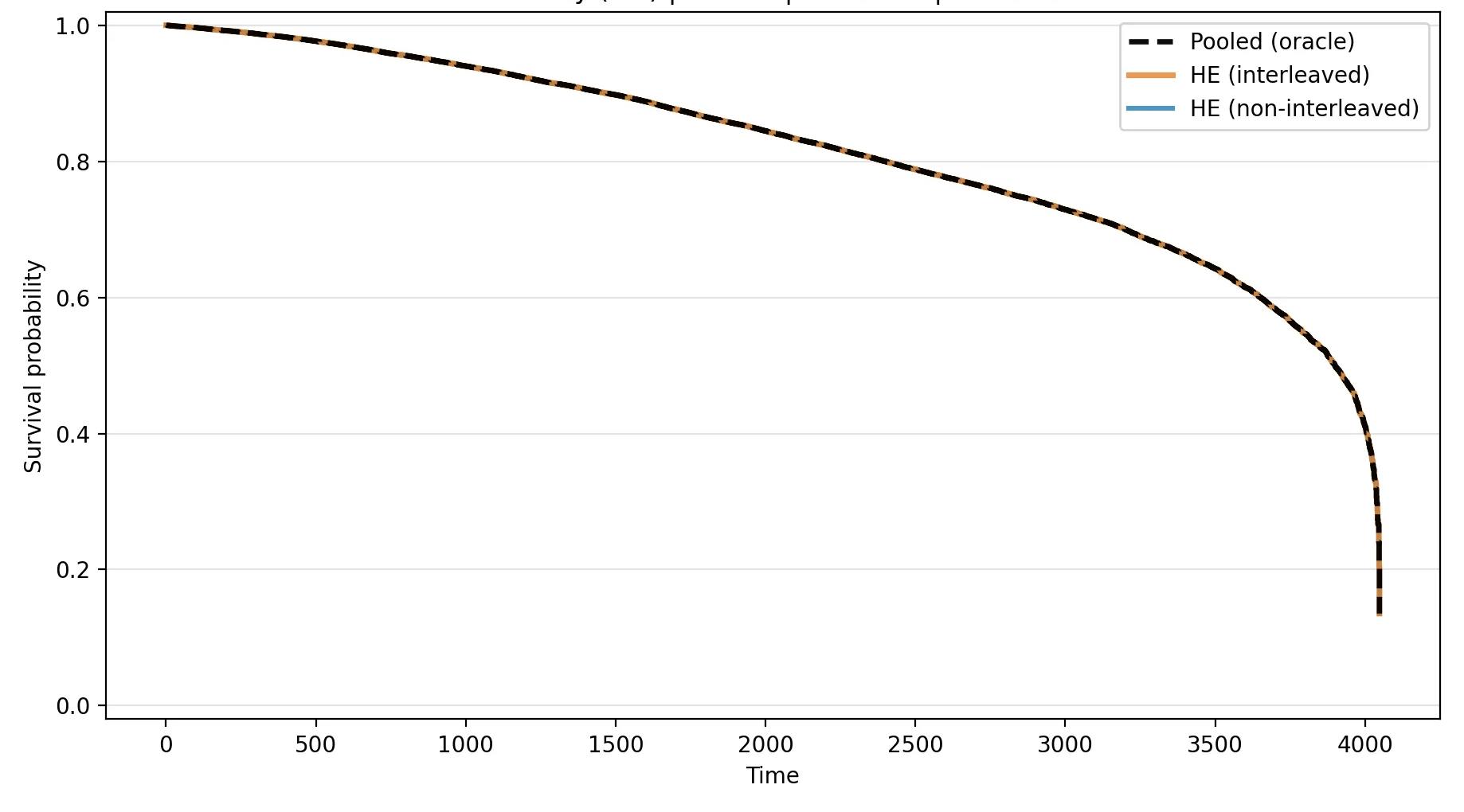}
  \caption{Kaplan--Meier survival curves for $K=500$ sites demonstrating visual indistinguishability between the pooled oracle (dashed black line), HE-federated with interleaved packing (orange), and HE-federated with non-interleaved/separate packing (blue). All three curves overlap almost perfectly, confirming that homomorphic encryption preserves statistical accuracy even at large scale.}
  \label{fig:visual}
\end{figure}
 
\emph{Do KM curves computed via our CKKS-based federated pipeline agree with the pooled (oracle) estimator as $K$ grows?} (\textbf{RQ2})
 
\noindent\textit{Setup:}
For each \(K \in \{2,10,30,50,70,90,100,200,300,500\}\), we run the full
protocol twice: once with \emph{interleaved} packing and once with
\emph{separate} packing. The oracle curve is computed on the pooled plaintext
data under identical time grids. We report means with 95\% CIs across 10
repeated runs; interleaved and separate lines are overlaid in each panel.
 
\noindent\textit{Metrics and equivalence tolerances (decision logic):}
We pre-specified equivalence metrics to assess numerical agreement between the
HE pipeline and the pooled oracle across survival, RMST, band, and hazard
spaces (Table~\ref{tab:equivalence-metrics}). Table~\ref{tab:equivalence-metrics}
lists each metric, its formula, and the pass threshold used to declare ``no
practical difference.'' Continuous-time definitions are used unless noted;
thresholds are chosen to sit well above floating-point epsilon while remaining
below CKKS rounding noise for our parameterization.
 
\noindent\textit{Findings:}
Figure~\ref{fig:nf-grid} summarizes eight diagnostics in a $2{\times}4$ grid.
Across all $K$, both HE configurations track the pooled oracle to within the
numerical noise floor. Visually (Figure~\ref{fig:visual}), survival curves and
bands are indistinguishable; numerically, all equivalence tests, including
TAAE and IAE on $S$, absolute and relative RMST gaps, sup-$Z$ and coverage,
and hazard-based metrics, pass for every $K$ and both packings. Packing choice
affects efficiency, not accuracy: interleaving reduces communication and
runtime, while fidelity remains unchanged.
 
\begin{itemize}
  \item \emph{Survival.} \(\mathrm{IAE}(S)\) remains at the \(10^{-8}\) scale
  (typically \(1.3\times10^{-8}\) to \(2.6\times10^{-8}\));
  \(\mathrm{TAAE}(S)\) is at \(10^{-12}\) scale (typically
  \(3\times10^{-12}\) to \(6\times10^{-12}\)). Interleaved and separate curves
  overlap within noise for every \(K\).
 
  \item \emph{RMST.} Absolute gaps are \(\mathcal{O}(10^{-8})\) and relative
  gaps \(\mathcal{O}(10^{-12})\) to \(\mathcal{O}(10^{-8})\), indicating machine-precision
  agreement with no practical \(K\)-dependence.
 
  \item \emph{Bands.} Coverage is essentially 1.0 (0.99975 to 0.99993) and
  \(\mathrm{supZ}\) is \(10^{-10}\) to \(10^{-9}\) across \(K\); all runs
  satisfy the pre-specified \(\mathrm{supZ}\) criterion.
 
  \item \emph{Hazard.} \(\max_t |H_{\mathrm{HE}}-H_{\mathrm{oracle}}|\) is
  \(10^{-10}\) to \(10^{-9}\) and \(\mathrm{IAE}_H\) is
  \(2\times10^{-8}\) to \(4.5\times10^{-8}\), with negligible sensitivity to
  \(K\) or packing.
\end{itemize}
 
\noindent
Across all \(K\) and both packings, the HE-federated KM curves \emph{match} the
pooled oracle to numerical precision in survival, RMST, band coverage,
supremum gap, and hazard, indicating that encryption and packing have no
observable effect on statistical fidelity.
 
\emph{Are the HE curves statistically indistinguishable from the pooled estimator?} (\textbf{RQ3})
 
We assess statistical equivalence using the same setup and figure as in
Numerical Fidelity, namely the coverage and standardized sup-gap panels in
Figure~\ref{fig:nf-grid}. Specifically, we report: (i) empirical 95\%
coverage of the pooled Greenwood bands evaluated at $\hat S_{\mathrm{HE}}(t)$,
and (ii) the standardized sup-gap
\[
\sup_{t}\frac{\big|\hat S_{\mathrm{HE}}(t)-\hat S_{\mathrm{oracle}}(t)\big|}
{\widehat{\mathrm{SE}}\!\left[\hat S_{\mathrm{oracle}}(t)\right]}.
\]
We also evaluate two pre-specified equivalence tests: total
time-averaged absolute error within its threshold
(Table~\ref{tab:equivalence-metrics}) and relative RMST gap within its
threshold (Table~\ref{tab:equivalence-metrics}).
 
\noindent\textit{Findings:}
Across all $K$ and for both CKKS packings (interleaved and separate), the
pooled 95\% Greenwood bands \emph{contain} $\hat S_{\mathrm{HE}}$ essentially
everywhere (mean containment $>0.999$), and the standardized sup-gap is
numerically $\approx 0$ (median $<10^{-8}$). Both equivalence checks pass for
100\% of replicates at every $K$. Interleaving does not change inference:
HE-interleaved and HE-separate overlay within Monte Carlo noise.
 
\noindent
Within our pre-specified tolerances, $\hat S_{\mathrm{HE}}(t)$ is
statistically indistinguishable from the pooled oracle
$\hat S_{\mathrm{oracle}}(t)$ for all $K$ and packing choices; the HE pipeline
preserves the statistical behavior of the KM estimator.
 
\emph{Do interleaved vs.\ non-interleaved CKKS packings change accuracy or only efficiency/communication?} (\textbf{RQ4})
 
\noindent\textit{Setup:}
We compare two encodings of the per-time-point statistics $(d_t,n_t)$ under
identical CKKS parameters and the same federated add--decrypt workflow:
(i) \emph{interleaved}, where one ciphertext stream co-packs
$[n_1,d_1,n_2,d_2,\ldots]$, and (ii) \emph{separate}, where distinct
ciphertext streams pack $[d_1,\ldots]$ and $[n_1,\ldots]$. Any accuracy
differences therefore arise solely from packing layout. Figure~\ref{fig:nf-grid}
compares both packings under identical CKKS parameters.
 
\noindent\textit{Findings:}
Across all $K$ tested (2--500), interleaved and separate packings are
\emph{indistinguishable in accuracy} within numerical noise: the two curves
overlap in every panel of Figure~\ref{fig:nf-grid}, their point estimates
track across $K$, and their 95\% CIs are nearly identical. IAE and TAAE remain
at machine-precision levels in both modes; absolute and relative RMST
differences are negligible; empirical coverage stays essentially $1.0$; and
the standardized sup-gap is $\approx 0$. All pre-specified equivalence checks
pass for both packings at every $K$.
 
\noindent
Packing serves as an efficiency knob rather than an accuracy knob, since
interleaving preserves numerical and statistical fidelity to the pooled oracle
just as well as separate packing (Figure~\ref{fig:nf-grid}). Choose
interleaving to gain efficiency, not to trade off accuracy.
 
\begin{figure}[htbp]
  \centering
  \includegraphics[width=0.8\linewidth]{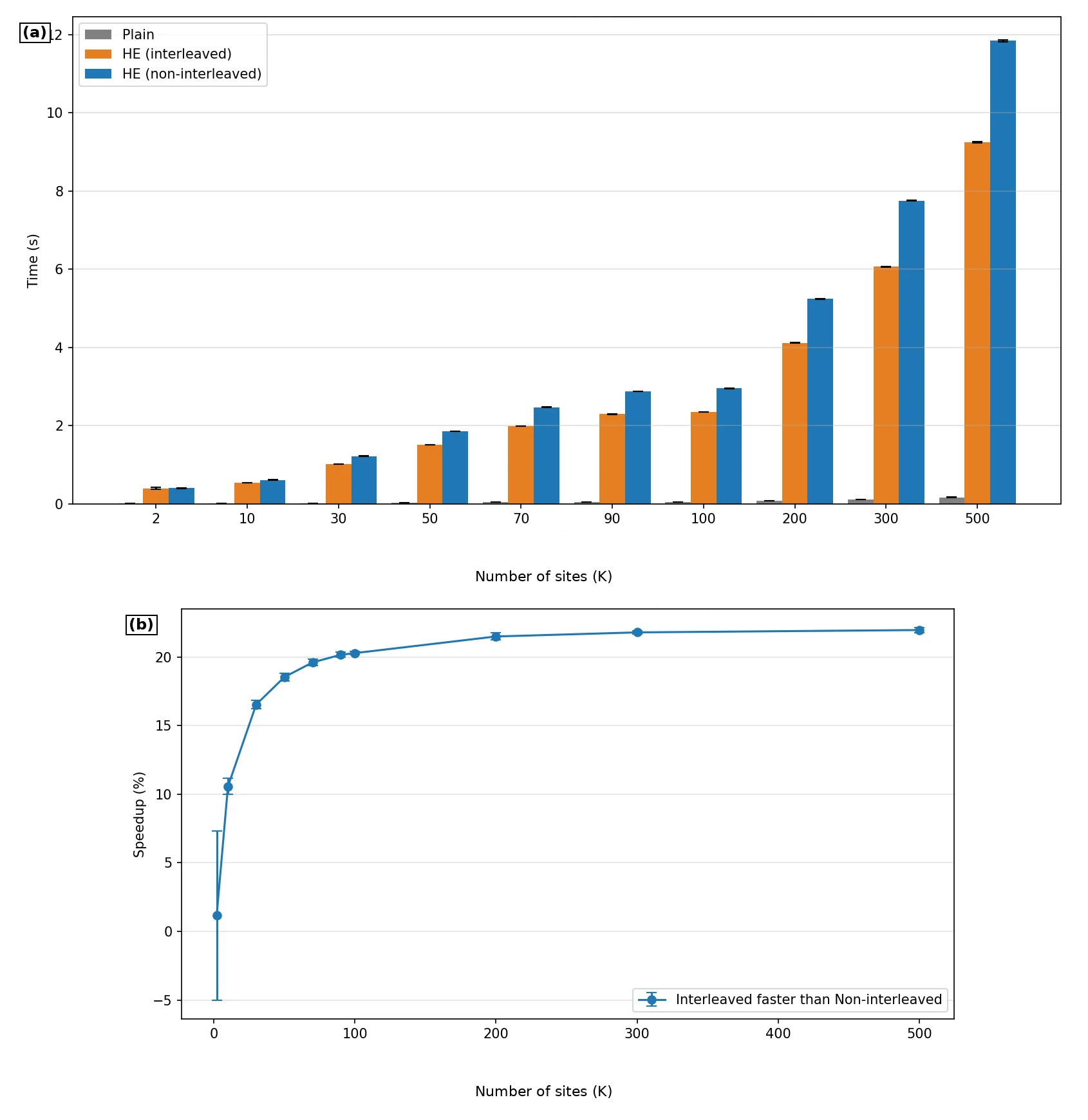}
  \caption{Computational scalability of HE-federated Kaplan--Meier on the Synthetic Breast dataset. (a) End-to-end runtime vs.\ number of sites $K$, showing approximately linear growth. Plain pooled oracle (gray) serves as a baseline reference, while HE with interleaved packing (orange) and non-interleaved/separate packing (blue) are shown with means and 95\% confidence intervals. (b) Percentage speedup of interleaved over non-interleaved packing demonstrates increasing computational gains as $K$ rises, plateauing at approximately 22\% improvement for large-scale federations.}
  \label{fig:comp-scale}
\end{figure}
 
\emph{How does end-to-end runtime scale with the number of sites $K$ for plain vs.\ HE modes?} (\textbf{RQ5})
 
\noindent\textit{Setup:}
All implementation measurements in this subsection use the deployed alignment
grid $T_{\mathrm{all}}$, so the packed length is $L=|T_{\mathrm{all}}|$.
We measure wall-clock time per federated round from the moment clients begin
encrypting their local per-time-point counts through aggregator homomorphic
accumulation and final threshold decryption. We compare: (i) a \emph{Plain}
pooled oracle (no encryption), (ii) HE with \emph{interleaved} CKKS packing
(co-packed slots), and (iii) HE with \emph{separate} packing (distinct
streams). All CKKS parameters, data, and code paths are identical across
conditions except for packing. We sweep
$K\in\{2,10,30,50,70,90,100,200,300,500\}$ and report means with 95\% CIs.
 
\noindent\textit{Metric:}
End-to-end wall-clock runtime (seconds) is used as the evaluation metric
(Figure~\ref{fig:comp-scale}a). We also report the percentage speedup of
interleaved over separate packing (Figure~\ref{fig:comp-scale}b),
\[
\text{speedup}~[\%] \;=\; 100\cdot\frac{W_{\text{sep}}-W_{\text{int}}}{W_{\text{sep}}}\, ,
\]
where $W_{\text{sep}}$ and $W_{\text{int}}$ denote the mean wall-clock
runtimes under separate and interleaved packing, respectively.
 
\noindent\textit{Findings:}
Figure~\ref{fig:comp-scale} places runtime trends side-by-side. Panel~(a)
plots mean runtime vs.\ $K$ for \textit{Plain}, \textit{HE-Interleaved}, and
\textit{HE-Separate} with 95\% CIs; Panel~(b) plots the percentage speedup of
interleaved over separate vs.\ $K$. Curves are monotone and roughly linear for
the two HE modes, with a widening gap in favor of interleaving as $K$ grows.
 
\begin{itemize}
  \item \emph{Near-linear scaling in $K$ for HE.} HE runtimes grow
  approximately linearly with the number of sites, consistent with an $O(K)$
  aggregator-addition cost per ciphertext. For interleaved packing, mean time
  rises from $\approx 0.39$s at $K{=}2$ to $\approx 9.24$s at $K{=}500$;
  separate packing rises from $\approx 0.40$s to $\approx 11.84$s over the
  same range. CIs are tight across all $K$.
 
  \item \emph{Interleaving accelerates compute without affecting accuracy.}
  Interleaved packing is consistently faster, with the speedup \emph{increasing}
  with $K$: about $1$--$2$\% at $K{=}2$, approximately $10.6$\% at $K{=}10$,
  approximately $16.5$\% at $K{=}30$, and approximately $22$\% by
  $K{=}300$--$500$. This matches the reduction in ciphertext counts and
  additions per round when $(n_t,d_t)$ are co-packed (see
  Corollary~\ref{cor:interleaving-adv}).
 
  \item \emph{Pooled-oracle baseline.} The pooled-oracle baseline
  remains in the millisecond regime (for example, approximately $5$--$160$\,ms
  from $K{=}2$ to $K{=}500$) since it omits encryption, transmission, and
  decryption. It serves as a lower bound rather than a deployable
  privacy-preserving alternative.
\end{itemize}
 
\noindent
HE-federated KM is computationally predictable: end-to-end runtime scales
roughly linearly with the number of sites, and interleaved packing yields a
consistent $10$--$22$\% speedup at moderate-to-large $K$ while maintaining
identical accuracy (see Proposition~\ref{prop:comp-law}).
 
\begin{figure}[htbp]
  \centering
  \includegraphics[width=0.95\linewidth]{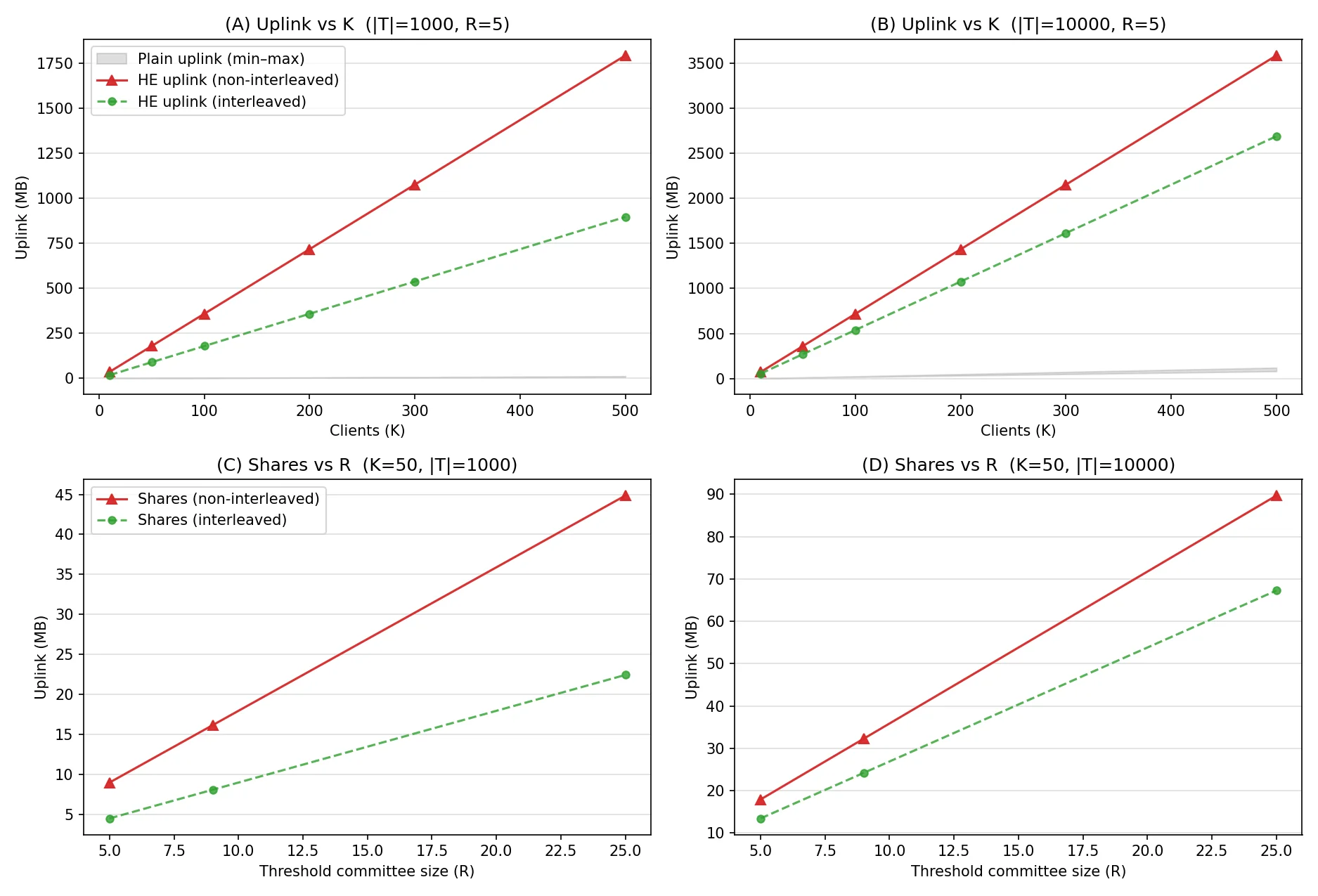}
  \caption{Communication scaling. (A) Uplink vs.\ $K$ with $|T|{=}1000$ and $R{=}5$; (B) Uplink vs.\ $K$ with $|T|{=}10000$ and $R{=}5$; (C) decryption-share uplink vs.\ $R$ with $K{=}50$ and $|T|{=}1000$; (D) decryption-share uplink vs.\ $R$ with $K{=}50$ and $|T|{=}10000$.}
  \label{fig:comm-grid-1}
\end{figure}
 
\begin{figure}[htbp]
  \centering
  \includegraphics[width=0.95\linewidth]{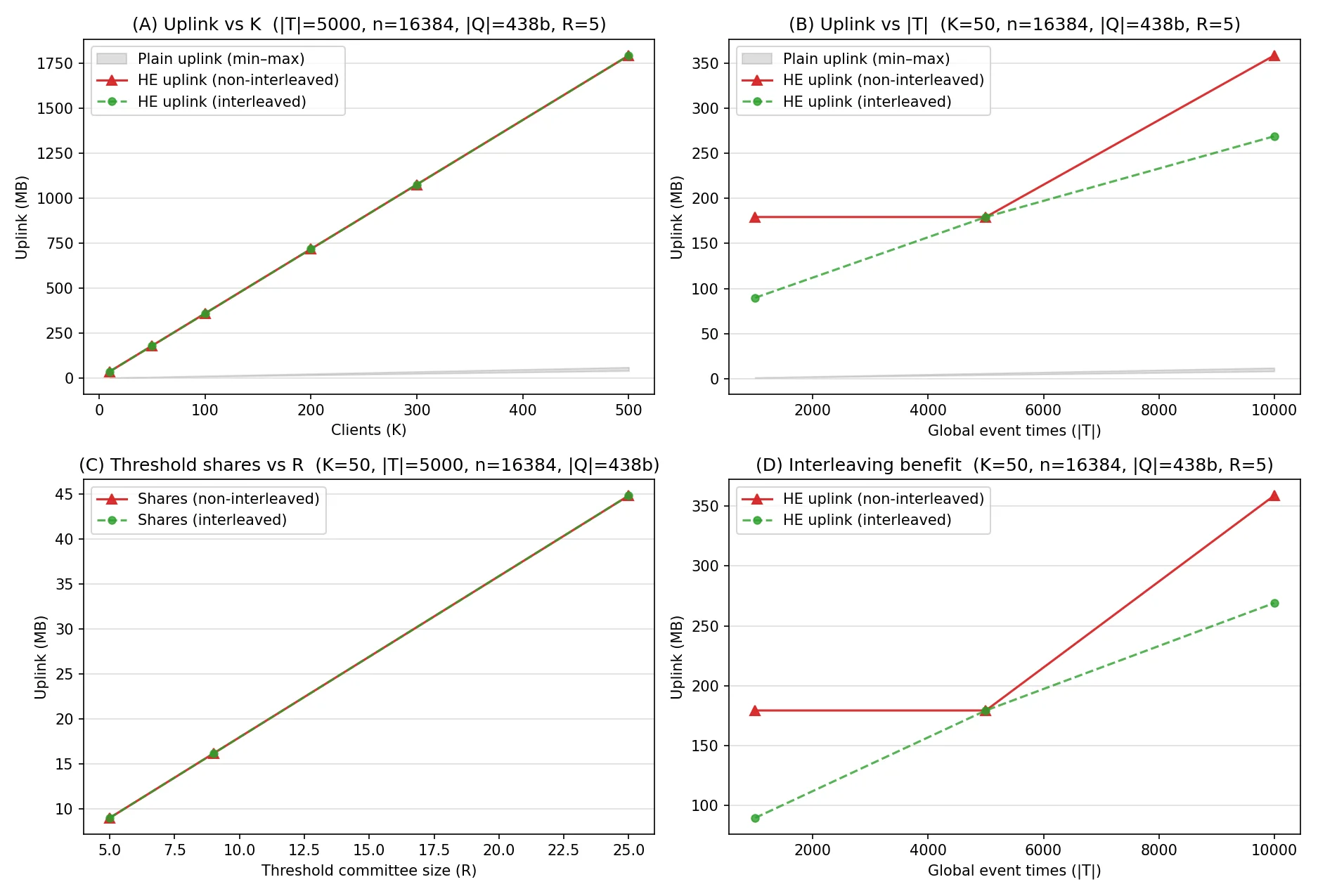}
  \caption{Communication scaling under CKKS packing. (A) HE uplink vs.\ $K$ at fixed $|T|{=}5000$ and $R{=}5$; (B) HE uplink vs.\ $|T|$ at fixed $K{=}50$ and $R{=}5$; (C) decryption-share uplink vs.\ $R$ at fixed $K{=}50$ and $|T|{=}5000$; (D) interleaving benefit shown as HE uplink for non-interleaved and interleaved packing vs.\ $|T|$, at fixed $K{=}50$ and $R{=}5$.}
  \label{fig:comm-grid-2}
\end{figure}
 
\begin{figure}[htbp]
  \centering
  \includegraphics[width=0.95\linewidth]{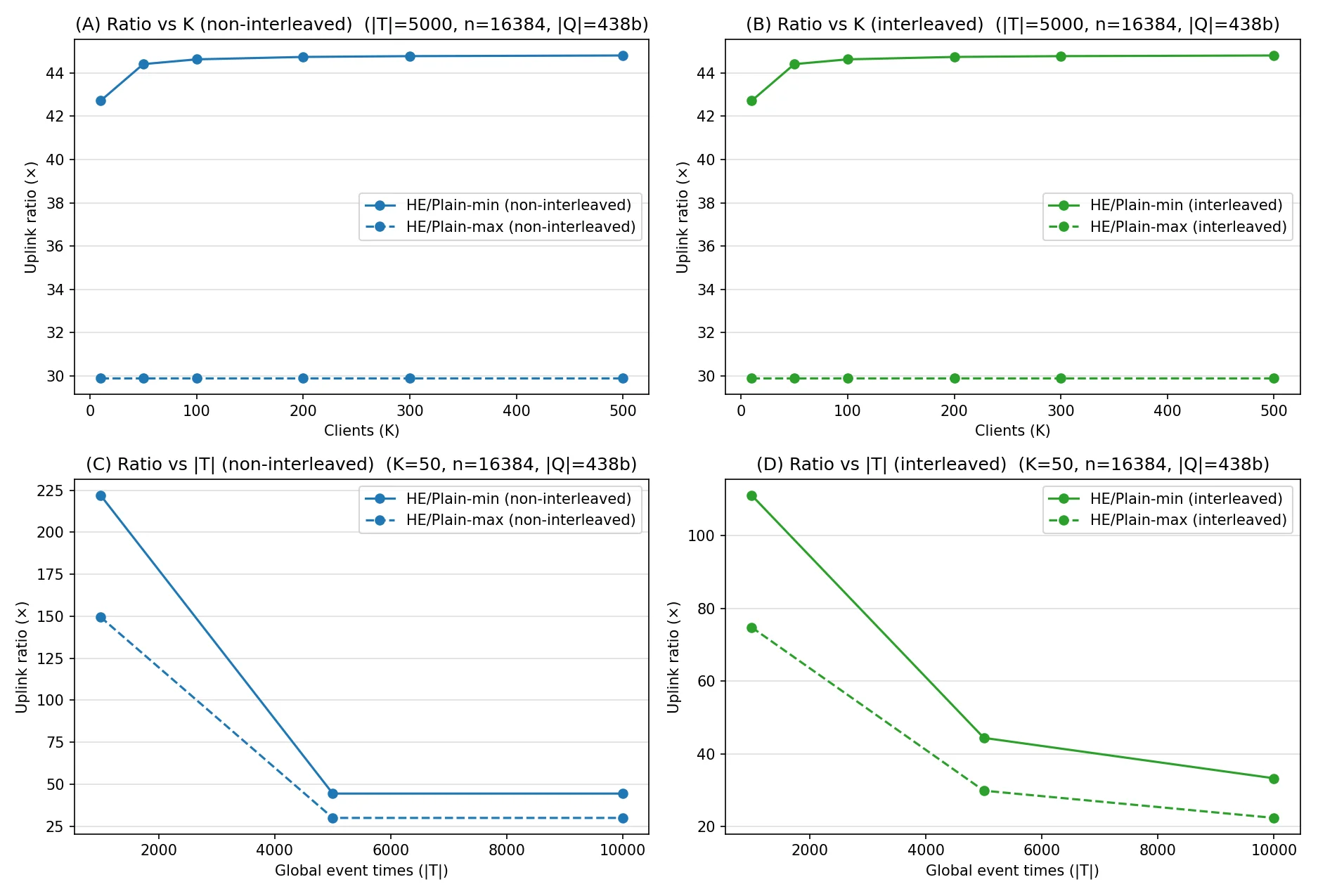}
  \caption{HE-versus-plaintext uplink ratio. (A) Non-interleaved HE/plain uplink ratio vs.\ $K$ at fixed $|T|{=}5000$; (B) interleaved HE/plain uplink ratio vs.\ $K$ at fixed $|T|{=}5000$; (C) non-interleaved HE/plain uplink ratio vs.\ $|T|$ at fixed $K{=}50$; (D) interleaved HE/plain uplink ratio vs.\ $|T|$ at fixed $K{=}50$. Solid lines use the plaintext minimum-uplink denominator and dashed lines use the plaintext maximum-uplink denominator.}
  \label{fig:comm-grid-3}
\end{figure}
 
\emph{How do uplink/downlink costs scale with $K$, the number of event-time points $|T|$, and CKKS packing; and how large is the interleaving benefit?} (\textbf{RQ6})
 
\noindent\textit{Setup:}
This subsection reports formula-based communication simulations rather than
measurements from the current implementation. Unless stated otherwise, these
simulations instantiate the generic formulas with $L=|T|$ to model an
event-time-only packing variant. By contrast, the deployed pipeline packs
over $T_{\mathrm{all}}$, i.e., $L=|T_{\mathrm{all}}|$.
 
We evaluate communication using formula-based simulations implemented in
our experimental evaluation. Each round, every site encrypts and uploads
aligned per-time-point vectors over a packed grid of length $L$ (in the
communication simulations presented here, $L=|T|$);
the aggregator sums and triggers threshold decryption with $R$ decryptors.
We compare \emph{separate} packing (two ciphertext streams for $d_\bullet$
and $n_\bullet$) versus \emph{interleaved} packing (co-packing $(n_t,d_t)$
across slots). All runs use the same CKKS parameters ($n{=}16384$,
$|Q|{\approx}438$ bits), giving $B{=}8192$ complex slots and ciphertext size
$\texttt{ct\_bytes}{\approx}1{,}794{,}048$ bytes and partial-share size
$\texttt{share\_bytes}{\approx}897{,}024$ bytes.
 
All communication quantities are parameterized by the number of event-time
points $|T|$, consistent with the simulation code used to generate
Figures~\ref{fig:comm-grid-1}--\ref{fig:comm-grid-3}.
 
\noindent\textit{Metrics and expectations:}
Let $L := |T|$ denote the packed vector length in the communication simulations. The number of ciphertexts per
client is
\[
M_{\mathrm{int}}=\Big\lceil\frac{2L}{B}\Big\rceil, \qquad
M_{\mathrm{sep}}=2\Big\lceil\frac{L}{B}\Big\rceil.
\]
 
Per-client uplink is $M\times\texttt{ct\_bytes}$, so total
uplink scales linearly in $K$. Decryption-share traffic scales as
\[
R \times M \times \texttt{share\_bytes}.
\]
 
The \emph{interleaving benefit} is defined as
\[
\text{benefit}=\frac{M_{\mathrm{sep}}}{M_{\mathrm{int}}},
\]
which equals $2\times$ when $L\le B/2$, $1\times$ when both packings use the
same number of ciphertexts, and intermediate values such as $4/3\times$ when
$(M_{\mathrm{sep}},M_{\mathrm{int}})=(4,3)$.
 
We also report the \emph{HE/Plain uplink ratio}. The plaintext protocol uploads
aligned counts over $T$ using $2K|T|$ scalars, whereas the HE protocol uploads
ciphertexts. This ratio is therefore independent of $K$ but depends on $|T|$
through the stepwise behavior of $M$.
 
\noindent\textit{Findings:}
Figures~\ref{fig:comm-grid-1}--\ref{fig:comm-grid-3} show communication scaling
as functions of $K$, $|T|$, and $R$.
 
\begin{itemize}
  \item \emph{Linear scaling in $K$.} For fixed $|T|$, each client uploads a
  constant number of ciphertexts, so total uplink grows linearly in $K$.
 
  \item \emph{Stepwise scaling in $|T|$.} As $|T|$ increases,
  $M$ changes only when $|T|$ crosses slot-capacity thresholds,
  producing staircase behavior in communication.
 
  \item \emph{Shares scale with $R$.} Decryption-share traffic grows linearly
  in $R$ and is independent of $K$.
 
  \item \emph{Interleaving reduces communication.} Interleaving lowers the
  number of ciphertexts whenever $\lceil 2|T|/B\rceil < 2\lceil |T|/B\rceil$,
  yielding up to a $2\times$ reduction in uplink and share traffic.
 
  \item \emph{HE vs.\ plaintext.} The HE/Plain uplink ratio is constant in $K$
  but varies with $|T|$, reflecting linear plaintext scaling versus stepwise
  HE scaling.
\end{itemize}
 
\noindent
In the current prototype, encrypted alignment may be performed over
the full grid $T_{\mathrm{all}}$. This affects measured byte counts in a live
deployment but does not affect the formula-based communication results
reported here, which are parameterized by $|T|$.
 
\noindent
Communication behavior follows directly from CKKS packing arithmetic.
Total HE uplink scales linearly in $K$ and stepwise in $|T|$, while
interleaving deterministically reduces communication by a factor
$M_{\mathrm{sep}}/M_{\mathrm{int}}$ (up to $2\times$).
 
\subsection{Theoretical Results}
\label{sec:theory-results}
 
We provide (i) model identities and basic invariance results supporting
RQ2--RQ3, (ii) CKKS perturbation bounds explaining numerical behavior
(RQ2--RQ3), (iii) identifiability from public outputs (privacy context for
RQ1), (iv) packing optimality (RQ4), and (v) communication and computational
complexity laws (RQ5--RQ6).
 
We use the notation of Table~\ref{tab:notation-hekm}. Throughout, we assume a
fixed centralized multiset of rows is horizontally partitioned among sites
(IID or non-IID, with or without overlap); that all parties use a shared
alignment grid $T_{\mathrm{all}}$, derive the event-time subset
$T \subseteq T_{\mathrm{all}}$, and apply the same tie convention; and that
the Kaplan--Meier (KM) product only changes at event times, so using event
times to form $T$ is sufficient for the estimator, whereas using
$T_{\mathrm{all}}$ merely refines the computational grid by inserting
zero-event time points. In theorem statements, we write $\hat S^\star(t)$ for
the pooled Kaplan--Meier estimator computed from the aggregate counts on the
event-time grid $T$; execution-mode-specific estimators carry explicit
subscripts ($\hat S_{\mathrm{plain}}$, $\hat S_{\mathrm{HE\text{-}exact}}$,
$\hat S_{\mathrm{HE\text{-}ckks}}$). In the protocol section the released
output is denoted $\hat S_{\mathrm{HE}}(t)$.
 
\subsection{Model and basic identities}
\label{subsec:model-notation}
 
There are $K$ sites, indexed by $k\in\{1,\dots,K\}$. Site $k$ has a local set
of unique observed survival times $T_k^{\mathrm{all}}$, including both event
and censoring times. Let
\[
T_{\mathrm{all}} := \bigcup_{k=1}^K T_k^{\mathrm{all}}
\]
denote the union of all unique survival times across sites. Let
$T_k^{\mathrm{event}} \subseteq T_k^{\mathrm{all}}$ denote the unique event
times at site $k$. The event-time grid used by the Kaplan--Meier estimator is
\[
T := \{\, t \in T_{\mathrm{all}} : \exists \text{ an event at } t \,\}
\;=\; \bigcup_{k=1}^K T_k^{\mathrm{event}},
\]
with cardinality $|T|$.
 
\noindent
The sets $T_k^{\mathrm{event}}$ and $T$ are theoretical objects used in KM
analysis and theorem statements. The implementation uploads only
$T_k^{\mathrm{all}}$ and constructs only $T_{\mathrm{all}}$; the event-time
subset $T$ is handled implicitly because $d_i = 0$ at censoring-only time
points and such factors do not alter the Kaplan--Meier product
(Lemma~\ref{lem:grid-refinement}).
 
For each $t_i \in T$, let $d_i^{(k)}$ denote the number of events at site $k$
occurring at time $t_i$, and let $n_i^{(k)}$ denote the number at risk at
site $k$ just before $t_i$. Global counts at $t_i$ are
\[
d_i=\sum_{k=1}^K d_i^{(k)}, \qquad n_i=\sum_{k=1}^K n_i^{(k)}.
\]
The KM estimator on the pooled aggregate counts is
\[
\hat S^\star(t)=\prod_{t_i\le t}\Bigl(1-\frac{d_i}{n_i}\Bigr),\qquad n_i>0.
\]
 
\medskip
\noindent\textit{Discrete and cumulative hazards.}
We write $h_i := d_i/n_i$ for the discrete hazard at $t_i$ and
\[
H(t) := \sum_{t_i \le t} h_i
\]
for the cumulative hazard evaluated on $T$.
 
\noindent\textit{Tie convention at event times.}
When events and censorings occur at the same time $t_i$, we adopt the standard
Kaplan--Meier convention: the at-risk count $n_i$ is taken just \emph{before}
$t_i$, events are applied at $t_i$, and censorings at $t_i$ affect only
subsequent risk sets, not $d_i/n_i$ at $t_i$.
 
\noindent\textit{HE aggregation and threshold decryption.}
Sites encrypt local contributions under a joint public key $\mathsf{pk}$.
Conceptually, the server forms homomorphic sums of aligned local count vectors,
and threshold fusion reconstructs packed plaintext blocks that decode to the
aggregated counts. In the present implementation, the encrypted count vectors
are aligned over $T_{\mathrm{all}}$; entries corresponding to times in
$T_{\mathrm{all}}\setminus T$ have event count zero and therefore do not alter
the Kaplan--Meier product. By Lemma~\ref{lem:grid-refinement}, exact
aggregation on $T_{\mathrm{all}}$ therefore yields the same KM estimator as
exact aggregation on $T$.
 
\noindent\textit{CKKS parameters (used below).}
CKKS with ring degree $n$ packs $B=n/2$ complex slots per ciphertext; the RNS
modulus chain is $Q=\prod_j q_j$ with bitlength
$|Q|=\sum_j \log_2 q_j$. We write $q_{\max}:=\max_j q_j$ for the largest
RNS limb modulus. A coarse size model is
$\textsf{ct\_bytes}\approx 2n|Q|/8$ per ciphertext and
$\textsf{share\_bytes}\approx n|Q|/8$ per partial share.
 
\medskip
 
\begin{lemma}[Global grid invariance]
\label{lem:grid}
For any horizontal partition of a fixed centralized multiset of rows
(IID or non-IID, with or without overlap), the union of all unique observed
times
\[
T_{\mathrm{all}}=\bigcup_{k=1}^{K} T_k^{\mathrm{all}}
\]
is invariant. The derived event-time grid
\[
T=\{\,t\in T_{\mathrm{all}}:\exists\ \text{an event at }t\}
\]
and its cardinality $|T|$ are therefore also invariant.
\end{lemma}
 
\begin{lemma}[Additivity of counts]
\label{lem:add}
For every $t_i\in T$, if local counting rules are aligned to the same global
grid and tie convention, then
\[
d_i=\sum_{k=1}^K d_i^{(k)} \qquad\text{and}\qquad n_i=\sum_{k=1}^K n_i^{(k)}.
\]
\end{lemma}
 
\begin{lemma}[Grid refinement invariance]
\label{lem:grid-refinement}
Let $T$ be the set of event times and let $\tilde T \supseteq T$ be any
refinement obtained by inserting time points with zero events. With consistent
tie conventions, the KM estimator evaluated on $\tilde T$ equals the KM
estimator on $T$ at all $t\in T$.
\end{lemma}
 
\begin{theorem}[Plain federated KM equals pooled oracle]
\label{thm:plain-invariance}
Running R1--R2 in plaintext yields
$\hat S_{\mathrm{plain}}(t)\equiv \hat S^\star(t)$ for any partition of the same
centralized multiset of rows.
\end{theorem}
 
\begin{theorem}[Exact HE yields identical KM]
\label{thm:exactHE}
If R2 uses exact homomorphic addition and threshold decryption returns the
exact aggregated counts, then
$\hat S_{\mathrm{HE\text{-}exact}}(t)\equiv \hat S^\star(t)$ for any partition.
This remains true whether alignment is performed directly on $T$ or on any
refinement $\tilde T \supseteq T$, including $T_{\mathrm{all}}$.
\end{theorem}
 
For \emph{proofs of Lemmas~\ref{lem:grid}--\ref{lem:grid-refinement} and
Theorems~\ref{thm:plain-invariance}--\ref{thm:exactHE}}, see Sections~S1--S2 of the
\SItext.
 
\subsection{Approximate HE (CKKS): perturbation of the KM product}
 
We model CKKS and threshold outputs on the event-time grid as
$\widehat d_i=d_i+\Delta d_i$ and $\widehat n_i=n_i+\Delta n_i$ with
deterministic error envelopes $\varepsilon_d(i)\ge|\Delta d_i|$ and
$\varepsilon_n(i)\ge|\Delta n_i|$. The bound below follows from a
multiplicative product inequality applied to the ratio
$\hat S_{\mathrm{HE\text{-}ckks}}(t)/\hat S^\star(t)$.
 
\begin{theorem}[CKKS perturbation bound]
\label{thm:ckks-bound}
Assume $0\le \varepsilon_n(i) < n_i$ (so $\widehat n_i>0$); define
$h_i=d_i/n_i$ and $h_{\max}=\max_{t_j\le t} h_j < 1$. Set
\[
\delta_i \;:=\; \frac{\varepsilon_d(i)}{n_i-\varepsilon_n(i)}\;+\;\frac{d_i}{n_i}\cdot\frac{\varepsilon_n(i)}{\,n_i-\varepsilon_n(i)\,}.
\]
Then, for all $t$,
\[
\bigl|\hat S_{\mathrm{HE\text{-}ckks}}(t)-\hat S^\star(t)\bigr|
\ \le\
\hat S^\star(t)\left(\exp\!\Big(\tfrac{1}{1-h_{\max}}\sum_{t_i\le t}\delta_i\Big)-1\right).
\]
\end{theorem}
 
\begin{remark}[Terminal jump with $h_i=1$]
If $d_i=n_i$ at a terminal time point (so $h_i=1$ and $\hat S^\star$ drops to $0$),
the stated bound applies on indices with $h_j<1$. At and after the terminal
jump, the trivial bound
$|\hat S_{\mathrm{HE\text{-}ckks}}(t)-\hat S^\star(t)|=\hat S_{\mathrm{HE\text{-}ckks}}(t)$
holds.
\end{remark}
 
\begin{corollary}[Uniform convergence on $T$]
\label{cor:convergence}
If $\varepsilon_d(i),\varepsilon_n(i)\to 0$ for all $t_i\in T$, then
$\hat S_{\mathrm{HE\text{-}ckks}}(t)\to \hat S^\star(t)$ uniformly on $T$.
\end{corollary}
 
For \emph{proofs}, see Section~S3 of the \SItext.
 
\subsection{Identifiability from Public Outputs}
\label{subsec:identifiability}
 
\begin{proposition}[Hazards are identified; counts are not]
\label{prop:identifiability}
Assume $T$ contains the event times and $\hat S(0)=1$ is known. Then the
per-time-point hazard ratios are determined by
\[
h_i \;=\; 1 - \frac{\hat S(t_i)}{\hat S(t_{i-1})}\, .
\]
However, without additional side information (e.g., initial at-risk size,
censoring schedule), the integer pairs $(d_i,n_i)$ consistent with these
ratios are not unique; there exist infinitely many real-valued $(d_i,n_i)$
solutions and, typically, multiple integer completions. Hence publishing
$\hat S(t)$ alone reveals $d_i/n_i$ but not $(d_i,n_i)$ nor any per-site split.
\end{proposition}
 
For \emph{proofs}, see Section~S4 of the \SItext.
 
\subsection{Noise envelopes and utility under a simple model}
 
We separate exact identities above from implementation-dependent error scaling.
The following assumption captures our add-only pipeline and the empirical
behavior of share fusion.
 
\begin{assumption}[Add-only CKKS noise and share aggregation]
\label{ass:ckks-noise}
(i) For a ciphertext contributed by site $k$ using $M_k$ ciphertext blocks at
fixed scale, the encode--encrypt perturbation satisfies
$\|\Delta_k(\cdot)\|=O(M_k\log q_{\max})$,
where $q_{\max}:=\max_{\ell} q_{\ell}$ is the largest RNS limb modulus.
(ii) Threshold-share fusion aggregates zero-mean, independent share errors
across decryptors with variance proxy $\sigma^2$, contributing
$O(\sigma\sqrt{R})$ to each block's error.
\end{assumption}
 
\noindent In our protocol $M_k = M$ for all $k$ since all sites pack over the
same grid $T_{\mathrm{all}}$; the propositions below are stated for the general
case.
 
\begin{proposition}[Utility envelope: coarse worst case]
\label{prop:utility-worst}
Under Assumption~\ref{ass:ckks-noise}, there exist nonnegative
$\varepsilon_d(i),\varepsilon_n(i)$ with
\[
\varepsilon_d(i),\varepsilon_n(i)\in O\!\big(\log(q_{\max})\sum_{k=1}^K M_k\big)+O(\sigma\sqrt{R})
\]
for which
\[
\Delta S(t):=\bigl|\hat S_{\mathrm{oracle}}(t)-\hat S_{\mathrm{HE}}(t)\bigr|
\ \le\
\prod_{t_i\le t}\Bigl(1-\tfrac{d_i}{n_i}\Bigr)\cdot
\sum_{t_i\le t}\frac{O\!\big(\log(q_{\max})\sum_k M_k\big)+O(\sigma\sqrt{R})}{\,n_i-\varepsilon_n(i)\,},
\]
reducing to the large-$n_i$ form with denominator $n_i$.
\end{proposition}
 
\begin{proposition}[Bounded at-risk counts]
\label{prop:bounded-risk}
If $n_i\ge c>0$ for all $t_i$, then
\[
\Delta S(t)\ \le\ \frac{1}{c}\sum_{t_i\le t}
\frac{O\!\big(\log(q_{\max})\sum_k M_k\big)+O(\sigma\sqrt{R})}{1-\big(O\!\big(\log(q_{\max})\sum_k M_k\big)+O(\sigma\sqrt{R})\big)/n_i},
\]
which further reduces to
\[
\frac{1}{c}\sum_{t_i\le t}\big(O\!\big(\log(q_{\max})\sum_k M_k\big)+O(\sigma\sqrt{R})\big)
\]
for large $n_i$.
\end{proposition}
 
\begin{proposition}[Error scaling with client count]
\label{prop:noise-scale}
Under Assumption~\ref{ass:ckks-noise},
\[
\varepsilon_{\mathrm{total}}=\sum_{k=1}^K O(M_k\log q_{\max})+O(\sigma\sqrt{R}) .
\]
\end{proposition}
 
For \emph{proofs}, see Section~S5 of the \SItext.
 
\subsection{Packing optimality}
\label{subsec:packing-opt}
 
Let $B=n/2$ be the CKKS slot budget and consider two aligned streams
$(d_\bullet,n_\bullet)$ of length $L$ (in the implementation $L=|T_{\mathrm{all}}|$;
in the communication simulations of Section~\ref{sec:exp_results}, $L=|T|$).
 
\begin{lemma}[Packing lower bound and interleaving optimality]
\label{lem:packing-opt}
Any packing that avoids cross-ciphertext multiplies and carries both streams
over length $L$ requires at least $\big\lceil \frac{2L}{B}\big\rceil$
ciphertexts. Interleaved co-packing attains this bound with
$M_{\mathrm{int}}=\lceil 2L/B\rceil$, while separate packing uses
$M_{\mathrm{sep}}=2\lceil L/B\rceil\ge \lceil 2L/B\rceil$.
\end{lemma}
 
\noindent\emph{Proof idea:} pigeonhole principle on total required slots; see
Section~S6 of the \SItext.
 
\begin{corollary}[Interleaving advantage]
\label{cor:interleaving-adv}
Let $B=n/2$ be the CKKS slot budget and let $L$ denote the length of the two
streams $(d_\bullet,n_\bullet)$. Separate packing uses
$M_{\mathrm{sep}}=2\lceil L/B\rceil$ ciphertexts; interleaving uses
$M_{\mathrm{int}}=\lceil 2L/B\rceil$. Then
$M_{\mathrm{int}}\le M_{\mathrm{sep}}$ with
$M_{\mathrm{sep}}-M_{\mathrm{int}}\in\{0,1\}$ and
\[
\frac{M_{\mathrm{sep}}}{M_{\mathrm{int}}} \in [1,\, 2].
\]
More precisely: (i) $\frac{M_{\mathrm{sep}}}{M_{\mathrm{int}}}=2$ when
$L\le B/2$; (ii) $\frac{M_{\mathrm{sep}}}{M_{\mathrm{int}}}=\frac{2c+2}{2c+1}\in(1,4/3]$
when $L=cB+r$ with $c\ge 1$ and $0<r\le B/2$; and
(iii) $\frac{M_{\mathrm{sep}}}{M_{\mathrm{int}}}=1$ when $L=cB+r$ with $c\ge 1$
and $B/2<r< B$.
\end{corollary}
 
For \emph{proofs}, see Section~S6 of the \SItext.
 
\subsection{Communication complexity}
\label{subsec:theory-comm}
 
In R1 (plaintext), each site uploads its local unique observed-time set
$T_k^{\mathrm{all}}$ and the coordinator broadcasts the resulting global
alignment grid $T_{\mathrm{all}}$. In R2, the implementation uploads aligned
at-risk and event counts over $T_{\mathrm{all}}$, giving $2K|T_{\mathrm{all}}|$
scalars. An event-time-only variant that filters to the event-time subset $T$
before uploading would use $2K|T|$ scalars instead.
 
With CKKS packing ($B=n/2$ slots), a length-$L$ vector packs into
$\lceil L/B\rceil$ ciphertexts. The formulas below use a generic
packed length $L$: in the communication simulations
(Figures~\ref{fig:comm-grid-1}--\ref{fig:comm-grid-3}) $L=|T|$; in the
implementation $L=|T_{\mathrm{all}}|$.
 
Under \emph{separate} packing,
\[
M_{\mathrm{sep}}=2\Big\lceil \frac{L}{B}\Big\rceil,
\]
and under \emph{interleaved} packing,
\[
M_{\mathrm{int}}=\Big\lceil \frac{2L}{B}\Big\rceil.
\]
 
\begin{theorem}[Communication: implemented plain federated KM]
\label{thm:comm-plain}
In the implemented protocol:
R1 uplink: $\sum_{k=1}^K |T_k^{\mathrm{all}}|$ scalars;\quad
R1 downlink: $K\,|T_{\mathrm{all}}|$ scalars;\quad
R2 uplink: $2K\,|T_{\mathrm{all}}|$ scalars.
In an event-time-only variant where R2 counts are aligned over $T$ instead
of $T_{\mathrm{all}}$, R2 uplink reduces to $2K\,|T|$ scalars.
\end{theorem}
 
\begin{theorem}[Communication: CKKS + threshold]
\label{thm:comm-ckks}
If each client uploads $M\in\{M_{\mathrm{sep}},M_{\mathrm{int}}\}$ ciphertexts
in R2 (the value of $M$ is identical across sites since all pack over the same
global alignment grid $T_{\mathrm{all}}$), the total R2 uplink is $KM$
ciphertexts. Threshold
decryption with a committee of size $R$ requires $RM$ partial shares,
where $M\in\{M_{\mathrm{sep}},M_{\mathrm{int}}\}$ is the number of
aggregated blocks to be decrypted. Equivalently,
\[
M_{\mathrm{int}}=\Big\lceil \frac{2L}{B}\Big\rceil,
\qquad
M_{\mathrm{sep}}=2\Big\lceil \frac{L}{B}\Big\rceil,
\]
where $L$ is the packed grid length.
\end{theorem}
 
\begin{corollary}[Scaling laws]
\label{cor:scaling}
Implemented plain: $\Theta(K)$ for fixed $|T_{\mathrm{all}}|$ and
$\Theta(|T_{\mathrm{all}}|)$ for fixed $K$; in an event-time-only variant,
the same asymptotics hold with $|T|$ replacing $|T_{\mathrm{all}}|$.\\
CKKS: R2 uplink is linear in $K$ and in $\lceil L/B\rceil$; share traffic is
linear in $R$ and in $\lceil L/B\rceil$, where $L$ is the packed grid length.
Interleaving satisfies $\lceil 2L/B\rceil \le 2\lceil L/B\rceil$.
\end{corollary}
 
The communication simulations in Section~\ref{sec:exp_results} instantiate
$L=|T|$; the implementation uses $L=|T_{\mathrm{all}}|$, and the same
formulas apply.
 
For \emph{proofs}, see Section~S7 of the \SItext.
 
\subsection{Computational complexity}
\label{subsec:work-law}
 
\begin{proposition}[Computational cost law]
\label{prop:comp-law}
The per-round work satisfies
\[
W(\text{mode}) \;=\; \alpha K M_\star \;+\; \beta R M_\star \;+\; \gamma M_\star \;+\; \delta_1 L \;+\; \delta_2 |T|,
\]
where $M_\star=M_{\mathrm{int}}$ or $M_{\mathrm{sep}}$ depending on packing,
$L$ is the packed grid length, and
$\alpha,\beta,\gamma,\delta_1,\delta_2>0$ are mode-independent constants. In
the current implementation, $L=|T_{\mathrm{all}}|$; in an event-time-only
packing variant, $L=|T|$.
\end{proposition}
 
The first term is homomorphic addition at the server; the second is
partial-share creation by the committee; the third is fusion; the fourth
captures packed-vector decode or alignment work on the packed grid; and the
fifth is the Kaplan--Meier pass on the event-time grid. This mirrors the
empirical near-linear trends in $K$ and the interleaving advantage observed in
Section~\ref{sec:exp_results}.
 
\begin{corollary}[Runtime impact of interleaving]
\label{cor:runtime-interleaving}
With $W_{\mathrm{sep}}$ and $W_{\mathrm{int}}$ the total work under separate
vs.\ interleaved packing,
\[
\frac{W_{\mathrm{sep}}}{W_{\mathrm{int}}}
\;=\;
\frac{(\alpha K+\beta R+\gamma) M_{\mathrm{sep}} + \delta_1 L + \delta_2 |T|}
     {(\alpha K+\beta R+\gamma) M_{\mathrm{int}} + \delta_1 L + \delta_2 |T|}
\;\le\;
\frac{M_{\mathrm{sep}}}{M_{\mathrm{int}}}
\;\le\; 2,
\]
where the last inequality follows from Corollary~\ref{cor:interleaving-adv}.
In the three regimes:
\begin{enumerate}[label=(\alph*), leftmargin=6mm]
\item If $L\le B/2$, then $\frac{W_{\mathrm{sep}}}{W_{\mathrm{int}}}\le 2$
(idealized $50\%$ speedup upper bound).
\item If $(M_{\mathrm{sep}},M_{\mathrm{int}})=(4,3)$, then
$\frac{W_{\mathrm{sep}}}{W_{\mathrm{int}}}\le \frac{4}{3}$ (up to $25\%$).
\item If $M_{\mathrm{sep}}=M_{\mathrm{int}}$, packing does not change runtime.
\end{enumerate}
Moreover, the fixed terms $\delta_1 L + \delta_2 |T|$ strictly reduce the
realized ratio below $M_{\mathrm{sep}}/M_{\mathrm{int}}$ whenever they are
positive.
\end{corollary}
 
For \emph{proofs}, see Section~S8 of the \SItext.
 
\section{Discussion}
\label{sec:discussion}
 
\begin{table}[htbp]
\centering
\scriptsize
\setlength{\tabcolsep}{3pt}
\renewcommand{\arraystretch}{1.08}
\begin{tabularx}{\linewidth}{@{}>{\raggedright\arraybackslash}p{0.16\linewidth} >{\raggedright\arraybackslash}p{0.17\linewidth} X@{}}
\toprule
\textbf{Knob} & \textbf{Default} & \textbf{Rationale / When to change}\\
\midrule
Ring degree $n$ & $16384$ & Meets approximately 128-bit RLWE security with our $|Q|$ while giving $B{=}8192$ slots; often places the packed grid length $L$ on favorable packing steps. Increase to $32768$ only if it moves $L$ to a lower ciphertext plateau.\\
 
Packing & \textbf{Interleaved} & Co-pack $(n_t,d_t)$ to minimize ciphertext count $M$; accuracy remains unchanged. Use \textit{separate} only if downstream plaintext postprocessing requires distinct streams.\\
 
CKKS scale & $2^{40}$ & Preserves approximately 9 to 10 decimal digits under additions and rotations; stable across sites for add-only KM.\\
 
Modulus chain $|Q|$ & approximately $420$ to $450$ bits & Leveled use without bootstrapping. Provides headroom for additions and rotations at scale $2^{40}$.\\
 
Galois keys & Powers of two & Minimal set for slot rotations required by packing; no relinearization keys required.\\
 
Committee size $R$ & $R\in\{9,25\}$ & Fix $R$ independent of $K$ to bound share traffic; $R{=}9$ is lightweight and $R{=}25$ provides a stronger quorum.\\
 
Threshold $\theta$ & $\theta{=}R$ (strict) & Maximizes confidentiality. Deployments supporting flexible thresholds may use $\theta{=}\lceil \phi R\rceil$ with $\phi\in[2/3,0.7]$ to balance confidentiality and liveness.\\
 
Decryptors vs.\ clients & $R\le K$ & Use all clients as decryptors ($R{=}K$) or a fixed subset; fixing $R$ decouples share traffic from $K$.\\
 
Packed grid length $L$ & $L{=}|T_{\mathrm{all}}|$ (implementation); $L{=}|T|$ (communication simulations) & The current implementation aligns encrypted counts over $T_{\mathrm{all}}$ (all observed times). The KM estimator depends only on event times $T$, since $d_t=0$ for censoring-only times. Filtering to $T$ before encryption reduces ciphertext count and communication when supported; the communication simulations in the Results section use $L{=}|T|$.\\
 
Tie or censoring rule & Consistent across sites & Ensures additivity and partition-invariant Kaplan--Meier estimation.\\
\bottomrule
\end{tabularx}
\caption{Recommended defaults for CKKS federated Kaplan--Meier computation with add-only aggregation. }
\label{tab:defaults}
\end{table}
Our framework advances privacy-preserving federated Kaplan--Meier (KM) analysis
by adopting CKKS-based multiparty homomorphic encryption with multiparty
decryption and strict output gating. Unlike integer-only
systems~\cite{Froelicher2021} that require fixed-point coordination for real
workloads, CKKS natively supports approximate arithmetic over real values. This
capability enables faithful time-to-event analytics while constraining privacy
risk in multi-institution collaborations~\cite{geva2023collaborative,ckks}.
 
We provide estimator-level guarantees including equality results for plaintext
and homomorphic settings, a CKKS perturbation bound with uniform convergence,
and an identifiability statement. These results are complemented by
Kaplan--Meier specific communication and computational scaling laws together
with a packing optimality result based on interleaving $(n_t,d_t)$. Empirically,
encrypted federated curves are numerically indistinguishable from pooled oracle
estimates across survival, RMST, band coverage, supremum gap, and hazard
diagnostics over large synthetic cohorts (up to $K=500$) and on the NCCTG lung
dataset (see Section~S9 of the \SItext; \SIFig{1}--\SIFig{3}). Interleaved and separate
packings match in accuracy, while interleaving reduces ciphertext count and
therefore bandwidth and runtime in the regimes predicted by the packing
arithmetic. Although homomorphic encryption introduces overhead relative to
plaintext computation, end-to-end performance remains feasible for moderate and
large federations. The resource model developed here provides concrete byte and
memory envelopes that support practical deployment planning.
 
We quantify a subtraction-based reconstruction attack in plaintext two-round
Kaplan--Meier protocols. A site can recover the residual contributions of other
participants by subtracting its own $(n_t^{(A)},d_t^{(A)})$ from broadcast totals
$(n_t^{(\mathrm{Fed})},d_t^{(\mathrm{Fed})})$. This leakage is exact when
$K=2$ and remains substantial for small overlapping coalitions. Our protocol
removes this attack surface because per-time-point tables are never released.
Only the survival curve $\hat S_{\mathrm{HE}}(t)$ and optional confidence bands are revealed.
By Proposition~\ref{prop:identifiability}, publishing $\hat S_{\mathrm{HE}}(t)$ identifies
per-time-point hazard ratios $h_i$ but does not determine integer totals
$(d_i,n_i)$ or any per-site decomposition. Consequently the subtraction
channel is closed even for honest-but-curious participants.
 
Estimator fidelity is established both theoretically and empirically. Plain
federated Kaplan--Meier equals the pooled oracle
(Theorem~\ref{thm:plain-invariance}). Exact homomorphic evaluation reproduces
the oracle (Theorem~\ref{thm:exactHE}). Approximate homomorphic evaluation
under CKKS yields a perturbation bound for the Kaplan--Meier product together
with uniform convergence as noise approaches zero
(Theorem~\ref{thm:ckks-bound}, Corollary~\ref{cor:convergence}). Empirical
evaluation confirms that survival errors, RMST differences, coverage
diagnostics, and hazard metrics remain within predefined tolerances.
 
Communication and computation follow simple laws specialized to
Kaplan--Meier estimation. With slot capacity $B=n/2$, interleaved packing
requires $M_{\mathrm{int}}=\lceil 2L/B\rceil$ ciphertexts and separate packing
requires $M_{\mathrm{sep}}=2\lceil L/B\rceil$, where $L$ is the packed grid
length. Encrypted uplink in Phase~R2 is therefore linear in $K$ and stepwise
in $L$, while decryption share traffic is linear in committee size $R$.
Computational effort decomposes into server additions $\Theta(M_\star K)$,
share generation $\Theta(RM_\star)$, fusion $\Theta(M_\star)$, packed-grid
decode or alignment work, and the final Kaplan--Meier pass $\Theta(|T|)$ on
the event-time grid, where $M_\star\in\{M_{\mathrm{int}},M_{\mathrm{sep}}\}$.
These scaling laws match the empirical behavior observed in
Section~\ref{sec:exp_results} (the implementation uses $L=|T_{\mathrm{all}}|$;
the communication simulations use $L=|T|$).
 
Our security posture reflects both the workload and the capabilities of current
homomorphic encryption systems. We instantiate CKKS with parameters providing
indistinguishability under chosen-plaintext attack (IND-CPA) under the RLWE
assumption and enforce aggregate-only decryption through a multiparty
committee. We do not claim IND-CCA2 security for CKKS. Instead we rely on
operational safeguards including output gating, quorum-based multiparty
decryption, authenticated transport, audit logging, and key rotation. These
controls reduce practical risk in clinical data collaborations while
preserving estimator fidelity.
 
In the present implementation, multiparty decryption shares are fused in an
all-of-$R$ configuration to recover aggregated counts required for
Kaplan--Meier evaluation. If fusion is performed at the coordinator, the
coordinator may transiently observe aggregated counts aligned over
$T_{\mathrm{all}}$, including zero-event entries at censoring-only times.
However, the coordinator never observes per-site contributions and only the
final survival curve $\hat S_{\mathrm{HE}}(t)$ is released to participating sites.
 
We deploy CKKS without decryption-time noise flooding in order to preserve
numerical precision, achieving agreement with the pooled oracle at the
$10^{-8}$ level. Instead of asserting IND-CPA$^{D}$ security, we constrain
potential oracle exposure operationally through four mechanisms. First,
multiparty decryption is used with $\theta = R$. Second, policy gating ties
each computation to a precommitted cohort and alignment grid
$T_{\mathrm{all}}$, from which the event-time grid $T$ is derived. Third, the
system exposes a restricted and non-adaptive interface with a bounded number of
decryptions. Fourth, strict output gating ensures that only
$\hat S_{\mathrm{HE}}(t)$ and optional derived public summaries are published;
the per-time-point aggregate table is not released.
 
Resource usage follows directly from the packing arithmetic. With
$(n=16384, |Q|\approx438\text{ bits})$, one ciphertext occupies approximately
$1.794$ MB and one partial share approximately $0.897$ MB. With slot capacity
$B=8192$, the ciphertext counts $(M_{\mathrm{int}},M_{\mathrm{sep}})$ equal
$(1,2)$ for packed length $L=1000$, $(2,2)$ for $L=5000$, and $(3,4)$ for
$L=10000$. In the current implementation, $L=|T_{\mathrm{all}}|$; in the
communication simulations reported in Section~\ref{sec:exp_results},
$L=|T|$. These values determine the bandwidth envelopes reported in the
Results. Reliability improvements such as idempotent uploads, bounded retry
strategies, reassignment of the lead role when a decryptor times out, and
chunked transfers improve resilience on unreliable networks without altering
the cryptographic design.
 
Operational parameters that influence deployment include the slot capacity
$B=n/2$, packing layout, committee size $R$, threshold $\theta$, and the
choice of packed grid length $L$. Interleaved packing is generally preferable
because it preserves accuracy while minimizing ciphertext count. Fixing a
modest committee size, such as $R=9$ or $R=25$, decouples share traffic from
federation size $K$. Selecting a ring degree $n$ that places $L$ on a
favorable packing step can reduce ciphertext counts more than it increases
per-ciphertext size. Under add-only Kaplan--Meier computation at scale
$2^{40}$, numerical stability remains largely insensitive to the specific
choices of $n$, $R$, and $\theta$. Table~\ref{tab:defaults} summarizes
recommended defaults.
 
Real-world multi-institutional deployments may experience network instability
or slow participants (``stragglers''). The protocol structure naturally limits
the impact of such delays. Phase~B aggregation requires only a single
ciphertext upload from each site, so network latency affects only the
completion time of the aggregation round rather than its correctness. Because
aggregation consists solely of homomorphic additions, the coordinator can
accumulate ciphertexts incrementally as they arrive and defer Phase~C until
all uploads are complete.
 
Phase~C decryption requires contributions from all $R$ decryptors in the
current all-of-$R$ configuration ($\theta = R$). In practice this stage can be
made resilient to stragglers using standard distributed systems techniques such
as bounded retry policies, checkpointing, and temporary lead reassignment
among decryptors. A deployment requiring stronger liveness guarantees could
instead adopt a $\theta$-of-$R$ configuration with $\theta < R$, allowing
fusion once a quorum of decryptors has responded. These operational mechanisms
do not change the communication or security analysis presented in this work but
improve robustness in heterogeneous network environments.
 
Secure multi-party computation (MPC) protocols provide an alternative framework
for privacy-preserving survival analysis. In MPC-based Kaplan--Meier
implementations, sites jointly compute the survival function using secret
sharing and interactive protocols~\cite{von2021privacy}. While MPC offers
strong cryptographic guarantees, it typically requires multiple rounds of
interaction among participating sites, with the number of rounds depending on
the multiplicative depth of the computation~\cite{evans2018pragmatic}.
 
In contrast, the homomorphic encryption approach used here requires a single
encrypted upload from each site (Phase~B), followed by server-side homomorphic
aggregation and a single round of partial decryption shares from the committee
(Phase~C). The total number of communication rounds is therefore small and
does not grow with the number of participating sites. This simplifies
deployment across institutions with heterogeneous connectivity.
 
Furthermore, CKKS natively supports approximate arithmetic over real values,
which aligns well with statistical estimators such as Kaplan--Meier and
restricted mean survival time (RMST). MPC-based solutions often require
fixed-point representations and additional coordination to maintain numerical
consistency across sites.
 
MPC-based approaches do offer complementary strengths. They can provide exact
arithmetic with no approximation error, may support stronger security models
that avoid assumptions related to approximate decryption in
CKKS~\cite{li2021security}, and can be designed without a trusted aggregation
coordinator. For settings where exact counts are required or where the
coordinator cannot be assumed honest-but-curious, an MPC design may therefore
be preferable.
 
For the federated survival analysis setting considered here, however, the CKKS
approximation error on the survival curve $\hat S_{\mathrm{HE}}(t)$ evaluated at event times
remains below $10^{-8}$ across all experimental configurations
(Section~\ref{sec:exp_results}), well within typical clinical reporting
tolerances. Under the threat model adopted in
Section~\ref{subsec:threat-key}, and with output gating ensuring that only
$\hat S_{\mathrm{HE}}(t)$ is released to sites, the threshold HE design provides a
practical balance between security, numerical fidelity, scalability, and
operational simplicity for large federations up to $K = 500$ sites.
 
Several limitations remain. The protocol currently focuses on add-only encrypted
aggregation for Kaplan--Meier estimation. Extensions to weighted estimators,
interval censoring, or Cox-type models will require careful management of
multiplicative depth or hybrid techniques combining homomorphic encryption with
differential privacy. Additional directions include verifiable aggregation to
protect against malicious coordinators, deployments supporting flexible
$\theta$-of-$R$ thresholds under participant churn, and broader stress testing
across heterogeneous institutional networks. Together these directions extend
the foundation established here for secure and scalable multi-institution
survival analysis.
\section{Methods}
\label{sec:methods}
\subsection{Notation and preliminaries}
\label{subsec:notation-preliminaries}
 
\begin{table}[htbp]
\centering
\small
\setlength{\tabcolsep}{6pt}
\renewcommand{\arraystretch}{1.1}
\begin{tabularx}{\linewidth}{@{}lX@{}}
\toprule
\textbf{Symbol} & \textbf{Meaning} \\
\midrule
$K$ & Number of sites (clients). \\
$\mathcal{R}$ & Set of parties authorized to participate in threshold decryption. \\
$R$ & Decryptor committee size, i.e.\ $R=|\mathcal{R}|$. \\
$r$ & Index for an individual decryptor ($r \in \mathcal{R}$). \\
$n$ & CKKS ring degree. \\
$B$ & Slot capacity per ciphertext ($B=n/2$ for CKKS). \\
$|Q|$ & Total modulus bitlength controlling ciphertext and share sizes. \\
$q_j$ & The $j$-th RNS limb modulus; $q_{\max}:=\max_j q_j$. \\
$T_{\mathrm{all}}$ & Global alignment grid: union of all unique observed survival times across sites (including event and censoring times). \\
$T=\{t_i\}_{i=1}^{|T|}$ & Event-time grid: subset of $T_{\mathrm{all}}$ containing only times with at least one observed event; $|T|$ denotes its cardinality. \\
$T_k^{\mathrm{all}}$ & Local unique survival times at site $k$ (including event and censoring times). \\
$T_k^{\mathrm{event}}$ & Local unique event times at site $k$; satisfies $T_k^{\mathrm{event}} \subseteq T_k^{\mathrm{all}}$. Theoretical; the implementation does not upload $T_k^{\mathrm{event}}$ separately (see Section~\ref{subsec:model-notation}). \\
$d_i^{(k)}$ & Site-$k$ event count at time $t_i$ (number of rows with $t=t_i$ and $e=1$). \\
$n_i^{(k)}$ & Site-$k$ at-risk count just before $t_i$ (number of rows with $t \ge t_i$). \\
$d_i$, $n_i$ & Aggregated event and at-risk counts at time $t_i$. \\
$L$ & Generic packed grid length appearing in layout, communication, and runtime formulas; instantiated explicitly in each section. \\
$M_0$ & Per-stream ciphertext count, $M_0=\lceil L/B\rceil$. \\
$M_{\mathrm{int}}$ & Ciphertext count under interleaved packing, $M_{\mathrm{int}}=\lceil 2L/B\rceil$. \\
$M_{\mathrm{sep}}$ & Total ciphertext count under separate packing, $M_{\mathrm{sep}}=2M_0=2\lceil L/B\rceil$. \\
$\mathcal{I}$ & Public ciphertext-index set induced by the packing layout: $\mathcal{I}=[M_{\mathrm{int}}]$ for interleaving, $\mathcal{I}=[M_0]\times\{d,n\}$ for separate packing. \\
$\mathbf{z}_{\iota}^{(k)}$ & Packed plaintext block at index $\iota\in\mathcal{I}$ for site $k$. For interleaving, $\iota=j$ indexes interleaved blocks; for separate packing, $\iota=(j,s)$ pairs a block index $j\in[M_0]$ with a stream label $s\in\{d,n\}$. We treat count vectors as real-valued slot payloads embedded in the CKKS plaintext space over $B$ complex slots. \\
$\mathsf{pk}$ & Joint public key produced by distributed key generation. \\
$\mathsf{sk}_k$ & Secret-key share held by site $k$. \\
$E_{\mathsf{pk}}(\cdot)$ & Public-key encryption under $\mathsf{pk}$ (CKKS). \\
$\mathsf{ct}_{\iota}^{(k)}$ & Ciphertext block uploaded by site $k$ at public index $\iota\in\mathcal{I}$. \\
$\mathsf{ct}_{\Sigma,\iota}$ & Aggregated ciphertext block at public index $\iota\in\mathcal{I}$: homomorphic sum $\bigoplus_{k=1}^K \mathsf{ct}_{\iota}^{(k)}$. \\
$\Delta_{\iota}^{(r)}$ & Partial decryption share produced by decryptor $r$ for aggregated block $\mathsf{ct}_{\Sigma,\iota}$. \\
$pt_{\iota}$ & Fused plaintext block after \textsc{MultipartyDecryptFusion} at index $\iota$. \\
$\mathbf{u}_{\iota}$ & Decoded slot vector extracted from $pt_{\iota}$. \\
$\hat S^\star(t)$ & Pooled Kaplan--Meier estimator on aggregate counts; used as the target in theorem statements. \\
$\hat S_{\mathrm{oracle}}(t)$ & Synonym for $\hat S^\star(t)$, used in experimental metrics and figures. \\
$\hat S_{\mathrm{plain}}(t)$ & Plaintext federated Kaplan--Meier estimator (Theorem~\ref{thm:plain-invariance}). \\
$\hat S_{\mathrm{HE}}(t)$ & HE-federated Kaplan--Meier estimator; subscripted as $\hat S_{\mathrm{HE\text{-}exact}}$ for exact HE (Theorem~\ref{thm:exactHE}) and $\hat S_{\mathrm{HE\text{-}ckks}}$ for CKKS approximate HE (Theorem~\ref{thm:ckks-bound}). \\
$\#\{\cdot\}$ & Set cardinality operator. \\
$\theta$ & Decryption threshold (minimum number of decryptors required for fusion); in our implementation and experiments $\theta = R$. \\
$\bigoplus$ & Homomorphic addition in CKKS (ciphertext addition corresponding to slotwise plaintext addition). \\
\bottomrule
\end{tabularx}
\caption{Notation used in the federated CKKS Kaplan--Meier protocol.}
\label{tab:notation-hekm}
\end{table}
 
\noindent\textit{Notation.}
Throughout the paper,
$T_{\mathrm{all}}$ denotes the global alignment grid of all distinct observed
times, $T \subseteq T_{\mathrm{all}}$ denotes the event-time grid used by
the Kaplan--Meier product, and $L$ denotes a generic packed grid length.
In the implementation, $L=|T_{\mathrm{all}}|$; in an event-time-only
packing variant, $L=|T|$. Each section that uses $L$ states the instantiation
explicitly.
 
For ciphertext layout, we write
\[
M_0 := \left\lceil L/B \right\rceil,\qquad
M_{\mathrm{int}} := \left\lceil 2L/B \right\rceil,\qquad
M_{\mathrm{sep}} := 2M_0.
\]
We let $\mathcal{I}$ denote the public ciphertext-index set induced by the
chosen layout: $\mathcal{I}=[M_{\mathrm{int}}]$ for interleaving and
$\mathcal{I}=[M_0]\times\{d,n\}$ for separate packing.
For interleaving, each $\iota\in\mathcal{I}$ is a block index;
for separate packing, each $\iota=(j,s)$ pairs a block index
$j\in[M_0]$ with a stream label $s\in\{d,n\}$.
 
In the implementation, $\mathcal{R}=[K]$ and hence $R=K$. When studying
communication scaling, we also consider hypothetical settings with fixed
committee size $R$ independent of $K$.
 
Modern Learning With Errors (LWE) and Ring Learning With Errors (RLWE) based
encryption schemes use noisy ciphertexts together with a linear-style
decryption relation determined by a secret key. In threshold variants, the
secret key is distributed into shares so that no single party can complete
decryption alone; instead, each decryptor contributes a \emph{partial
decryption share}, and a combiner fuses a qualified set of shares to recover
the plaintext. General constructions and security proofs for threshold
LWE/RLWE-style encryption appear in Asharov et al.~\cite{10.1007/978-3-642-29011-4_29}
and Cramer et al.~\cite{Cramer_Damgard_Nielsen_2015}.
 
The protocol notation in this work allows a general $\theta$-of-$R$ threshold
among decryptors. In our implementation, however, we instantiate the strict
all-of-$R$ configuration $\theta = R$ using OpenFHE's multiparty
Lead/Main/Fusion APIs. Supporting a more general $\theta$-of-$R$ threshold is
orthogonal to the Kaplan--Meier pipeline itself and depends on the underlying
multiparty HE configuration rather than on the survival-analysis logic.
 
We use CKKS for approximate packed arithmetic on real-valued vectors. The later
security analysis treats the protocol phase by phase. In particular, the Phase~B
ciphertext privacy claim is stated for the distributed threshold setting used
here, where an adversary may already hold a strict subset of secret-key shares
from the distributed key-generation phase. For that reason, the formal Phase~B
privacy theorem is stated under an explicit threshold-aware ciphertext privacy
assumption rather than under vanilla single-key IND-CPA alone.
 
We distinguish between the full alignment grid $T_{\mathrm{all}}$ and the
event-time grid $T \subseteq T_{\mathrm{all}}$. The protocol and current
implementation align encrypted counts on $T_{\mathrm{all}}$, while the
Kaplan--Meier estimator depends only on event times in $T$. Inserting time
points with zero events does not change the estimator (Lemma~\ref{lem:grid-refinement}), so this distinction
preserves consistency between theory and implementation.
 
\subsection{Federated KM with threshold CKKS: protocol and phases}
\label{subsec:protocol-phases}
 
\begin{table*}[htbp]
  \centering
  \scriptsize
  \setlength{\tabcolsep}{3pt}
  \renewcommand{\arraystretch}{1.05}
  \begin{tabularx}{\textwidth}{@{}T{0.30\textwidth} C{0.05\textwidth} L{0.20\textwidth} Y@{}}
    \toprule
    Keyword & Phase & Role & Purpose  \\
    \midrule
 
    InitCKKS & A & Coordinator & Initialize CKKS parameters and context (Alg.~\ref{alg:phaseA}, l.~1). \\
    ChooseOrder & A & Coordinator & Fix deterministic DKG order of sites (Alg.~\ref{alg:phaseA}, l.~2). \\
    KeyGen & A & First site & Create the initial public key and the first secret-key share (Alg.~\ref{alg:phaseA}, l.~6). \\
    MultipartyKeyGen & A & Each subsequent site & Mix in the site's secret share to update the joint $\mathsf{pk}$; each site keeps its local $\mathsf{sk}_k$ (Alg.~\ref{alg:phaseA}, l.~8). \\
    Broadcast($\mathsf{pk}$) & A & Coordinator & Distribute the joint public key to all sites (Alg.~\ref{alg:phaseA}, l.~11). \\
    UniqueTimes & A & Each site & Extract and upload local unique observed survival times $T_k^{\mathrm{all}}$ (Alg.~\ref{alg:phaseA}, l.~12). \\
    Union \& Sort & A & Coordinator & Form and sort the global alignment grid $T_{\mathrm{all}}=\bigcup_k T_k^{\mathrm{all}}$ and broadcast it to all sites (Alg.~\ref{alg:phaseA}, l.~12). The event-time subset $T=\{t\in T_{\mathrm{all}}:\exists\text{ an event at }t\}$ is not constructed explicitly; it is handled implicitly during KM evaluation because $d_i=0$ at censoring-only times. \\
 
    Pack (interleaved) & B & Site $k$ & Fill slots as $[n_1,d_1,n_2,d_2,\dots]$ with $M=\lceil 2|T_{\mathrm{all}}|/B\rceil$ (Alg.~\ref{alg:phaseB}, ll.~6--7). \\
    Pack (separate) & B & Site $k$ & Pack $[d_\bullet]$ and $[n_\bullet]$ separately; $M=2\lceil |T_{\mathrm{all}}|/B\rceil$ (Alg.~\ref{alg:phaseB}, ll.~9--10). \\
    Enc / $E_{\mathsf{pk}}(\cdot)$ & B & Site $k$ & Encrypt packed vectors under the joint public key (Alg.~\ref{alg:phaseB}, ll.~8,11). \\
    Agg:Add & B & Coordinator & Perform slotwise homomorphic addition (Alg.~\ref{alg:phaseB}, ll.~14--20). \\
 
    Multiparty\-Decrypt\-Lead & C & One decryptor & Produce the lead partial decryption share (Alg.~\ref{alg:phaseC}, l.~4). \\
    Multiparty\-Decrypt\-Main & C & Other decryptors & Produce the remaining partial decryption shares (Alg.~\ref{alg:phaseC}, l.~6). \\
    Multiparty\-Decrypt\-Fusion & C & Combiner & Fuse the collected valid shares; in our implementation this is all-of-$R$ (Alg.~\ref{alg:phaseC}, l.~9). \\
    GetCKKS\-Packed\-Value & C & Combiner & Decode CKKS plaintext slots (Alg.~\ref{alg:phaseC}, l.~10). \\
    ReconstructCounts & C & Combiner & Reconstruct the aggregate table $\{(n_i,d_i)\}_{i=1}^{L}$ from fused plaintext blocks (Alg.~\ref{alg:phaseC}). \\
    ComputeKM & C & Combiner & Evaluate the Kaplan--Meier estimator on the alignment grid $T_{\mathrm{all}}$ from the reconstructed aggregate table (equivalent to evaluation on the event-time subset $T$ since $d_i=0$ at non-event times; Alg.~\ref{alg:phaseC}). \\
    Reveal & C & Coordinator & Release only the final survival curve $\hat S_{\mathrm{HE}}(t)$ (and optional public summaries), not the per-time-point aggregate table (Alg.~\ref{alg:phaseC}). \\
    Broadcast($\hat S_{\mathrm{HE}}$) & C & Coordinator & Disseminate the final public output (Alg.~\ref{alg:phaseC}). \\
 
    \bottomrule
  \end{tabularx}
  \caption{Protocol keywords grouped by phase: Phase~A performs setup and time-grid construction; Phase~B performs encrypted aggregation; Phase~C performs multiparty decryption and output reveal. The keywords map to the abstract protocol tuple as follows: \textsc{Setup}~$\leftrightarrow$~Phase~A (InitCKKS through Union~\&~Sort); \textsc{Pack}+\textsc{Enc}+\textsc{Agg}~$\leftrightarrow$~Phase~B; \textsc{PartDec}~$\leftrightarrow$~MultipartyDecryptLead/Main; \textsc{Fuse}~$\leftrightarrow$~MultipartyDecryptFusion; \textsc{Reveal}~$\leftrightarrow$~ReconstructCounts+ComputeKM+Reveal+Broadcast($\hat S_{\mathrm{HE}}$).}
  \label{tab:protocol-keywords}
\end{table*}
 
We use CKKS~\cite{ckks} with batching to support approximate real arithmetic on
packed vectors. Our protocol organizes the federated Kaplan--Meier workflow into
three phases that separate \emph{key setup}, \emph{encrypted aggregation}, and
\emph{threshold decryption with output gating}. The coordinator orchestrates the
round structure and performs add-only homomorphic aggregation; sites compute local per-time Kaplan--Meier counts aligned to the shared global alignment grid $T_{\mathrm{all}}$ and encrypt them under a joint public key, while the final Kaplan--Meier product is evaluated on the event-time subset $T \subseteq T_{\mathrm{all}}$; a decryptor committee contributes
partial decryption shares that are fused into aggregate plaintexts from which
the public survival output is computed. Algorithms~\ref{alg:phaseA}--\ref{alg:phaseC}
formalize the steps, and Table~\ref{tab:protocol-keywords} groups the protocol
keywords by phase.
 
The protocol description and the security claims address different aspects of
the system. In particular, Phase~B uses CKKS encryption under a joint public key,
but the formal privacy theorem for distributed encrypted aggregation is stated
under the threshold-aware auxiliary-input ciphertext privacy assumption
introduced in Section~\ref{subsec:security-by-phase}, rather than being derived
here directly from a standalone vanilla IND-CPA theorem for single-key CKKS.
 
Let $\mathsf{Params}$ denote the CKKS parameter set. The protocol
\[
\Pi_{\mathrm{KM\text{-}Fed\text{-}CKKS}}
=\bigl(\textsc{Setup},\textsc{Pack},\textsc{Enc},\textsc{Agg},
\textsc{PartDec},\textsc{Fuse},\textsc{Reveal}\bigr)
\]
comprises: \textsc{Setup} (Phase~A),
\textsc{Pack}+\textsc{Enc}+\textsc{Agg} (Phase~B), and
\textsc{PartDec}+\textsc{Fuse}+\textsc{Reveal} (Phase~C). The protocol
notation allows a general $\theta$-of-$R$ threshold among decryptors, but in
our implementation and experiments we instantiate the strict all-of-$R$ setting
$\theta = R$. Only the released survival curve $\hat S_{\mathrm{HE}}(t)$ (and optional
derived summaries such as confidence bands, if computed) is published to sites.
 
\begin{algorithm}[H]
\caption{Phase A: Key Initialization and Alignment Grid}
\label{alg:phaseA}
\begin{algorithmic}[1]
\State $\mathsf{cc} \gets \textsc{InitCKKS}(n,|Q|,\text{depth},\dots)$ \Comment{initialize CKKS parameters/context}
\State $\pi \gets \textsc{ChooseOrder}(\{1,\dots,K\})$ \Comment{fixed DKG order}
\State $\mathsf{pk} \gets \bot$
\For{\textbf{each} $k \in \pi$}
  \If{$k$ is first in $\pi$}
    \State $(\mathsf{sk}_k,\mathsf{pk}) \gets \textsc{KeyGen}()$
  \Else
    \State $(\mathsf{sk}_k,\mathsf{pk}) \gets \textsc{MultipartyKeyGen}(\mathsf{pk})$
  \EndIf
\EndFor
\State \textsc{Broadcast} $\mathsf{pk}$ to all sites
\State $T_{\mathrm{all}} \gets \textsc{Sort}\!\left(\bigcup_k \textsc{UniqueTimes}(k)\right)$ \Comment{global alignment grid}
\end{algorithmic}
\end{algorithm}
 
\begin{algorithm}[H]
\caption{Phase B: Encrypted Survival Counts \& Server Aggregation}
\label{alg:phaseB}
\begin{algorithmic}[1]
\State $L \gets |T_{\mathrm{all}}|$ \Comment{current implementation}
\State $M_0 \gets \lceil L/B \rceil$
\For{each site $k$}
  \For{$i \gets 1$ to $L$}
    \State $d_i^{(k)} \gets \#\{t=t_i,\ e=1\}$;\quad $n_i^{(k)} \gets \#\{t\ge t_i\}$ \Comment{$t_i \in T_{\mathrm{all}}$}
  \EndFor
\EndFor
\If{\texttt{he\_mode} = \texttt{interleaved}}
  \State $M \gets \lceil 2L/B \rceil$ \Comment{$=M_{\mathrm{int}}$}
  \For{each site $k$}
    \For{$j \gets 1$ to $M$}
      \State form interleaved plaintext block $\mathbf{z}^{(k)}_j \gets [\,n_{\cdot},d_{\cdot},\ldots]$
      \State $\mathsf{ct}^{(k)}_j \gets E_{\mathsf{pk}}(\mathbf{z}^{(k)}_j)$
    \EndFor
  \EndFor
  \For{$j \gets 1$ to $M$} \Comment{coordinator aggregation}
    \State $\mathsf{ct}_{\Sigma,j}\gets \bigoplus_{k=1}^K \mathsf{ct}^{(k)}_j$
  \EndFor
\Else \Comment{separate packing}
  \State $M \gets 2M_0$ \Comment{$=M_{\mathrm{sep}}$}
  \For{each site $k$}
    \For{$j \gets 1$ to $M_0$}
      \State form event-count block $\mathbf{z}^{(k)}_{j,d}\gets[d_\bullet^{(k)}]$
      \State form at-risk block $\mathbf{z}^{(k)}_{j,n}\gets[n_\bullet^{(k)}]$
      \State $\mathsf{ct}^{(k)}_{j,d}\gets E_{\mathsf{pk}}(\mathbf{z}^{(k)}_{j,d})$;\quad $\mathsf{ct}^{(k)}_{j,n}\gets E_{\mathsf{pk}}(\mathbf{z}^{(k)}_{j,n})$
    \EndFor
  \EndFor
  \For{$j \gets 1$ to $M_0$} \Comment{coordinator aggregation}
    \State $\mathsf{ct}_{\Sigma,j,d}\gets \bigoplus_{k=1}^K \mathsf{ct}^{(k)}_{j,d}$;\quad
           $\mathsf{ct}_{\Sigma,j,n}\gets \bigoplus_{k=1}^K \mathsf{ct}^{(k)}_{j,n}$
  \EndFor
\EndIf
\end{algorithmic}
\end{algorithm}
 
\begin{algorithm}[H]
\caption{Phase C: Multiparty Decryption \& Output Gating}
\label{alg:phaseC}
\begin{algorithmic}[1]
\If{\texttt{he\_mode} = \texttt{interleaved}}
  \For{$j \gets 1$ to $M_{\mathrm{int}}$}
    \For{each $r \in \mathcal{R}$} \Comment{decryptor committee}
      \If{$r$ is lead}
        \State $\Delta^{(r)}_j \gets \textsc{MultipartyDecryptLead}(\mathsf{ct}_{\Sigma,j})$
      \Else
        \State $\Delta^{(r)}_j \gets \textsc{MultipartyDecryptMain}(\mathsf{ct}_{\Sigma,j})$
      \EndIf
    \EndFor
    \State $pt_j \gets \textsc{MultipartyDecryptFusion}(\{\Delta^{(r)}_j\}_{r\in\mathcal{R}})$
    \State $\mathbf{u}_j \gets \textsc{GetCKKSPackedValue}(pt_j)$
  \EndFor
  \State Reconstruct $\{(n_i,d_i)\}_{i=1}^{L}$ from $\{\mathbf{u}_j\}_{j=1}^{M_{\mathrm{int}}}$
\Else \Comment{separate packing}
  \For{$j \gets 1$ to $M_0$}
    \For{each $r \in \mathcal{R}$}
      \If{$r$ is lead}
        \State $\Delta^{(r)}_{(j,d)} \gets \textsc{MultipartyDecryptLead}(\mathsf{ct}_{\Sigma,j,d})$
        \State $\Delta^{(r)}_{(j,n)} \gets \textsc{MultipartyDecryptLead}(\mathsf{ct}_{\Sigma,j,n})$
      \Else
        \State $\Delta^{(r)}_{(j,d)} \gets \textsc{MultipartyDecryptMain}(\mathsf{ct}_{\Sigma,j,d})$
        \State $\Delta^{(r)}_{(j,n)} \gets \textsc{MultipartyDecryptMain}(\mathsf{ct}_{\Sigma,j,n})$
      \EndIf
    \EndFor
    \State $pt_{(j,d)} \gets \textsc{MultipartyDecryptFusion}(\{\Delta^{(r)}_{(j,d)}\}_{r\in\mathcal{R}})$
    \State $pt_{(j,n)} \gets \textsc{MultipartyDecryptFusion}(\{\Delta^{(r)}_{(j,n)}\}_{r\in\mathcal{R}})$
    \State $\mathbf{u}_{(j,d)} \gets \textsc{GetCKKSPackedValue}(pt_{(j,d)})$
    \State $\mathbf{u}_{(j,n)} \gets \textsc{GetCKKSPackedValue}(pt_{(j,n)})$
  \EndFor
  \State Reconstruct $\{(n_i,d_i)\}_{i=1}^{L}$ from $\{\mathbf{u}_{(j,d)},\mathbf{u}_{(j,n)}\}_{j=1}^{M_0}$
\EndIf
\State $S[0]\gets 1$
\For{each $t_i \in T$} \Comment{KM product on event times only}
  \State $S[i]\gets S[i{-}1]\cdot(1-d_i/n_i)$
\EndFor
\State \textbf{Return and broadcast} $\hat S_{\mathrm{HE}}(t)$ \textbf{only}
\end{algorithmic}
\end{algorithm}
 
\textit{Phase A} performs key initialization and construction of the global
alignment grid. The coordinator initializes the CKKS context, fixes a deterministic
site order $\pi$, and runs iterative multiparty key generation: the first site
calls \textsc{KeyGen}, and each subsequent site calls
\textsc{MultipartyKeyGen} to incorporate its contribution into the joint public
key. Each site retains its own secret-key share $\mathsf{sk}_k$; the
coordinator learns only the resulting joint public key $\mathsf{pk}$. Measured
in $K$, this phase performs one \textsc{KeyGen}, $K-1$
\textsc{MultipartyKeyGen} calls, and one broadcast.
 
In the same phase, each site uploads its local unique observed survival
times $T_k^{\mathrm{all}}$ (including both event and censoring times). The
coordinator forms the global alignment grid by union:
\[
T_{\mathrm{all}} = \mathrm{Sort}\!\left(\bigcup_{k=1}^K T_k^{\mathrm{all}}\right).
\]
The event-time subset $T = \{t \in T_{\mathrm{all}} : \exists\text{ an event at }t\}$
is not constructed explicitly in the deployed pipeline; it is handled
implicitly during aligned counting and KM evaluation, since $d_i = 0$ at
censoring-only time points and such factors do not alter the Kaplan--Meier
product (Lemma~\ref{lem:grid-refinement}).
The protocol operates over $T_{\mathrm{all}}$ for alignment and
encrypted aggregation, while the Kaplan--Meier estimator depends only on the
event-time subset $T$ in theoretical statements.
 
The time-grid union runs in plaintext and reveals
$T_{\mathrm{all}}$ by design. Deployments that require hiding local time
sets $T_k^{\mathrm{all}}$ would need to replace the plaintext union with a
private-set-union or related encrypted set-operation primitive.
 
\textit{Phase B} performs encrypted aggregation of Kaplan--Meier counts. Given
the shared alignment grid $T_{\mathrm{all}}$ and joint public key $\mathsf{pk}$,
each site computes local event counts $d_i^{(k)}$ and at-risk counts
$n_i^{(k)}$ at each $t_i \in T_{\mathrm{all}}$, packs them into CKKS plaintext
blocks, encrypts the packed vectors, and uploads the resulting ciphertexts.
For non-event times, $d_i^{(k)} = 0$, so inclusion of these indices does not
affect the Kaplan--Meier product.
 
In the \emph{interleaved} layout, slots are packed as
$[n_1,d_1,n_2,d_2,\ldots]$, giving
\[
M_{\mathrm{int}}=\left\lceil \frac{2|T_{\mathrm{all}}|}{B}\right\rceil,
\qquad B=n/2.
\]
In the \emph{separate} layout, the $d_\bullet$ and $n_\bullet$ streams are
packed independently, giving
\[
M_{\mathrm{sep}}=2\left\lceil \frac{|T_{\mathrm{all}}|}{B}\right\rceil.
\]
The coordinator performs only elementwise homomorphic addition to form
aggregate ciphertext blocks. Server work is $O(MK)$ homomorphic additions, and
per-site uplink is $\Theta(M)$. No plaintext is revealed in this phase. The
formal privacy guarantee for this phase is given later as an
assumption-based ciphertext-content indistinguishability result in the
distributed threshold setting.
 
\textit{Phase C} performs multiparty decryption and output gating. The
aggregated ciphertext blocks are sent to the decryptor committee. In the
current implementation, all $R$ decryptors participate, so fusion is
all-of-$R$. For each block $j$, exactly one decryptor produces the lead share
via \textsc{MultipartyDecryptLead}, and the remaining decryptors produce main
shares via \textsc{MultipartyDecryptMain}. The combiner calls
\textsc{MultipartyDecryptFusion} once per block to obtain plaintext block
$pt_j$, decodes the CKKS slots, and reconstructs the aggregate count table
\[
\{(n_i,d_i)\}_{i=1}^{|T_{\mathrm{all}}|}.
\]
It then evaluates the Kaplan--Meier estimator over the event-time grid $T$:
\[
\hat S_{\mathrm{HE}}(t)=\prod_{\substack{t_i \le t \\ t_i \in T}}
\left(1-\frac{d_i}{n_i}\right).
\]
Because $d_i = 0$ at non-event times, evaluating the product over $T$ is
equivalent to evaluating it over $T_{\mathrm{all}}$.
 
Only $\hat S_{\mathrm{HE}}(t)$, and optionally derived public summaries, is returned to
sites; the aggregate count table is not released by the protocol interface.
Costs scale as $O(RM)$ partial-share computations, $O(M)$ fusions, and
$O(|T|)$ for the Kaplan--Meier pass.
 
If the combiner is co-located with the coordinator, then the coordinator may
transiently observe the aggregate plaintext counts $(n_i,d_i)$ needed to
evaluate $\hat S_{\mathrm{HE}}(t)$. This does not change what is \emph{released} to sites,
but it does enlarge the coordinator's transient operational visibility and is
therefore treated explicitly in the later deployment qualification of the
security analysis.
\subsection{Threat model, threshold, and key management}
\label{subsec:threat-key}
 
The system assumes a semi-honest coordinator and semi-honest sites. All parties
follow the protocol as specified but may attempt to infer additional
information from the transcripts they observe. Transport is protected by
mutually authenticated, encrypted channels. Malicious deviations, adaptive
corruption, and coordinator-site collusion are outside the present model and
would require additional mechanisms and a separate analysis.
 
Sites perform distributed key generation as follows: the first site calls
\textsc{KeyGen}, and each subsequent site calls
\textsc{MultipartyKeyGen}, producing a joint public key $\mathsf{pk}$ while
each site retains its own secret-key share $\mathsf{sk}_k$. The protocol
notation allows a general $\theta$-of-$R$ decryption threshold, where $\theta$
is the minimum number of decryptors required for successful fusion. In our
implementation and experiments we instantiate the strict all-of-$R$ setting
$\theta = R$, meaning all decrypting parties contribute shares before
decryption completes.
 
Phase~A reveals public alignment metadata: each site discloses its local
unique observed survival times, from which the coordinator
forms the global alignment grid $T_{\mathrm{all}}$. The event-time
grid $T$ is a theoretical subset of $T_{\mathrm{all}}$ and is not
constructed explicitly in the implementation. The alignment grid is treated as
non-sensitive and is broadcast to all sites.
 
The decryptor committee is intended to enforce an \emph{aggregate-only}
decryption policy bound to a precommitted cohort and alignment grid $T_{\mathrm{all}}$, from which the event-time grid $T$ is derived. In
particular, the intended interface does not offer a general adaptive decryption
oracle. Decryption is restricted to ciphertexts produced by the fixed
aggregation pipeline, tied to the committed cohort and grid, and used only to
recover the aggregate plaintext required for the Kaplan--Meier output. This
restriction is important because our formal Phase~C claim is output-relative:
it characterizes what a corrupted decryptor learns relative to the fused
plaintext and the released curve, rather than asserting a generic adaptive
chosen-ciphertext guarantee for CKKS.
 
The following deployment controls are recommended:
\begin{enumerate}[label=(\roman*),nosep]
    \item time- or usage-based key rotation with fresh DKG;
    \item authorization logs binding each decryption request to the committed
    cohort and alignment grid $T_{\mathrm{all}}$;
    \item attestation and audit of decryptor hosts; and
    \item mutually authenticated encrypted transport for all protocol traffic.
\end{enumerate}
 
If the combiner operates within the decryptor committee and is not co-located
with the coordinator, then the coordinator never sees aggregate plaintext
counts. If the combiner is co-located with the coordinator, then the
coordinator may transiently observe the aggregate plaintext table
$\{(n_i,d_i)\}_{i=1}^{|T_{\mathrm{all}}|}$ required for Kaplan--Meier evaluation; however,
only the final survival curve $\hat S_{\mathrm{HE}}(t)$ is released to sites.

\subsection{Security model and claims}
\label{subsec:security-model-claims}
 
We analyze
\[
\Pi_{\mathrm{KM\text{-}Fed\text{-}CKKS}}
\]
in a \emph{semi-honest} (honest-but-curious), \emph{static} adversarial model.
Corruption sets are fixed before protocol execution begins, and corrupted
parties follow the protocol faithfully while attempting to infer additional
information from their observed transcripts. The view of a corrupted party
includes its local input, its full random tape, and all messages it receives.
Adaptive corruption and malicious deviations are outside the scope of this
analysis.
 
The coordinator is modeled as a semi-honest, non-colluding aggregator: it
performs homomorphic operations faithfully but may attempt to infer information
from ciphertexts and metadata. Collusion between the coordinator and corrupted
sites would enlarge the adversarial view and is excluded from the present
model. Transport is assumed to be protected by mutually authenticated,
encrypted channels.
 
Let $\mathcal{C}_S \subsetneq [K]$ denote the set of corrupted sites. Let
$\mathcal{R} \subseteq [K]$ denote the set of parties authorized to
participate in threshold decryption, with $R = |\mathcal{R}|$ and decryption
threshold $\theta$. In our implementation and experiments, every site
participates in decryption, so $\mathcal{R} = [K]$ and we instantiate the
strict all-of-$R$ setting $\theta = R$. Let
$\mathcal{C}_R \subsetneq \mathcal{R}$ denote the set of corrupted
decryptors. The corruption sets $\mathcal{C}_S$ and $\mathcal{C}_R$ are
indexed over different protocol roles and may overlap.
 
\paragraph{Oracle surface and decryption policy.}
Standard IND-CPA for CKKS governs ciphertext indistinguishability under chosen
plaintext attack, but does not cover a setting in which an adversary can issue
adaptive decryption queries. In principle, approximate decryption behavior can
become relevant under stronger oracle models. Rather than claiming a generic
adaptive decryption guarantee for CKKS, our protocol constrains the oracle
surface operationally:
\begin{enumerate}[label=(\roman*),nosep]
    \item no single party can decrypt aggregated ciphertexts on its own;
    \item decryption is restricted to ciphertexts bound to a precommitted
    cohort and alignment grid $T_{\mathrm{all}}$;
    \item the computation interface is fixed to add-only aggregation followed
    by one Kaplan--Meier evaluation pass; and
    \item only the released survival output $\hat S_{\mathrm{HE}}(t)$ is published to sites,
    while the aggregated plaintext table
    $\{(n_i,d_i)\}_{i=1}^{|T_{\mathrm{all}}|}$ remains internal to the
    decryption workflow unless the deployment explicitly co-locates fusion with
    the coordinator.
\end{enumerate}
Accordingly, the formal Phase~C claim below is \emph{simulation relative to
the fused plaintext and released output}, not a generic adaptive decryption
security statement. The Kaplan--Meier estimator itself depends only on the
event-time subset $T \subseteq T_{\mathrm{all}}$, since $d_i = 0$ at
censoring-only time points.
 
\paragraph{Scope of the security claims.}
Our formal guarantees are phase-specific and transcript-based.
\begin{enumerate}[label=(C\arabic*), leftmargin=8mm]
    \item \textbf{Phase A setup privacy.} The distributed key-generation
    transcript is simulatable from the corrupted parties' local DKG state and
    the final global alignment grid $T_{\mathrm{all}}$ and event-time grid $T$. This does \emph{not}
    hide the plaintext union steps beyond what is revealed by
    $T_{\mathrm{all}}$ and $T$ themselves.
    \item \textbf{Phase B encrypted aggregation privacy.} The encrypted
    aggregation transcript is computationally indistinguishable from a
    simulation that does not use the honest sites' local Kaplan--Meier count
    vectors, \emph{assuming} threshold-aware auxiliary-input ciphertext privacy
    for the distributed CKKS layer. This is an assumption-based claim; we do
    not prove that assumption from RLWE for our specific threshold CKKS
    instantiation in this work.
    \item \textbf{Phase C output-relative privacy.} The decryption transcript
    seen by a corrupted decryptor is simulatable from the aggregate plaintext
    table $P$, the signed metadata $\mathsf{md}$, the released curve
    $\hat S_{\mathrm{HE}}(t)$, and the corrupted decryptors' local state,
    under the stated threshold decryption assumptions. This is a protocol-level
    reduction to the underlying threshold decryption mechanism, not a derivation
    from RLWE alone.
    \item \textbf{Output gating.} Only $\hat S_{\mathrm{HE}}(t)$ is released
    to sites. The aggregated plaintext table $P=\{(n_i,d_i)\}_{i=1}^{L}$ is
    not released by the protocol interface. This closes the direct subtraction
    channel that would arise if per-time aggregate counts were published in
    plaintext.
\end{enumerate}
 
\paragraph{Relation to the phase-wise theorems.}
The formal statements proved in
Section~\ref{subsec:security-by-phase} establish:
\begin{enumerate}[label=(\roman*),nosep]
    \item identical-distribution simulation for the DKG transcript in
    Phase~A, relative to the exposed alignment grid $T_{\mathrm{all}}$ and event-time grid $T$;
    \item computational indistinguishability of the Phase~B encrypted
    aggregation transcript from a zero-encryption simulation, under
    Assumption~\ref{asm:threshold-indcpa}; and
    \item computational indistinguishability of the Phase~C decryption
    transcript from a simulation relative to the fused plaintext and released
    output, under the threshold-decryption assumptions in
    Theorem~\ref{thm:phaseC-privacy}.
\end{enumerate}
These are the precise claims supported by the proofs. In particular, the
Phase~B claim is not a standalone theorem about vanilla CKKS IND-CPA alone; it
is a theorem about our distributed setting under
Assumption~\ref{asm:threshold-indcpa}. Likewise, the Phase~C claim is
output-relative and committee-relative, not a statement that the decryption
phase hides the fused plaintext from committee members who legitimately
participate in fusion.
 
\paragraph{Operational authenticity and integrity.}
Our security analysis focuses on confidentiality and transcript privacy in the
semi-honest model. Authenticity and integrity are provided operationally by the
deployment assumptions: mutually authenticated channels, per-message integrity
protection, and signed metadata binding each decryption request to its cohort,
alignment grid $T_{\mathrm{all}}$, and ciphertext identifiers. We do not prove
cryptographic soundness against an actively malicious coordinator or malicious
sites in this work. Preventing malformed ciphertext uploads, replayed shares,
or incorrect aggregation under malicious deviations would require additional
mechanisms such as ciphertext validity checks, verifiable aggregation, and
verifiable threshold decryption.
 
\paragraph{Residual leakage and limitations.}
The guarantees above are transcript-based and do not eliminate all side
channels. In particular:
\begin{enumerate}[label=(\roman*), leftmargin=8mm]
    \item the plaintext union step in Phase~A reveals the global alignment grid
    $T_{\mathrm{all}}$; the theoretical event-time grid $T \subseteq T_{\mathrm{all}}$ is not transmitted explicitly in the implementation but is included in the security model for theoretical generality;
    \item metadata such as ciphertext count, ciphertext size, upload timing,
    and committee size may reveal coarse-grained system information;
    \item the current analysis excludes coordinator-site collusion; and
    \item if fusion is co-located with the coordinator, then the coordinator
    may transiently observe the aggregate plaintext table
    $\{(n_i,d_i)\}_{i=1}^{|T_{\mathrm{all}}|}$ needed to evaluate $\hat S_{\mathrm{HE}}(t)$,
    even though only $\hat S_{\mathrm{HE}}(t)$ is released to sites.
\end{enumerate}
These limitations are reflected explicitly in the phase-wise theorems and in
the deployment qualifications stated there.
 
\paragraph{Summary.}
Under the stated semi-honest, static model, our protocol provides:
(i) setup privacy relative to the intentionally exposed alignment grid
$T_{\mathrm{all}}$ and event-time grid $T$;
(ii) assumption-based ciphertext content privacy during encrypted aggregation;
and (iii) output-relative privacy during threshold decryption and release.
Together, these characterize what the protocol hides, what it intentionally
reveals, and what remains dependent on deployment choices and system-level
controls.
\subsection{Security analysis by phase}
\label{subsec:security-by-phase}
 
Let
\[
\Pi := \Pi_{\mathrm{KM\text{-}Fed\text{-}CKKS}}
\]
denote the protocol from Section~\ref{sec:methods}. Let $\lambda$ denote the
security parameter throughout. We analyze the protocol phase by phase, since
the three phases have different security goals and rely on different
assumptions.
 
\paragraph{Scope of the phase-wise claims.}
The guarantees proved below are intentionally phase-specific.
\begin{enumerate}[label=(\roman*),nosep]
    \item \textbf{Phase A} establishes privacy of the distributed key-generation
    transcript, but does \emph{not} claim privacy for the plaintext time-grid
    union steps beyond what is revealed by the final global alignment grid
    $T_{\mathrm{all}}$ and event-time grid $T$.
    \item \textbf{Phase B} establishes computational indistinguishability of the
    encrypted aggregation transcript: honest-site ciphertext content is
    computationally hidden, subject to the public metadata exclusion stated in
    Theorem~\ref{thm:phaseB-privacy}. The guarantee is assumption-based and
    relies on Assumption~\ref{asm:threshold-indcpa} for the distributed CKKS
    layer.
    \item \textbf{Phase C} establishes simulation relative to the fused
    plaintext and released output: the decryption transcript of a corrupted
    committee member is simulatable from the aggregate plaintext table
    $\{(n_i,d_i)\}_{i=1}^{|T_{\mathrm{all}}|}$, the signed metadata
    $\mathsf{md}$, the released curve $\hat S_{\mathrm{HE}}(t)$, and the adversary's local
    state, under the threshold-decryption assumptions stated in
    Theorem~\ref{thm:phaseC-privacy}. Informally, ordinary sites receiving only
    $\hat S_{\mathrm{HE}}(t)$ are exposed to less information than committee members, but
    this intuition is not formalized as a separate theorem.
\end{enumerate}
 
\paragraph{Adversarial model.}
Throughout this analysis we assume a \emph{semi-honest} (honest-but-curious),
\emph{static} adversary: corruption sets are fixed before protocol execution
begins, and corrupted parties follow the protocol faithfully while attempting
to infer additional information from their observed transcripts. The
adversarial view of a corrupted party includes its input, its full local random
tape, and all received messages. Adaptive corruption and malicious deviations
are outside the scope of this analysis.
 
The coordinator is modeled as a semi-honest, non-colluding aggregator: it
performs homomorphic operations faithfully but may attempt to infer information
from the ciphertexts and metadata it observes. Collusion between the
coordinator and corrupted sites would enlarge the adversary's view and is
outside the present model.
 
\paragraph{Corruption notation.}
Let $\mathcal{C}_S \subsetneq [K]$ denote the set of corrupted sites. Let
$\mathcal{R} \subseteq [K]$ denote the set of parties authorized to participate
in threshold decryption; in our implementation every site participates, so
$\mathcal{R} = [K]$ and $R = |\mathcal{R}| = K$. Let
$\mathcal{C}_R \subsetneq \mathcal{R}$ denote the set of corrupted decryptors.
The protocol supports a general $\theta$-of-$R$ threshold; our implementation
uses $\theta = R$. Since $\mathcal{C}_S$ and $\mathcal{C}_R$ are indexed over
different protocol roles, they are treated as independent corruption sets even
when they overlap. When $\mathcal{C}_S \cap \mathcal{C}_R \neq \emptyset$, the
shared secret-key shares are carried as persistent adversary state across
phases; the composition discussion below addresses this explicitly.
 
\paragraph{Composition note.}
The simulators defined for the three phases may be composed into a single
global simulator. The phases execute sequentially, and the inter-phase
interfaces relevant to simulation consist only of public outputs:
$(\mathsf{pk}, T_{\mathrm{all}})$ from Phase~A and the ciphertext blocks from
Phase~B. Corrupted parties' secret-key shares persist as auxiliary adversary
state and are threaded unchanged through the composition; they do not alter the
public inter-phase outputs to which the sequential composition argument
applies.
 
Table~\ref{tab:security-notation} summarizes the security-specific notation
used throughout the phase-wise proofs.
 
\begin{table}[htbp]
\centering
\footnotesize
\setlength{\tabcolsep}{4pt}
\renewcommand{\arraystretch}{1.02}
\begin{tabularx}{\linewidth}{@{}>{\raggedright\arraybackslash}p{0.28\linewidth}X@{}}
\toprule
\textbf{Symbol} & \textbf{Meaning} \\
\midrule
\multicolumn{2}{@{}l}{\textit{Security parameters and models}} \\
$\lambda$ & Security parameter. \\
PPT & Probabilistic polynomial-time (adversary, simulator, or algorithm). \\
$\mathcal{A}$ & PPT adversary. \\
$\mathsf{negl}(\lambda)$ & A negligible function of $\lambda$. \\
$=_d$ & Identical distribution between two random variables or ensembles. \\
$\stackrel{c}{\approx}$ & Computational indistinguishability (advantage bounded by $\mathsf{negl}(\lambda)$). \\
Transcript & The full set of messages and local state observed by a party (or adversary) during protocol execution, formalized as $\mathsf{View}_{\mathcal{A}}^{(\cdot)}$. \\
\midrule
\multicolumn{2}{@{}l}{\textit{Protocol and corruption}} \\
$\Pi$ & The protocol $\Pi_{\mathrm{KM\text{-}Fed\text{-}CKKS}}$ (shorthand used in proofs). \\
$\mathcal{C}_S \subsetneq [K]$ & Static set of corrupted sites (strict subset); fixed before execution begins. \\
$\mathcal{C}_R \subsetneq \mathcal{R}$ & Static set of corrupted decryptors (strict subset); fixed before execution begins. \\
$\rho_k$ & Local key-generation randomness of site $k$ (revealed if $k \in \mathcal{C}_S$), part of its full random tape. \\
$\rho_k^{\mathrm{enc}}$ & Encryption randomness used by site $k$ in Phase~B (revealed if $k \in \mathcal{C}_S$), part of its full random tape. \\
\midrule
\multicolumn{2}{@{}l}{\textit{Adversarial views}} \\
$\mathsf{View}_{\mathcal{A}}^{(A)}(\Pi)$ & Phase~A view: $\mathsf{Params}$, corrupted DKG state $\{\mathsf{sk}_k, \rho_k\}_{k \in \mathcal{C}_S}$, joint $\mathsf{pk}$, complete DKG transcript, global alignment grid $T_{\mathrm{all}}$, event-time grid $T$. \\
$\mathsf{View}_{\mathcal{A}}^{(B)}(\Pi)$ & Phase~B view: public setup; corrupted sites' inputs $\{(d_i^{(k)}, n_i^{(k)})_{i=1}^{L}\}_{k \in \mathcal{C}_S}$, secret-key shares, and encryption randomness; all uploaded ciphertexts $\{\mathsf{ct}_{\iota}^{(k)}\}_{k \in [K],\, \iota \in \mathcal{I}}$ and aggregated ciphertexts $\{\mathsf{ct}_{\Sigma,\iota}\}_{\iota \in \mathcal{I}}$. \\
$\mathsf{View}_{\mathcal{A}}^{(C)}(\Pi)$ & Phase~C view: aggregated ciphertexts $\{\mathsf{ct}_{\Sigma,\iota}\}_{\iota \in \mathcal{I}}$; corrupted secret-key shares $\{\mathsf{sk}_r\}_{r \in \mathcal{C}_R}$ and partial decryption shares $\{\Delta_{\iota}^{(r)}\}_{r \in \mathcal{C}_R,\, \iota \in \mathcal{I}}$; signed metadata $\mathsf{md}$; aggregate plaintext table $P:=\{(n_i, d_i)\}_{i=1}^{L}$; released curve $\hat S_{\mathrm{HE}}(t)$. \\
\midrule
\multicolumn{2}{@{}l}{\textit{Simulators}} \\
$\mathcal{S}_A$ & PPT simulator for Phase~A; input: $(\mathsf{Params},\ \{\mathsf{sk}_k, \rho_k\}_{k \in \mathcal{C}_S},\ T_{\mathrm{all}},\ T)$. \\
$\mathcal{S}_B$ & PPT simulator for Phase~B; input: $(\mathsf{Params},\ \{(d_i^{(k)}, n_i^{(k)})_{i=1}^{L},\ \mathsf{sk}_k,\ \rho_k^{\mathrm{enc}}\}_{k \in \mathcal{C}_S},\ \mathsf{pk})$. \\
$\mathcal{S}_C$ & PPT simulator for Phase~C; input: $(\{\mathsf{sk}_r\}_{r \in \mathcal{C}_R},\ \{\mathsf{ct}_{\Sigma,\iota}\}_{\iota \in \mathcal{I}},\ \mathsf{md},\ P,\ \hat S_{\mathrm{HE}}(t))$. \\
$\mathcal{S}_{\mathrm{share}}$ & PPT share-privacy simulator guaranteed by Assumption~(ii) of Theorem~\ref{thm:phaseC-privacy}. \\
\midrule
\multicolumn{2}{@{}l}{\textit{Phase~B proof objects}} \\
$|\mathcal{I}|$ & Total number of ciphertext blocks per site under the chosen layout; $|\mathcal{I}|=M_{\mathrm{int}}$ for interleaving and $|\mathcal{I}|=M_{\mathrm{sep}}$ for separate packing. Public and identical across sites. \\
$N$ & Total number of honest-site ciphertext blocks in the hybrid argument; $N = (K - |\mathcal{C}_S|) \cdot |\mathcal{I}|$. \\
$H^{(\ell)}$ & Hybrid distribution in the Phase~B proof: first $\ell$ honest-site ciphertext blocks replaced by encryptions of $\mathbf{0}_B$ (the zero vector in $\mathbb{R}^B$), remainder real; $H^{(0)}$ is the real execution, $H^{(N)}$ is the simulator output. \\
$\mathsf{Real}$,\,$\mathsf{Real}'$ & Real execution ($\mathsf{Real}$) and a re-randomized equivalent execution ($\mathsf{Real}'$) used in the initial hybrid step of the Phase~B proof. \\
\midrule
\multicolumn{2}{@{}l}{\textit{Phase~C proof objects}} \\
$\mathsf{md}$ & Signed metadata blob: coordinator-authenticated statement attesting the cohort identifier, committed alignment grid $T_{\mathrm{all}}$, and ciphertext block identifiers $\{\mathsf{ct}_{\Sigma,\iota}\}_{\iota \in \mathcal{I}}$. \\
$P$ & Fused plaintext aggregate table $\{(n_i, d_i)\}_{i=1}^{L}$ obtained after threshold decryption; used as input in the share-privacy assumption of Theorem~\ref{thm:phaseC-privacy}. \\
\midrule
\multicolumn{2}{@{}l}{\textit{Assumptions}} \\
Assumption~\ref{asm:threshold-indcpa} & Threshold-aware auxiliary-input IND-CPA for the distributed CKKS layer: ciphertext indistinguishability holds even when the adversary holds a strict subset of secret-key shares and polynomially many auxiliary ciphertexts under the joint public key. \\
\bottomrule
\end{tabularx}
\caption{Security-specific notation used in the phase-wise proofs (Theorems~\ref{thm:phaseA-privacy}--\ref{thm:phaseC-privacy} and Corollary~\ref{cor:end-to-end}). Symbols shared with the main protocol are defined in Table~\ref{tab:notation-hekm}; this table contains only notation that is specific to or has a refined meaning within the security analysis.}
\label{tab:security-notation}
\end{table}
 
\subsubsection{Phase A: setup and time-grid union}
\label{subsubsec:phaseA-security}
 
Phase~A consists of two substeps: (a) distributed key generation, which
produces the joint public key $\mathsf{pk}$ and local secret-key shares
$\{\mathsf{sk}_k\}$; and (b) construction of the global alignment grid
\[
T_{\mathrm{all}} = \mathrm{Sort}\Bigl(\bigcup_{k=1}^K T_k^{\mathrm{all}}\Bigr).
\]
The event-time subset $T = \{t \in T_{\mathrm{all}} : \exists\text{ an event at }t\}$
is a theoretical object; it is not constructed explicitly in the implementation.
The first substep is data-independent; the second intentionally reveals the
union of local observed-time sets.
 
\paragraph{Adversarial view in Phase A.}
Let $\mathsf{View}_{\mathcal{A}}^{(A)}(\Pi)$ denote the Phase~A view of a
PPT adversary $\mathcal{A}$ corrupting $\mathcal{C}_S \subsetneq [K]$. This
view consists of:
\begin{enumerate}[label=(\alph*),nosep]
    \item the public CKKS parameters $\mathsf{Params}$ and all public setup
    values;
    \item the corrupted sites' secret-key shares
    $\{\mathsf{sk}_k\}_{k \in \mathcal{C}_S}$ and their local key-generation
    randomness $\{\rho_k\}_{k \in \mathcal{C}_S}$;
    \item the joint public key $\mathsf{pk}$ and the complete DKG transcript; and
    \item the final global alignment grid $T_{\mathrm{all}}$ and event-time grid $T$.
\end{enumerate}
\noindent In the implementation used here, only $T_{\mathrm{all}}$ is explicitly
disclosed; the event-time grid $T$ is included in the adversarial view for
theoretical generality, since $T \subseteq T_{\mathrm{all}}$ is derivable from
the data. The theorem therefore provides a conservative (overgenerous in
leakage) guarantee.
 
\begin{theorem}[Phase~A setup privacy]
\label{thm:phaseA-privacy}
Assume the distributed key-generation mechanism used in $\Pi$ is correctly
specified and data-independent: for any fixed corrupted-party state
$\{\mathsf{sk}_k, \rho_k\}_{k \in \mathcal{C}_S}$, the joint distribution of the
DKG transcript and joint public key depends only on the parties'
key-generation randomness and secret-key material, not on any site's plaintext
dataset. Then there exists a PPT simulator $\mathcal{S}_A$ such that
\[
\mathcal{S}_A\bigl(
\mathsf{Params},\,
\{\mathsf{sk}_k, \rho_k\}_{k \in \mathcal{C}_S},\,
T_{\mathrm{all}},\, T
\bigr)
\;=_d\;
\mathsf{View}_{\mathcal{A}}^{(A)}(\Pi),
\]
where $=_d$ denotes identical distribution.
\end{theorem}
 
\begin{proof}
The simulator $\mathcal{S}_A$ operates as follows. For each honest party
$k \notin \mathcal{C}_S$, it samples fresh key-generation randomness
$\tilde\rho_k$ independently from the same distribution specified by the DKG
protocol. It then runs the DKG protocol using the real corrupted-party inputs
$\{\mathsf{sk}_k, \rho_k\}_{k \in \mathcal{C}_S}$ and the freshly sampled
$\{\tilde\rho_k\}_{k \notin \mathcal{C}_S}$, producing a simulated DKG
transcript and joint public key $\tilde{\mathsf{pk}}$. It appends
$T_{\mathrm{all}}$ and $T$.
 
By the data-independence assumption, the joint distribution of
$(\text{DKG transcript}, \mathsf{pk})$ depends only on key-generation
randomness and secret-key material. In the real execution, the honest parties'
randomness $\{\rho_k\}_{k \notin \mathcal{C}_S}$ is drawn independently from the
prescribed distribution; the simulator draws
$\{\tilde\rho_k\}_{k \notin \mathcal{C}_S}$ from the identical distribution.
Since the corrupted-party state is identical in both cases, and the honest
randomness has the same distribution, the joint distribution of the DKG
transcript, $\tilde{\mathsf{pk}}$, $T_{\mathrm{all}}$, and $T$ in the simulation is
identical to that of the DKG transcript, $\mathsf{pk}$, $T_{\mathrm{all}}$, and $T$
in a real execution.
\end{proof}
 
\paragraph{Qualification for the time-grid union.}
Theorem~\ref{thm:phaseA-privacy} does not claim privacy for the individual
sets $T_k^{\mathrm{all}}$ beyond what $T_{\mathrm{all}}$ reveals. In the present
implementation the union is computed in plaintext; timepoints are treated as
non-sensitive metadata. If a stricter deployment requires hiding $T_k^{\mathrm{all}}$, the
union step must be replaced by a private-set-union protocol analyzed
separately.
 
\subsubsection{Phase B: encrypted aggregation}
\label{subsubsec:phaseB-security}
 
Each site $k$ computes local Kaplan--Meier count pairs
$(d_i^{(k)}, n_i^{(k)})_{i=1}^{|T_{\mathrm{all}}|}$ over the shared alignment
grid $T_{\mathrm{all}}$, packs them into CKKS plaintext vectors, encrypts them
under $\mathsf{pk}$, and uploads the resulting ciphertext blocks. Sites with
no event at a given grid point contribute $d_i^{(k)} = 0$ for that position.
The coordinator performs only homomorphic addition on received ciphertexts.
 
\paragraph{Packing notation.}
Let $B$ denote the number of plaintext slots in a single CKKS ciphertext, a
public parameter (with $B = n/2$ for CKKS). Each site $k$ packs its count
pairs over the alignment grid $T_{\mathrm{all}}$ into CKKS plaintext vectors
using a fixed, publicly known packing algorithm. With packed grid length
$L=|T_{\mathrm{all}}|$, the public ciphertext-index set $\mathcal{I}$ has
$|\mathcal{I}|=M_{\mathrm{int}}=\lceil 2L/B\rceil$ for interleaving and
$|\mathcal{I}|=M_{\mathrm{sep}}=2\lceil L/B\rceil$ for separate packing.
We write $\mathbf{z}_{\iota}^{(k)} \in \mathbb{R}^{B}$ for the packed vector
of site $k$ at index $\iota \in \mathcal{I}$. Since $B$, $L$, and
$\mathcal{I}$ are public, all ciphertext block dimensions are derivable without
accessing any site's private data. Because every site packs counts over the
same global alignment grid $T_{\mathrm{all}}$, the index set $\mathcal{I}$
is identical for all sites. The proof below is stated for a generic
$\iota$; for interleaving each $\iota$ is a block index, while for separate
packing each $\iota=(j,s)$ pairs a block index $j\in[M_0]$ with a stream
label $s\in\{d,n\}$.
 
\paragraph{Threshold-aware ciphertext privacy assumption.}
Standard IND-CPA for public-key encryption is stated for adversaries that see
the public key but not any secret-key material. In our distributed setting, a
Phase~B adversary may additionally hold a strict subset of secret-key shares
produced by the DKG. To make the security claim precise, we isolate the
following assumption for the distributed CKKS layer used by
$\Pi_{\mathrm{KM\text{-}Fed\text{-}CKKS}}$.
 
\begin{assumption}[Threshold-aware auxiliary-input IND-CPA]
\label{asm:threshold-indcpa}
Let $\mathsf{pk}$ be a joint public key produced by the distributed
key-generation procedure, and let
$\{\mathsf{sk}_k\}_{k \in \mathcal{C}_S}$ be any strict subset of secret-key
shares with $|\mathcal{C}_S| < \theta$. Then ciphertext privacy remains
computationally secure against any PPT distinguisher given:
\begin{enumerate}[label=(\roman*),nosep]
    \item the public key $\mathsf{pk}$,
    \item the corrupted shares $\{\mathsf{sk}_k\}_{k \in \mathcal{C}_S}$, and
    \item polynomially many auxiliary ciphertexts under $\mathsf{pk}$,
\end{enumerate}
in the following sense: for any equal-dimension plaintexts
$\mathbf{m}_0, \mathbf{m}_1$, the distributions
\[
(\mathsf{pk}, \{\mathsf{sk}_k\}_{k \in \mathcal{C}_S}, \mathsf{aux},
E_{\mathsf{pk}}(\mathbf{m}_0))
\quad\text{and}\quad
(\mathsf{pk}, \{\mathsf{sk}_k\}_{k \in \mathcal{C}_S}, \mathsf{aux},
E_{\mathsf{pk}}(\mathbf{m}_1))
\]
are computationally indistinguishable for every PPT-computable auxiliary input
$\mathsf{aux}$ consisting of polynomially many ciphertexts under
$\mathsf{pk}$.
\end{assumption}
 
\noindent\emph{Remark.}
Assumption~\ref{asm:threshold-indcpa} is stronger than standard IND-CPA because
the distinguisher is additionally given a strict subset of secret-key shares and
auxiliary ciphertexts. We therefore state it explicitly as an assumption on the
distributed CKKS layer used by our protocol; establishing it formally for a
specific threshold CKKS instantiation is outside the scope of this work.
 
\paragraph{Adversarial view in Phase B.}
Let $\mathsf{View}_{\mathcal{A}}^{(B)}(\Pi)$ denote the pre-decryption Phase~B
view of a PPT adversary $\mathcal{A}$ corrupting $\mathcal{C}_S \subsetneq [K]$.
This view consists of:
\begin{enumerate}[label=(\alph*),nosep]
  \item the public parameters $\mathsf{Params}$, the public key $\mathsf{pk}$,
        and public setup values from Phase~A;
  \item the corrupted sites' inputs
        $\{(d_i^{(k)}, n_i^{(k)})_{i=1}^{|T_{\mathrm{all}}|}\}_{k \in \mathcal{C}_S}$,
        secret-key shares $\{\mathsf{sk}_k\}_{k \in \mathcal{C}_S}$, and
        encryption randomness
        $\{\rho_k^{\mathrm{enc}}\}_{k \in \mathcal{C}_S}$;
  \item ciphertext uploads $\{\mathsf{ct}_j^{(k)}\}_{k \in [K],\, j \in [M]}$ from
        all sites; and
  \item aggregated ciphertexts $\{\mathsf{ct}_{\Sigma,j}\}_{j=1}^{M}$ produced
        by the coordinator.
\end{enumerate}
The coordinator is treated as semi-honest: it performs homomorphic aggregation
faithfully. The corrupted sites' encryption randomness
$\{\rho_k^{\mathrm{enc}}\}_{k \in \mathcal{C}_S}$ is included in the view because
the adversarial model includes each corrupted party's full local random tape.
 
\begin{theorem}[Transcript privacy of Phase~B encrypted aggregation]
\label{thm:phaseB-privacy}
Assume Assumption~\ref{asm:threshold-indcpa}. Then there exists a PPT simulator
$\mathcal{S}_B$ such that
\[
\mathcal{S}_B\!\Bigl(
\mathsf{Params},\,
\{(d_i^{(k)}, n_i^{(k)})_{i=1}^{|T_{\mathrm{all}}|},\, \mathsf{sk}_k,\,
\rho_k^{\mathrm{enc}}\}_{k \in \mathcal{C}_S},\,
\mathsf{pk}
\Bigr)
\;\stackrel{c}{\approx}\;
\mathsf{View}_{\mathcal{A}}^{(B)}(\Pi).
\]
That is, the Phase~B encrypted aggregation transcript is computationally
indistinguishable from a simulation that does not use the honest sites' local
Kaplan--Meier count vectors. This guarantee covers ciphertext \emph{content}
privacy only; public metadata such as the common block count $M$, ciphertext
sizes, and upload timing is excluded.
\end{theorem}
 
\begin{proof}
We construct $\mathcal{S}_B$ and prove indistinguishability by a standard
sequence-of-games argument.
 
\paragraph{Simulator construction.}
Given $\mathsf{Params}$, the corrupted sites' inputs, corrupted secret-key
shares, corrupted encryption randomness, and $\mathsf{pk}$, the simulator
$\mathcal{S}_B$ proceeds as follows.
For notational simplicity, $\rho_k^{\mathrm{enc}}$ denotes the full encryption
random tape of corrupted site $k$, whose per-block components are written as
$\rho_{k,j}^{\mathrm{enc}}$ when individual ciphertext blocks are referenced.
 
\begin{enumerate}[label=(\roman*),nosep]
\item It sets $\tilde{\mathsf{pk}} := \mathsf{pk}$.
 
\item For each corrupted site $k \in \mathcal{C}_S$, it computes the packed
plaintext vectors $\mathbf{z}_j^{(k)}$ from the given corrupted inputs and
encrypts them using the given randomness:
\[
\widetilde{\mathsf{ct}}_j^{(k)}
:=
E_{\tilde{\mathsf{pk}}}(\mathbf{z}_j^{(k)};\, \rho_{k,j}^{\mathrm{enc}}),
\qquad j=1,\dots,M.
\]
Hence the corrupted sites' ciphertexts are reproduced exactly.
 
\item Since every site packs over the same public alignment grid
$T_{\mathrm{all}}$, each site uploads exactly $M$ ciphertext blocks. For each
honest site $k \notin \mathcal{C}_S$ and each $j = 1,\dots,M$, the simulator
outputs
\[
\widetilde{\mathsf{ct}}_j^{(k)}
\leftarrow
E_{\tilde{\mathsf{pk}}}(\mathbf{0}_B),
\]
using fresh randomness.
 
\item For each block index $j$, it computes the aggregate ciphertext by CKKS
homomorphic addition:
\[
\widetilde{\mathsf{ct}}_{\Sigma,j}
=
\bigoplus_{k=1}^{K} \widetilde{\mathsf{ct}}_j^{(k)}.
\]
\end{enumerate}
 
\paragraph{Hybrid argument.}
Let
\[
N = (K - |\mathcal{C}_S|) \cdot M
\]
be the total number of honest-site ciphertext blocks. Fix an arbitrary but
deterministic ordering of these honest blocks, and define hybrids
\[
H^{(0)}, H^{(1)}, \dots, H^{(N)},
\]
where:
\begin{itemize}[nosep]
    \item $H^{(0)}$ is the real Phase~B transcript;
    \item in $H^{(\ell)}$, the first $\ell$ honest ciphertext blocks in the
    fixed ordering are replaced by encryptions of $\mathbf{0}_B$, while all
    remaining honest blocks are real encryptions; and
    \item $H^{(N)}$ is exactly the simulator output.
\end{itemize}
 
The corrupted sites' inputs, secret-key shares, encryption randomness, and
ciphertexts are identical in every hybrid. The only changes occur in the honest
ciphertext blocks.
 
Now consider adjacent hybrids $H^{(\ell-1)}$ and $H^{(\ell)}$. They differ in
exactly one honest ciphertext block, say block $(k,j)$. In $H^{(\ell-1)}$ that
block is
\[
E_{\mathsf{pk}}(\mathbf{z}_j^{(k)}),
\]
whereas in $H^{(\ell)}$ it is
\[
E_{\mathsf{pk}}(\mathbf{0}_B).
\]
All other ciphertexts, including all corrupted-site ciphertexts and the other
honest-site ciphertexts, are unchanged. Therefore they may be treated as
auxiliary input in Assumption~\ref{asm:threshold-indcpa}, together with the
corrupted secret-key shares.
 
If some PPT distinguisher separated $H^{(\ell-1)}$ from $H^{(\ell)}$ with
non-negligible advantage, then this would violate
Assumption~\ref{asm:threshold-indcpa} for the plaintext pair
\[
\bigl(\mathbf{z}_j^{(k)},\, \mathbf{0}_B\bigr),
\]
with the unchanged ciphertext blocks and corrupted shares supplied as auxiliary
input. Hence
\[
H^{(\ell-1)} \stackrel{c}{\approx} H^{(\ell)}
\qquad\text{for every } \ell = 1, \dots, N.
\]
 
Because $N = \mathrm{poly}(\lambda)$, a standard telescoping hybrid argument
gives
\[
H^{(0)} \stackrel{c}{\approx} H^{(N)}.
\]
 
Finally, the aggregate ciphertexts are deterministic efficient functions of the
uploaded ciphertext blocks. Computational indistinguishability is preserved
under such post-processing, so the aggregate ciphertext components remain
computationally indistinguishable as well.
 
Since $H^{(N)}$ is exactly the transcript output by $\mathcal{S}_B$, we obtain
\[
\mathcal{S}_B\!\Bigl(
\mathsf{Params},\,
\{(d_i^{(k)}, n_i^{(k)})_{i=1}^{|T_{\mathrm{all}}|},\, \mathsf{sk}_k,\,
\rho_k^{\mathrm{enc}}\}_{k \in \mathcal{C}_S},\,
\mathsf{pk}
\Bigr)
\stackrel{c}{\approx}
\mathsf{View}_{\mathcal{A}}^{(B)}(\Pi),
\]
as claimed.
\end{proof}
 
\paragraph{Remark (metadata leakage).}
The above argument covers ciphertext \emph{content} privacy. The block count
$M$, upload timing, and ciphertext sizes are observable metadata. Because all
sites pack over the same global alignment grid $T_{\mathrm{all}}$, the block
count $M$ is identical across sites and reveals only $|T_{\mathrm{all}}|$,
which is already public from Phase~A. Upload timing may reveal participation
ordering and is outside the cryptographic guarantee; it should be addressed at
the network layer if needed.
 
\subsubsection{Phase C: threshold decryption and output gating}
\label{subsubsec:phaseC-security}
 
Phase~C releases a public result, so the security goal is not full hiding but
simulation relative to the information legitimately revealed.
 
\paragraph{Scope of the Phase~C claim.}
The following Phase~C claim is output-relative: it simulates the view of a
corrupted decryptor coalition given the aggregate plaintext table recovered
inside the decryption committee and the released curve $\hat S_{\mathrm{HE}}(t)$.
It does \emph{not} assert that a subthreshold coalition can recover the
aggregate table on its own. The adversary is modeled relative to the internal
decryption outcome available within the committee execution. Informally,
ordinary sites receiving only $\hat S_{\mathrm{HE}}(t)$ are exposed to less
information than committee members, but this is not formalized as a separate
theorem. The theorem does not claim security against adaptive decryption
queries, which are operationally excluded by policy controls in
Section~\ref{subsec:threat-key}. The simulated leakage for Phase~C is
$\mathcal{L}_C = (\mathsf{md}, P, \hat S_{\mathrm{HE}}(t))$, together with
the corrupted parties' local keys and local randomness.
 
\paragraph{Signed metadata.}
The decryption request is accompanied by a signed metadata blob $\mathsf{md}$:
a coordinator-authenticated statement attesting the cohort identifier, the
committed alignment grid $T_{\mathrm{all}}$, and the ciphertext block identifiers
$\{\mathsf{ct}_{\Sigma,\iota}\}_{\iota \in \mathcal{I}}$. The signature is produced under a signing key
provisioned separately from the CKKS setup, for example via out-of-band PKI.
The Phase~C adversarial view includes $\mathsf{md}$ directly; the simulator
receives $\mathsf{md}$ as explicit input and copies it verbatim, requiring no
signing capability.
 
\paragraph{Adversarial view in Phase C.}
Let $\mathsf{View}_{\mathcal{A}}^{(C)}(\Pi)$ denote the Phase~C view of a
PPT adversary corrupting $\mathcal{C}_R \subsetneq \mathcal{R}$. This view
consists of:
\begin{enumerate}[label=(\alph*),nosep]
    \item the aggregated ciphertexts $\{\mathsf{ct}_{\Sigma,\iota}\}_{\iota\in\mathcal{I}}$;
    \item the corrupted decryptors' secret-key shares
    $\{\mathsf{sk}_r\}_{r \in \mathcal{C}_R}$;
    \item the corrupted decryptors' partial decryption shares
    $\{\Delta_{\iota}^{(r)}\}_{r \in \mathcal{C}_R,\, \iota\in\mathcal{I}}$,
    computed deterministically from
    $(\mathsf{sk}_r, \mathsf{ct}_{\Sigma,\iota})$ by the partial decryption
    algorithm;
    \item the signed metadata blob $\mathsf{md}$;
    \item the aggregate plaintext table
    $P:=\{(n_i, d_i)\}_{i=1}^{L}$ produced by threshold decryption; and
    \item the released survival curve $\hat S_{\mathrm{HE}}(t)$, computed
    deterministically from $P$ via the Kaplan--Meier formula.
\end{enumerate}
The aggregate table is the direct plaintext output of the decryption step and
is modeled relative to the internal decryption outcome available within the
committee execution. In deployments where only
$\hat S_{\mathrm{HE}}(t)$ is externally published, the table remains internal
to the decryptor committee. The Kaplan--Meier estimator depends only on the
event-time subset $T \subseteq T_{\mathrm{all}}$, since $d_i = 0$ for time
points without events.
 
\begin{theorem}[Phase~C privacy relative to the fused plaintext and released output]
\label{thm:phaseC-privacy}
Assume the threshold decryption mechanism satisfies:
\begin{enumerate}[label=(\roman*),nosep]
    \item \emph{Threshold confidentiality:} no PPT adversary holding fewer than
    $\theta$ secret-key shares can recover the plaintext of a well-formed
    ciphertext with non-negligible advantage in $\lambda$.
    \item \emph{Share privacy:} for any fixed ciphertext collection
    $\{\mathsf{ct}_{\Sigma,\iota}\}_{\iota\in\mathcal{I}}$, any strict subset
    $\mathcal{C}_R$ with $|\mathcal{C}_R| < \theta$, and the correct fused
    plaintext table $P = \{(n_i, d_i)\}_{i=1}^{L}$, there exists a PPT
    simulator $\mathcal{S}_{\mathrm{share}}$ such that the simulated honest
    partial shares are computationally indistinguishable from the real ones:
    \[
    \mathcal{S}_{\mathrm{share}}\!\bigl(
    \{\mathsf{sk}_r\}_{r \in \mathcal{C}_R},\,
    \{\mathsf{ct}_{\Sigma,\iota}\}_{\iota\in\mathcal{I}},\,
    P
    \bigr)
    \;\stackrel{c}{\approx}\;
    \{\Delta_{\iota}^{(r)}\}_{r \notin \mathcal{C}_R,\, \iota\in\mathcal{I}}.
    \]
    \item \emph{Fusion correctness:} a valid threshold set of partial shares
    combined by the fusion algorithm yields the correct plaintext.
    \item \emph{Deterministic partial decryption:} the partial decryption
    algorithm is deterministic given
    $(\mathsf{sk}_r, \mathsf{ct}_{\Sigma,\iota})$.
\end{enumerate}
Assumptions (ii)--(iv) are used directly in the simulation proof below.
Assumption~(i) is not used in the indistinguishability argument itself, but
justifies interpreting the simulator input as an upper bound on the information
available to any subthreshold coalition.
 
Then there exists a PPT simulator $\mathcal{S}_C$ such that
\[
\mathcal{S}_C\!\Bigl(
\{\mathsf{sk}_r\}_{r \in \mathcal{C}_R},\,
\{\mathsf{ct}_{\Sigma,\iota}\}_{\iota\in\mathcal{I}},\,
\mathsf{md},\,
P,\,
\hat S_{\mathrm{HE}}(t)
\Bigr)
\;\stackrel{c}{\approx}\;
\mathsf{View}_{\mathcal{A}}^{(C)}(\Pi).
\]
In the all-of-$R$ implementation, this holds for every strict subset
$\mathcal{C}_R \subsetneq \mathcal{R}$, since no incomplete coalition can
complete decryption independently.
\end{theorem}
 
\noindent\emph{Interpretation.}
Theorem~\ref{thm:phaseC-privacy} is a protocol-level reduction to the assumed
security of the threshold decryption mechanism, not a derivation from RLWE
alone.
 
\begin{proof}
Construct $\mathcal{S}_C$ as follows, given
\[
(\{\mathsf{sk}_r\}_{r \in \mathcal{C}_R},\,
\{\mathsf{ct}_{\Sigma,\iota}\}_{\iota\in\mathcal{I}},\,
\mathsf{md},\,
P,\,
\hat S_{\mathrm{HE}}(t)).
\]
\begin{enumerate}[label=(\alph*),nosep]
    \item \emph{Metadata.} Copy $\mathsf{md}$ verbatim; no signing key is
    needed.
    \item \emph{Corrupted partial shares.} For each $r \in \mathcal{C}_R$ and
    each $\iota \in \mathcal{I}$, compute $\Delta_{\iota}^{(r)}$ using
    $\mathsf{sk}_r$ and $\mathsf{ct}_{\Sigma,\iota}$ exactly as in the real
    protocol. By assumption~(iv), this matches the real partial shares with
    probability~1.
    \item \emph{Honest partial shares.} Invoke the share-privacy simulator
    $\mathcal{S}_{\mathrm{share}}$ from assumption~(ii) with inputs
    \[
    (\{\mathsf{sk}_r\}_{r \in \mathcal{C}_R},\,
    \{\mathsf{ct}_{\Sigma,\iota}\}_{\iota\in\mathcal{I}},\,
    P)
    \]
    to obtain simulated partial shares
    $\{\widetilde{\Delta}_{\iota}^{(r)}\}_{r \notin \mathcal{C}_R,\, \iota\in\mathcal{I}}$.
    \item \emph{Public outputs.} Append $P$ and $\hat S_{\mathrm{HE}}(t)$.
\end{enumerate}
 
\paragraph{Indistinguishability.}
We compare the simulated transcript $\mathcal{S}_C(\cdot)$ with
$\mathsf{View}_{\mathcal{A}}^{(C)}(\Pi)$ component by component.
 
\begin{itemize}[nosep]
\item \emph{Metadata:} copied exactly, hence identically distributed.
\item \emph{Corrupted partial shares:} computed identically to the real
protocol by assumption~(iv), hence identically distributed.
\item \emph{Honest partial shares:} computationally indistinguishable from the
real shares by share privacy, assumption~(ii).
\item \emph{Aggregate table and released curve:} given as simulator input; in
the real protocol these are computed from the ciphertexts via threshold
decryption, which by fusion correctness, assumption~(iii), yields the same
aggregate table. The curve $\hat S_{\mathrm{HE}}(t)$ is a deterministic
function of that table. Hence these components are identically distributed.
\end{itemize}
 
Any PPT distinguisher between the simulation and the real Phase~C view would
therefore distinguish simulated honest shares from real honest shares,
violating share privacy, assumption~(ii). Hence
\[
\mathcal{S}_C\!\Bigl(
\{\mathsf{sk}_r\}_{r \in \mathcal{C}_R},\,
\{\mathsf{ct}_{\Sigma,\iota}\}_{\iota\in\mathcal{I}},\,
\mathsf{md},\,
P,\,
\hat S_{\mathrm{HE}}(t)
\Bigr)
\stackrel{c}{\approx}
\mathsf{View}_{\mathcal{A}}^{(C)}(\Pi).
\qedhere
\]
\end{proof}
 
\paragraph{Deployment qualification.}
The theorem is strongest when fusion and Kaplan--Meier evaluation are performed
inside the decryptor committee and only $\hat S_{\mathrm{HE}}(t)$ is released
externally. If the combiner is co-located with the coordinator, the coordinator
transiently observes $P$; the theorem still characterizes the committee-member
transcript while ordinary sites observe only $\hat S_{\mathrm{HE}}(t)$. The
coordinator's transient access to the aggregate table is a potential disclosure
surface and should be mitigated operationally, for example through hardened
execution environments, audit logging, and access controls.
 
\begin{corollary}[End-to-end privacy of $\Pi_{\mathrm{KM\text{-}Fed\text{-}CKKS}}$: phase-level leakage summary]
\label{cor:end-to-end}
Under the assumptions of Theorems~\ref{thm:phaseA-privacy},
\ref{thm:phaseB-privacy}, and \ref{thm:phaseC-privacy}, the per-phase
privacy guarantees can be jointly interpreted as follows:
\begin{enumerate}[label=(\roman*),nosep]
    \item \textbf{Phase A:} the setup transcript is identically distributed to
    a simulation that uses no honest-site plaintext data, beyond the
    intentionally revealed alignment grid $T_{\mathrm{all}}$ and event-time grid $T$.
    \item \textbf{Phase B:} the encrypted aggregation transcript is
    computationally indistinguishable from a simulation that uses no honest-site
    Kaplan--Meier count vectors, beyond publicly observable metadata such as
    the ciphertext-index set $\mathcal{I}$, ciphertext sizes, and upload timing.
    \item \textbf{Phase C:} the decryption transcript of a corrupted committee
    member is computationally indistinguishable from a simulation that uses only
    the aggregate plaintext table $P$, the signed metadata, and the released
    curve $\hat S_{\mathrm{HE}}(t)$.
\end{enumerate}
\end{corollary}
 
\begin{proof}
Items (i)--(iii) are the direct phase-wise consequences of
Theorems~\ref{thm:phaseA-privacy},~\ref{thm:phaseB-privacy},
and~\ref{thm:phaseC-privacy}. The sequential composition discussion above
justifies viewing them jointly as an end-to-end phase-level leakage summary,
because the following conditions hold:
(1) the phases execute sequentially;
(2) each phase's public output, namely $(\mathsf{pk}, T_{\mathrm{all}}, T)$
from Phase~A and the ciphertext blocks from Phase~B, is the sole inter-phase
communication channel;
(3) corrupted parties' secret-key shares persist as auxiliary adversary state
threaded unchanged through the composition and do not affect the public
inter-phase outputs; and
(4) all individual-phase simulators run in PPT.
 
When $\mathcal{C}_S \cap \mathcal{C}_R \neq \emptyset$, the shared secret-key
shares appear in the adversary's input for both Phase~B and Phase~C. This does
not invalidate the per-phase claims: the Phase~B simulator takes these shares
as explicit input, and the Phase~C simulator independently takes its corrupted
shares as explicit input. Crucially,
Theorem~\ref{thm:phaseC-privacy} holds for \emph{any} fixed input ciphertext
collection $\{\mathsf{ct}_{\Sigma,\iota}\}_{\iota\in\mathcal{I}}$, so the
Phase~C guarantee remains valid even when the aggregated ciphertexts are
simulated outputs from Phase~B rather than the real aggregates.
\end{proof}
 
\paragraph{Security discussion and limitations.}
The above analysis establishes simulation-based privacy under a semi-honest,
static adversarial model. Extending to the malicious setting would require
additional mechanisms such as zero-knowledge proofs of correct encryption or
range proofs compatible with CKKS approximate arithmetic, ciphertext validity
checks, and verifiable threshold decryption, and is left for future work.
 
The guarantees are transcript-based and do not eliminate all leakage channels.
Upload timing and network-layer metadata may reveal participation ordering and
should be mitigated at the transport layer if needed. The plaintext time-grid
union reveals the union of local timepoint sets, treated as non-sensitive in
the present deployment but potentially requiring a private-set-union protocol in
stricter settings. The analysis also excludes coordinator-site collusion; such
collusion would enlarge the adversarial view beyond the present model, combining
the coordinator's observed ciphertexts and metadata with corrupted sites'
secret-key shares and plaintext inputs, and would require a separate analysis.
Our Phase~B claim therefore relies on
Assumption~\ref{asm:threshold-indcpa} for the distributed CKKS layer, rather
than on a standalone reduction proved in this work.
 
The Phase~C guarantee is relative to the fused plaintext and released output,
and depends on trusted execution of the threshold decryption workflow. If
intermediate aggregate plaintexts are exposed outside the decryptor committee,
for example due to co-location of the combiner with the coordinator, then
additional trust assumptions or secure execution environments are required.
These considerations highlight that the protocol provides strong cryptographic
protection for local Kaplan--Meier contributions, while leaving certain
system-level leakage paths and deployment choices to complementary safeguards.
 
\subsection{IID vs.\ non-IID (practical invariance)}
The cohort is fixed centrally (no duplicate counting), preprocessing is harmonized
(units, censoring/ties), and all sites align to the common global alignment grid
$T_{\mathrm{all}}$ in R1, from which the event-time grid
$T \subseteq T_{\mathrm{all}}$ is derived. Under these conditions, per-time-point
counts add exactly on the event-time grid
($d_i=\sum_k d_i^{(k)}$, $n_i=\sum_k n_i^{(k)}$ for $t_i \in T$), so the federated
KM is \emph{partition-invariant} (Theorem~\ref{thm:plain-invariance}): IID vs.\
non-IID does not change $\hat S_{\mathrm{HE}}(t)$. Differences arise only if harmonization or
cohort assumptions are violated (Section~\ref{sec:discussion}).
 
\subsection{Design choices and guarantees}
The number of ciphertext blocks $M$ grows stepwise with the length of the grid
used for encrypted alignment and depends on the packing layout. In the current
implementation, packing is performed over $T_{\mathrm{all}}$, so
\[
M_{\mathrm{int}}=\left\lceil \frac{2|T_{\mathrm{all}}|}{B}\right\rceil
\qquad\text{and}\qquad
M_{\mathrm{sep}}=2\left\lceil \frac{|T_{\mathrm{all}}|}{B}\right\rceil,
\]
which determine per-site uplink and server additions. If packing were instead
performed over the event-time subset $T$, the same formulas would hold with
$|T|$ in place of $|T_{\mathrm{all}}|$. Let $R$ denote the committee size
(number of decryptors), where $R\!\le\!K$; experiments use $R{=}K$ by default,
while fixed-$R$ variants are included for communication studies. The decryption
threshold $\theta$ determines how many decryptors must participate to recover
plaintext; we set $\theta = R$ by default (strict threshold).
 
Our protocol provides these guarantees under the stated threat model: no party
learns other sites' plaintext contributions. The coordinator sees only
ciphertexts in Phase~B. In Phase~C, if fusion is co-located with the
coordinator, it may transiently observe the aggregated plaintext count vectors
$(n_i,d_i)$ over the aligned grid required for Kaplan--Meier evaluation;
however, individual site contributions are never decrypted separately, and only
the final survival curve $\hat S_{\mathrm{HE}}(t)$ is released to sites, eliminating the
subtraction-based residualization that is present when $(d_i,n_i)$ are
disclosed. The computational complexity of the protocol is characterized as
follows: Phase~A builds $T_{\mathrm{all}}$ in
$O(\sum_k |T_k^{\mathrm{all}}|\log \sum_k |T_k^{\mathrm{all}}|)$ time; Phase~B performs $O(MK)$ homomorphic
additions with linear-in-$K$ uplink; and Phase~C performs $O(RM)$ share
generations, $O(M)$ fusions, and $O(|T_{\mathrm{all}}|)$ decoding work,
followed by an $O(|T|)$ Kaplan--Meier pass on the event-time grid.
 
With the same event-time grid $T$ and add-only aggregation, $\hat S_{\mathrm{HE}}(t)$ matches
the pooled oracle up to CKKS-level numerical noise; validation tolerances
appear in Section~\ref{sec:exp_results}.
 
\subsection{Datasets}
We evaluate our approach on two datasets:
\begin{itemize}
\item \textbf{NCCTG Lung Cancer} contains 228 observations with survival times
and censoring indicators (\texttt{status} $=1$ censored, $=2$ event).
\item \textbf{Synthetic Breast Cancer (IKNL)} contains 60{,}000 synthetic
observations designed to mimic real-world survival data.
\end{itemize}
 
\subsection{Partitioning and heterogeneity}
\label{subsubsec:datapartition}
Each dataset is horizontally partitioned among $K$ clients. We first shuffle
the full dataset (fixed seed for reproducibility) and split rows into $K$
subsets of approximately equal size while retaining all features; this
approximates an i.i.d.\ partition.
 
To study non-IID effects, we induce \emph{outcome heterogeneity} via the event
indicator. Before splitting we: (i) standardize the event/status column to
binary \texttt{event}$\in\{0,1\}$ (map $\{1,2\}$ as $1\!\to\!0$ censored,
$2\!\to\!1$ event; otherwise declared ``event'' values map to $1$ and all others
to $0$); (ii) enforce strictly positive survival times (drop non-positive
entries); and (iii) log configuration for reproducibility.
 
Let $K$ be the number of clients. We split by \texttt{event}$\in\{0,1\}$ and
allocate each group with symmetric Dirichlet weights (strong heterogeneity):
\[
\mathbf{w}^{(g)} \sim \mathrm{Dirichlet}(\underbrace{\alpha,\ldots,\alpha}_{K}),\quad
\alpha=0.2,\quad g\in\{0,1\}.
\]
Rows in group $g$ are randomly permuted and assigned multinomially with
probabilities $\mathbf{w}^{(g)}$; we then enforce a ``no-empty-clients''
constraint by reassigning surplus rows until every client has at least one row
(this may soften the strongest skews but preserves the heterogeneity type). As
$\alpha\!\to\!\infty$ the split tends to i.i.d.
 
This construction preserves the centralized multiset of rows and hence the
union of survival times $T_{\mathrm{all}}$. The event-time grid
$T \subseteq T_{\mathrm{all}}$ derived from that union is therefore also
preserved. At each $t_i\in T$, per-client at-risk and event counts sum to
centralized totals. Consequently, under exact aggregation/decryption the
federated KM equals the centralized estimator, and under CKKS it converges to
it with negligible numerical error (Section~\ref{sec:theory-results}). For
plaintext baselines we also vary client-level overlap (None/Small/Large) to
analyze reconstruction risk; these overlap settings do not affect the encrypted
protocol.
 
\subsection{Evaluation protocol}
\label{sec:exp-setup}
\begin{table}[htbp]
  \centering
  \setlength{\tabcolsep}{7pt}
  \renewcommand{\arraystretch}{1.1}
  \begin{tabular}{l l}
    \toprule
    \textbf{Parameter} & \textbf{Value / Rationale} \\
    \midrule
    Scheme & CKKS (OpenFHE, multiparty/threshold APIs) \\
    Ring degree $n$ & $16{,}384$ (meets $\approx 128$-bit RLWE security with our $|Q|$) \\
    Slot budget $B$ & $n/2 = 8{,}192$ complex slots per ciphertext \\
    Total modulus $|Q|$ & $\approx 438$ bits (NTT-friendly RNS primes) \\
    Initial scale & $2^{40}$ (preserves $\sim$9--10 decimal digits) \\
    Galois keys & Powers-of-two rotations (no relinearization keys) \\
    Ciphertext size & \texttt{ct\_bytes} $\approx 1{,}794{,}048$ bytes (measured) \\
    Share size & \texttt{share\_bytes} $\approx 897{,}024$ bytes (measured) \\
    \bottomrule
  \end{tabular}
  \caption{CKKS instantiation used throughout.}
  \label{tab:ckks-params}
\end{table}
We repeat each configuration $10$ times to visualize curve variability,
quantify numerical and statistical agreement between encrypted and pooled
(oracle) pipelines, and measure overhead. We explicitly include non-IID label
splits (Dirichlet $\alpha{=}0.2$) and overlap regimes to stress heterogeneity;
as predicted by partition invariance
(Theorem~\ref{thm:plain-invariance}), the HE-federated Kaplan--Meier (KM)
matches the pooled oracle under both IID and non-IID partitions. Data
partitioning follows Section~\ref{subsubsec:datapartition}.
 
\subsection{Single-machine environment and timing}
\label{subsec:env}
All experiments run on a MacBook Pro (Apple M3 Pro, 36\,GB RAM, macOS
Sonoma~14.5). The experiments were implemented in Python~3.12.2 using OpenFHE
(C++ back end) via Python bindings. Unless stated otherwise, the runtime we report for a round
starts when sites begin packing/encrypting their local vectors and ends
immediately after Fusion produces plaintext aggregates (i.e., includes
pack+encrypt, upload, server adds, share generation, and fusion; excludes
plotting and file I/O). Each point shows the sample mean with 95\% CIs
(Student's $t$ across 10 replicates).
 
\subsection{Homomorphic-encryption parameters}
\label{subsec:he-params}
We use OpenFHE~\cite{OpenFHE} multiparty (threshold) APIs for CKKS. Phase~B
(R2) performs slotwise additions (and occasional rotations) only; no
ciphertext--ciphertext multiplications are used, so precision is governed by
the initial scale (no rescaling/bootstrapping). Unless otherwise stated,
experiments use the strict all-of-$R$ setting $\theta = R$; for fidelity and
runtime studies we set $R = K$, while fixed-$R$ variants are used only in
communication-scaling studies.
 
In the current implementation, we encode at-risk counts and event counts over
the full alignment grid $T_{\mathrm{all}}$. Using two layouts:
\[
\begin{aligned}
\textbf{Interleaved:}\;&(n_1,d_1,n_2,d_2,\ldots)
&&\Rightarrow\quad M_{\mathrm{int}}=\Big\lceil \tfrac{2|T_{\mathrm{all}}|}{B} \Big\rceil,\\[-0.2em]
\textbf{Separate:}\;&(d_1,\ldots,d_{|T_{\mathrm{all}}|})\ \ \text{and}\ \ (n_1,\ldots,n_{|T_{\mathrm{all}}|})
&&\Rightarrow\quad M_{\mathrm{sep}}=2\Big\lceil \tfrac{|T_{\mathrm{all}}|}{B} \Big\rceil.
\end{aligned}
\]
If packing were instead performed on the event-time subset $T$, the same
formulas would hold with $|T|$ replacing $|T_{\mathrm{all}}|$.
 
With $B{=}8192$ the stepwise dependence on the packed grid length is:
\[
L{=}1000:\ (1,2),\quad L{=}5000:\ (2,2),\quad L{=}10{,}000:\ (3,4)
\ \ \text{for}\ \ (M_{\mathrm{int}},M_{\mathrm{sep}}),
\]
where $L$ denotes the grid length used for packing. In our implementation
$L=|T_{\mathrm{all}}|$. Interleaving hence gives a $2\times$ reduction when
$L\!\le\!B/2$, ties at $L\!\in(B/2,B]$, and a $4/3\times$ reduction on
$(B,1.5B]$ (as detailed in Section~S6 of the Supplementary Information).
 
We implement per-client uplink accounting. Encrypted payload per client in
Phase~B (R2) is
\[
\text{he\_uplink} \;=\; M\times \texttt{ct\_bytes}.
\]
Decryptor-share payload is
\[
\text{shares\_uplink} \;=\; M\times R \times \texttt{share\_bytes}.
\]
Thus total encrypted uplink to the coordinator is linear in $K$ (sites), while
share traffic is linear in $R$ (decryptors) and independent of $K$.
 
For fidelity/statistics/runtime experiments we set $R{=}K$ (all clients act as
decryptors). For the \emph{communication-scalability} study only, we fix
$R\in\{5,9,25\}$ independent of $K$ to expose the $R$-linearity of share
traffic. This changes communication tallies only; numerical accuracy is
unaffected.
 
We compute restricted mean survival time ($\mathrm{RMST}(\tau)$) as a
left-Riemann sum over the event-time grid $T$ up to horizon $\tau$:
\[
\mathrm{RMST}(\tau) = \sum_{t_i \le \tau} \hat S(t_{i-1})\,\Delta t_i,
\qquad
\Delta t_i = t_i - t_{i-1},
\qquad
t_0=0.
\]
Unless stated otherwise we set $\tau=\max T$.
 
\begin{table}[htbp]
\centering
\small
\setlength{\tabcolsep}{4pt}
\renewcommand{\arraystretch}{1.12}
\begin{tabularx}{\linewidth}{@{}l l >{\raggedright\arraybackslash}X >{\raggedleft\arraybackslash}p{0.16\linewidth}@{}}
\toprule
\textbf{Acr.} & \textbf{Metric} & \textbf{Definition} & \textbf{Threshold} \\
\midrule
TAAE($S$) & Time-avg $|\Delta S|$ & $\mathrm{TAAE}(S)=\mathrm{IAE}(S)/\tau$\textsuperscript{a} & $\le 10^{-9}$ \\
IAE($S$) & Int $|\Delta S|$ & $\mathrm{IAE}(S)=\int |\hat S_{\mathrm{HE}}(t)-\hat S_{\mathrm{oracle}}(t)|\,dt$ & $\le 10^{-6}$ \\
$|\Delta\!\mathrm{RMST}|$ & Abs RMST gap & $\big|\mathrm{RMST}_{\mathrm{HE}}(\tau)-\mathrm{RMST}_{\mathrm{oracle}}(\tau)\big|$ & $\le 10^{-6}$ \\
$\Delta\!\mathrm{RMST}_{\mathrm{rel}}$ & Rel RMST gap & $\frac{\big|\mathrm{RMST}_{\mathrm{HE}}(\tau)-\mathrm{RMST}_{\mathrm{oracle}}(\tau)\big|}{\mathrm{RMST}_{\mathrm{oracle}}(\tau)}$ & $\le 10^{-6}$ \\
supZ & Supremum $z$ & $\sup_{t\in T}\frac{|\hat S_{\mathrm{HE}}(t)-\hat S_{\mathrm{oracle}}(t)|}{\widehat{\mathrm{SE}}[\hat S_{\mathrm{oracle}}(t)]}$ & $\le 10^{-6}$ \\
Coverage & 95\% coverage & $\frac{1}{|T|}\sum_{t\in T}\mathbf{1}\{\hat S_{\mathrm{HE}}(t)\in [\hat S_{\mathrm{oracle}}(t)\pm1.96\,\widehat{\mathrm{SE}}(t)]\}$ & $\ge 0.99$ \\
$\max_t |\Delta H|$ & Max $|\Delta H|$ & $\max_{t\in T}|H_{\mathrm{HE}}(t)-H_{\mathrm{oracle}}(t)|$ & $\le 10^{-8}$ \\
$\mathrm{IAE}_H$ & Int $|\Delta H|$ & $\int |H_{\mathrm{HE}}(t)-H_{\mathrm{oracle}}(t)|\,dt$ & $\le 10^{-7}$ \\
\bottomrule
\end{tabularx}
\caption{Equivalence metrics, definitions, and pass thresholds used to judge agreement between HE-federated and pooled (oracle) estimators. Notes: $\tau$ is the horizon; Greenwood SE for bands.}
\label{tab:equivalence-metrics}
\end{table}
In addition to the metric-specific thresholds in
Table~\ref{tab:equivalence-metrics}, we report two summary tolerances:
\[
\max_{t\in T} \left| \hat S_{\mathrm{HE}}(t) - \hat S_{\mathrm{oracle}}(t) \right| \le 10^{-8},\qquad
\frac{\sum_{t\in T} \left|\hat S_{\mathrm{HE}}(t)-\hat S_{\mathrm{oracle}}(t)\right|}{\sum_{t\in T} \left|\hat S_{\mathrm{oracle}}(t)\right|} \le 10^{-9}.
\]
Bands use pooled Greenwood SE (coverage and standardized sup-gap); all figures
show mean with 95\% CIs across the 10 repeats (Student's $t$).
 
\subsection{Communication scalability}
\label{subsec:comm-scalability}
 
\begin{table}[htbp]
  \centering
  \renewcommand{\arraystretch}{1.2}
  \begin{tabular}{|c|l|l|}
    \hline
    \textbf{Symbol} & \textbf{Description} & \textbf{Values (grid)} \\
    \hline
    $K$   & Number of clients & $\{10,\,50,\,100,\,200,\,300,\,500\}$ \\
    \hline
    $|T_{\mathrm{all}}|$ & Number of global unique survival times & depends on dataset and censoring pattern \\
    \hline
    $|T|$ & Number of global event times & $\{1000,\,5000,\,10000\}$ \\
    \hline
    $R$   & Committee size & $\{5,\,9,\,25\}$ \\
    \hline
    $n$   & CKKS ring degree & $\{16384\}$ \\
    \hline
    $|Q|$ & Total modulus bitlength & $\{438\}\ \text{bits}$ \\
    \hline
  \end{tabular}
  \caption{Parameter grid used to instantiate the communication formulas.}
  \label{tab:comm-grid}
\end{table}
 
We analyze communication using closed-form counts consistent with
Section~\ref{subsec:theory-comm}. Let $K$ denote the number of sites,
$T_{\mathrm{all}}$ the union of all unique survival times across sites
(including censoring times), $T \subseteq T_{\mathrm{all}}$ the
event-time grid used for Kaplan--Meier aggregation, $R \le K$ the
decryptor committee size, $n$ the CKKS ring degree, and $B=n/2$ the slot
capacity per ciphertext. Unless stated otherwise, payload sizes are
reported in MB (decimal units, $10^6$ bytes).
 
\textbf{Plain protocol.}
Our plaintext two-round protocol operates as follows. In Round~1 (R1),
each site uploads its local set of unique observed survival times
$T_k^{\mathrm{all}}$ (including both event and censoring times). The coordinator
computes the union
\[
T_{\mathrm{all}} = \bigcup_{k=1}^{K} T_k^{\mathrm{all}}
\]
and broadcasts $T_{\mathrm{all}}$ to all sites. The event-time subset
$T = \{t \in T_{\mathrm{all}} : \exists\text{ an event at }t\}$ is not
constructed explicitly; it is handled implicitly because $d_i = 0$ at
censoring-only times.
 
In Round~2 (R2), each site uploads aligned counts
\[
\{(n_i^{(k)}, d_i^{(k)})\}_{t_i \in T_{\mathrm{all}}}.
\]
 
The total number of uploaded plaintext scalars is therefore
\[
\underbrace{\sum_{k=1}^{K} |T_k^{\mathrm{all}}|}_{\text{R1 upload}}
\;+\;
\underbrace{2K|T_{\mathrm{all}}|}_{\text{R2 counts (at-risk and events)}}.
\]
 
The downlink consists of broadcasting $T_{\mathrm{all}}$, giving
\[
K|T_{\mathrm{all}}|
\]
plaintext scalars in total.
 
The R1 uplink satisfies
\[
\sum_{k=1}^{K} |T_k^{\mathrm{all}}|
\;\le\;
K|T_{\mathrm{all}}|.
\]
The R1 uplink bounds (in bytes, with 8 bytes per scalar) are therefore
\[
\mathrm{uplink}^{\mathrm{R1}}_{\min}
= |T_{\mathrm{all}}| \cdot 8,
\qquad
\mathrm{uplink}^{\mathrm{R1}}_{\max}
= K|T_{\mathrm{all}}| \cdot 8.
\]
 
Combining R1 and R2 yields the min--max uplink bounds
\[
\mathrm{uplink}_{\min}
= (|T_{\mathrm{all}}| + 2K|T_{\mathrm{all}}|)\cdot 8,
\qquad
\mathrm{uplink}_{\max}
= (K|T_{\mathrm{all}}| + 2K|T_{\mathrm{all}}|)\cdot 8
= 3K|T_{\mathrm{all}}| \cdot 8.
\]
 
The total downlink is
\[
K|T_{\mathrm{all}}| \cdot 8.
\]
 
\noindent\textit{Event-time-only variant.}
If a deployment filters to the event-time subset $T$ before R2, the R2
uplink reduces to $2K|T|$ scalars; R1 communication is unchanged.
 
\textbf{CKKS with multiparty decryption.}
In the encrypted protocol we retain R1 in plaintext and encrypt only R2. In the
current implementation, each site encrypts two length-$|T_{\mathrm{all}}|$
vectors containing at-risk and event counts aligned to the full grid
$T_{\mathrm{all}}$, with $d_i=0$ at non-event times. With CKKS packing and slot
capacity $B=n/2$, the number of ciphertexts per client is
\[
M =
\begin{cases}
2\left\lceil |T_{\mathrm{all}}|/B \right\rceil, & \text{separate packing}, \\[4pt]
\left\lceil 2|T_{\mathrm{all}}|/B \right\rceil, & \text{interleaved packing}.
\end{cases}
\]
If encryption were applied only to the event-time subset $T$, the same formulas
would hold with $|T|$ replacing $|T_{\mathrm{all}}|$. This is not the current
implementation; rather, it is an event-time-only packing variant used for the
communication simulations in Section~\ref{sec:exp_results}, which instantiate
these formulas with $L=|T|$.
 
Using the conservative byte model
\[
\texttt{ct\_bytes} \approx \frac{2n|Q|}{8},
\qquad
\texttt{share\_bytes} \approx \frac{n|Q|}{8},
\]
the encrypted R2 uplink is
\[
\mathrm{HE\ R2\ uplink} = K M \cdot \texttt{ct\_bytes},
\]
and the decryption-share traffic is
\[
\mathrm{shares\ uplink} = R M \cdot \texttt{share\_bytes}.
\]
 
Dividing by $10^6$ converts these values to MB.
 
We instantiate these formulas on the grid in
Table~\ref{tab:comm-grid} and report:
(i) plaintext uplink and downlink including the R1 min--max band,
(ii) encrypted R2 uplink in MB, and
(iii) decryption-share traffic in MB.
 
\textbf{Assumptions.}
The analysis assumes:
(i) payload-only accounting with transport and TLS framing omitted,
(ii) R1 remains plaintext in both the plain and HE protocols,
(iii) R2 uses add-only encrypted aggregation,
(iv) ciphertext sizes follow the uncompressed model above, which gives a
conservative upper bound, and
(v) sites align on a common global alignment grid $T_{\mathrm{all}}$, from
which the event-time grid $T$ used for Kaplan--Meier evaluation is derived.
 
In the current implementation, R2 uses $T_{\mathrm{all}}$ as the alignment grid
passed to each client, so encrypted count vectors are packed over
$T_{\mathrm{all}}$ rather than over the event-time subset $T$. Counts are
therefore uploaded for every time in $T_{\mathrm{all}}$, with $d_i = 0$ at
non-event times. This does not change the Kaplan--Meier estimator because
inserting zero-event time points refines the grid without altering the product.
However, the encrypted R2 uplink in the implementation scales with
$M(|T_{\mathrm{all}}|)$ rather than $M(|T|)$. Filtering to event times
before encryption would therefore reduce the number of ciphertexts whenever
$|T_{\mathrm{all}}| > |T|$.
\backmatter
 
\section*{Data availability}
The NCCTG lung cancer dataset used in this study is publicly available in the R survival package and can be accessed at \href{https://rdrr.io/cran/survival/man/lung.html}{https://rdrr.io/cran/survival/man/lung.html}.  For our experiments, we downloaded the dataset as a CSV file and processed it using Python.

The synthetic breast cancer dataset used in this study is a controlled-access synthetic dataset
provided by IKNL. Information about access is available at
\url{https://iknl.nl/international/data-request}.

Researchers may request access from IKNL in accordance with its data request procedures.
The code to reproduce our experiments is available at \href{https://github.com/vnragavan/homomorphic-federated-km}{https://github.com/vnragavan/homomorphic-federated-km}
\section*{Acknowledgements}
This work has been partially carried out in the context of the Center for Research-based Innovation NORCICS, funded by the Research Council of Norway (grant number 310105/F40).

\section*{Competing interests}
The authors declare no competing interests.

\bibliographystyle{sn-nature}
\bibliography{sn-bibliography}

@inproceedings{BrakerskiGV12,
  author    = {Zvika Brakerski and Craig Gentry and Vinod Vaikuntanathan},
  title     = {({L}eveled) Fully Homomorphic Encryption Without Bootstrapping},
  editor    = {Shafi Goldwasser},
  booktitle = {Proceedings of the 3rd Innovations in Theoretical Computer Science Conference (ITCS '12)},
  year      = {2012},
  pages     = {309--325},
  publisher = {Association for Computing Machinery},
  address   = {New York, NY, USA},
  doi       = {10.1145/2090236.2090262},
}

@inproceedings{li2021security,
  author    = {Li, Baiyu and Micciancio, Daniele},
  title     = {On the Security of Homomorphic Encryption on Approximate Numbers},
  editor    = {Canteaut, Anne and Standaert, Fran{\c{c}}ois-Xavier},
  booktitle = {Advances in Cryptology -- EUROCRYPT 2021},
  series    = {Lecture Notes in Computer Science},
  volume    = {12696},
  pages     = {648--677},
  publisher = {Springer},
  address   = {Cham},
  year      = {2021},
  doi       = {10.1007/978-3-030-77870-5_23}
}

@article{von2021privacy,
  title={A Privacy-Preserving Log-Rank test for the Kaplan-Meier estimator with secure multiparty computation: algorithm development and validation},
  author={von Maltitz, Marcel and Ballhausen, Hendrik and Kaul, David and Fleischmann, Daniel F and Niyazi, Maximilian and Belka, Claus and Carle, Georg},
  journal={JMIR medical informatics},
  volume={9},
  number={1},
  pages={e22158},
  year={2021},
  publisher={JMIR Publications Toronto, Canada}
}

@article{evans2018pragmatic,
  title={A pragmatic introduction to secure multi-party computation},
  author={Evans, David and Kolesnikov, Vladimir and Rosulek, Mike},
  journal={Foundations and Trends{\textregistered} in Privacy and Security},
  volume={2},
  number={2-3},
  pages={70--246},
  year={2018},
  publisher={Emerald Publishing Limited}
}

@InProceedings{32009BZ,
author="Brakerski, Zvika",
editor="Safavi Naini, Reihaneh
and Canetti, Ran",
title="Fully Homomorphic Encryption without Modulus Switching from Classical GapSVP",
booktitle="Advances in Cryptology -- CRYPTO 2012",
year="2012",
publisher="Springer Berlin Heidelberg",
address="Berlin, Heidelberg",
pages="868--886",
isbn="978-3-642-32009-5"
}

@InProceedings{10.1007/978-3-662-53887-6_1,
author="Chillotti, Ilaria
and Gama, Nicolas
and Georgieva, Mariya
and Izabach{\`e}ne, Malika",
editor="Cheon, Jung Hee
and Takagi, Tsuyoshi",
title="Faster Fully Homomorphic Encryption: Bootstrapping in Less Than 0.1 Seconds",
booktitle="Advances in Cryptology -- ASIACRYPT 2016",
year="2016",
publisher="Springer Berlin Heidelberg",
address="Berlin, Heidelberg",
pages="3--33",
isbn="978-3-662-53887-6"
}

@misc{cryptoeprint:2012/144,
      author = {Junfeng Fan and Frederik Vercauteren},
      title = {Somewhat Practical Fully Homomorphic Encryption},
      howpublished = {Cryptology ePrint Archive, Paper 2012/144},
      year = {2012},
      note  = {accessed: 2024-10-01},
      url = {https://eprint.iacr.org/2012/144}
}

@InProceedings{10.1007/978-3-662-46800-5_24,
author="Ducas, L{\'e}o
and Micciancio, Daniele",
editor="Oswald, Elisabeth
and Fischlin, Marc",
title="FHEW: Bootstrapping Homomorphic Encryption in Less Than a Second",
booktitle="Advances in Cryptology -- EUROCRYPT 2015",
year="2015",
publisher="Springer Berlin Heidelberg",
address="Berlin, Heidelberg",
pages="617--640",
isbn="978-3-662-46800-5"
}

@InProceedings{ckks,
author="Cheon, Jung Hee
and Kim, Andrey
and Kim, Miran
and Song, Yongsoo",
editor="Takagi, Tsuyoshi
and Peyrin, Thomas",
title="Homomorphic Encryption for Arithmetic of Approximate Numbers",
booktitle="Advances in Cryptology -- ASIACRYPT 2017",
year="2017",
publisher="Springer International Publishing",
address="Cham",
pages="409--437",
isbn="978-3-319-70694-8"
}

@misc{OpenFHE,
      author = {Ahmad Al Badawi and Andreea Alexandru and Jack Bates and Flavio Bergamaschi and David Bruce Cousins and Saroja Erabelli and Nicholas Genise and Shai Halevi and Hamish Hunt and Andrey Kim and Yongwoo Lee and Zeyu Liu and Daniele Micciancio and Carlo Pascoe and Yuriy Polyakov and Ian Quah and Saraswathy R.V. and Kurt Rohloff and Jonathan Saylor and Dmitriy Suponitsky and Matthew Triplett and Vinod Vaikuntanathan and Vincent Zucca},
      title = {{OpenFHE}: Open-Source Fully Homomorphic Encryption Library},
      howpublished = {Cryptology ePrint Archive, Paper 2022/915},
      year = {2022},
      url = {https://eprint.iacr.org/2022/915},
      note  = {accessed: 2024-10-01},
}

@article{Froelicher2021,
  author = {Froelicher, David and Troncoso-Pastoriza, Juan Ramon and Raisaro, Jean Louis and Cuendet, M{\'a}rcio A. and Sousa, Jo{\~a}o S{\'a} and Cho, Hyunghoon and Berger, Bonnie and Fellay, Jacques and Hubaux, Jean-Pierre},
  title = {Truly Privacy-Preserving Federated Analytics for Precision Medicine with Multiparty Homomorphic Encryption},
  journal = {Nature Communications},
  volume = {12},
  number = {1},
  pages = {5910},
  year = {2021},
  doi = {10.1038/s41467-021-25972-y},
  PMID = {34635645},
  PMCID = {PMC8505638},
  erratum = {Erratum in: Nat Commun. 2021 Nov 11;12(1):6649. doi: 10.1038/s41467-021-26885-6.},
  date = {2021-10-11}
}

@InProceedings{10.1007/978-3-642-29011-4_29,
author="Asharov, Gilad
and Jain, Abhishek
and L{\'o}pez-Alt, Adriana
and Tromer, Eran
and Vaikuntanathan, Vinod
and Wichs, Daniel",
editor="Pointcheval, David
and Johansson, Thomas",
title="Multiparty Computation with Low Communication, Computation and Interaction via Threshold FHE",
booktitle="Advances in Cryptology -- EUROCRYPT 2012",
year="2012",
publisher="Springer Berlin Heidelberg",
address="Berlin, Heidelberg",
pages="483--501",
isbn="978-3-642-29011-4"
}

@article{geva2023collaborative,
  title={Collaborative privacy-preserving analysis of oncological data using multiparty homomorphic encryption},
  author={Geva, Ravit and Gusev, Alexander and Polyakov, Yuriy and Liram, Lior and Rosolio, Oded and Alexandru, Andreea and Genise, Nicholas and Blatt, Marcelo and Duchin, Zohar and Waissengrin, Barliz and others},
  journal={Proceedings of the National Academy of Sciences},
  volume={120},
  number={33},
  pages={e2304415120},
  year={2023},
  publisher={National Acad Sciences}
}

@inproceedings{rahimian2024private,
  author    = {Shadi Rahimian and Raouf Kerkouche and Ina Kurth and Mario Fritz},
  title     = {Private and Collaborative {Kaplan--Meier} Estimators},
  editor    = {Erman Ayday and Jaideep Vaidya},
  booktitle = {Proceedings of the 23rd Workshop on Privacy in the Electronic Society (WPES '24)},
  year      = {2024},
  pages     = {212--241},
  publisher = {Association for Computing Machinery},
  address   = {New York, NY, USA},
  doi       = {10.1145/3689943.3695039},
}

@article{VONMALTITZ2021,
title = {A Privacy-Preserving Log-Rank Test for the Kaplan-Meier Estimator With Secure Multiparty Computation: Algorithm Development and Validation},
journal = {JMIR Medical Informatics},
volume = {9},
number = {1},
year = {2021},
issn = {2291-9694},
doi = {https://doi.org/10.2196/22158},
url = {https://www.sciencedirect.com/science/article/pii/S2291969421000326},
author = {Marcel {von Maltitz} and Hendrik Ballhausen and David Kaul and Daniel F Fleischmann and Maximilian Niyazi and Claus Belka and Georg Carle},
}

@book{Cramer_Damgard_Nielsen_2015,
author    = {Cramer, Ronald and Damg{\aa}rd, Ivan Bjerre and Nielsen, Jesper Buus},
  title     = {Secure {M}ultiparty {C}omputation and {S}ecret {S}haring},
  year      = {2015},
  publisher = {Cambridge University Press},
  address   = {Cambridge}
}

@misc{cryptoeprint:2018/421,
      author = {Ilaria Chillotti and Nicolas Gama and Mariya Georgieva and Malika Izabach{\`e}ne},
      title = {{TFHE}: Fast Fully Homomorphic Encryption over the Torus},
      howpublished = {Cryptology {ePrint} Archive, Paper 2018/421},
      year = {2018},
      url = {https://eprint.iacr.org/2018/421}
}

@misc{veeraragavan2025,
      title={Federated Survival Analysis with Node-Level Differential Privacy: Private Kaplan-Meier Curves}, 
      author={Narasimha Raghavan Veeraragavan and Jan Franz Nyg{\aa}rd},
      year={2025},
      eprint={2509.00615},
      archivePrefix={arXiv},
      primaryClass={cs.CR},
      url={https://arxiv.org/abs/2509.00615}, 
}

\section*{Figure Legends}

\noindent\textbf{Figure 1.} Reconstruction metrics vs.\ number of sites $K$ for IID data partitioning on the Synthetic Breast dataset. Ten metrics are shown: (A) F1 score for event support, (B) precision, (C) recall, (D) Jaccard index, (E) relative $L_1$ error on $d_t$, (F) relative $L_\infty$ error on $d_t$, (G) relative $L_1$ error on $n_t$, (H) relative $L_\infty$ error on $n_t$, (I) exact-match rate on $d_t$, and (J) exact-match rate on $n_t$. Curves show means with 95\% confidence intervals. Three overlap conditions are compared: none (blue), small (orange), and large (red). All conditions yielded identical results, demonstrating exact reconstruction regardless of overlap.
 
\noindent\textbf{Figure 2.} Reconstruction metrics vs.\ number of sites $K$ for label non-IID data partitioning (Dirichlet $\alpha=0.2$) on the Synthetic Breast dataset. The layout is identical to Figure~1 and shows results under label-based non-IID partitioning. All overlap conditions yielded identical results, demonstrating exact reconstruction even under heterogeneous data distributions.
 
\noindent\textbf{Figure 3.} Per-site event-rate distributions for the Synthetic Breast dataset. Violin plots show the distribution of event rates (\#events/\#rows) across sites under (a) IID partitioning and (b) label non-IID partitioning. Wider, skewed violins under non-IID indicate heterogeneous site contributions but do not prevent exact residualization when plaintext per-time-point counts are disclosed.
 
\noindent\textbf{Figure 4.} Numerical fidelity metrics for HE-federated Kaplan--Meier vs.\ pooled oracle across number of sites $K$ on the Synthetic Breast dataset. Eight metrics are shown: (a) integrated absolute error on survival function $\mathrm{IAE}(S)$, (b) time-averaged absolute error $\mathrm{TAAE}(S)$, (c) absolute difference in restricted mean survival time $|\Delta\!\mathrm{RMST}|$, (d) relative RMST difference $\Delta\!\mathrm{RMST}_{\mathrm{rel}}$, (e) coverage probability of 95\% confidence intervals, (f) standardized supremum gap $\mathrm{supZ}$ measuring maximum standardized deviation, (g) maximum absolute difference in cumulative hazard $\max_t|H_{\mathrm{HE}}-H_{\mathrm{oracle}}|$, and (h) integrated absolute error on cumulative hazard $\mathrm{IAE}_H$. Each panel compares interleaved (blue) and non-interleaved/separate (orange) packing strategies with means and 95\% confidence intervals. The two packing strategies are numerically indistinguishable across all metrics as $K$ increases.
 
\noindent\textbf{Figure 5.} Kaplan--Meier survival curves for $K=500$ sites demonstrating visual indistinguishability between the pooled oracle (dashed black line), HE-federated with interleaved packing (orange), and HE-federated with non-interleaved/separate packing (blue). All three curves overlap almost perfectly, confirming that homomorphic encryption preserves statistical accuracy even at large scale.
 
\noindent\textbf{Figure 6.} Computational scalability of HE-federated Kaplan--Meier on the Synthetic Breast dataset. (a) End-to-end runtime vs.\ number of sites $K$, showing approximately linear growth. Plain pooled oracle (gray) serves as a baseline reference, while HE with interleaved packing (orange) and non-interleaved/separate packing (blue) are shown with means and 95\% confidence intervals. (b) Percentage speedup of interleaved over non-interleaved packing demonstrates increasing computational gains as $K$ rises, plateauing at approximately 22\% improvement for large-scale federations.
 
\noindent\textbf{Figure 7.} Communication scaling. (A) Uplink vs.\ $K$ with $|T|{=}1000$ and $R{=}5$; (B) Uplink vs.\ $K$ with $|T|{=}10000$ and $R{=}5$; (C) decryption-share uplink vs.\ $R$ with $K{=}50$ and $|T|{=}1000$; (D) decryption-share uplink vs.\ $R$ with $K{=}50$ and $|T|{=}10000$.
 
\noindent\textbf{Figure 8.} Communication scaling under CKKS packing. (A) HE uplink vs.\ $K$ at fixed $|T|{=}5000$ and $R{=}5$; (B) HE uplink vs.\ $|T|$ at fixed $K{=}50$ and $R{=}5$; (C) decryption-share uplink vs.\ $R$ at fixed $K{=}50$ and $|T|{=}5000$; (D) interleaving benefit shown as HE uplink for non-interleaved and interleaved packing vs.\ $|T|$, at fixed $K{=}50$ and $R{=}5$.
 
\noindent\textbf{Figure 9.} HE-versus-plaintext uplink ratio. (A) Non-interleaved HE/plain uplink ratio vs.\ $K$ at fixed $|T|{=}5000$; (B) interleaved HE/plain uplink ratio vs.\ $K$ at fixed $|T|{=}5000$; (C) non-interleaved HE/plain uplink ratio vs.\ $|T|$ at fixed $K{=}50$; (D) interleaved HE/plain uplink ratio vs.\ $|T|$ at fixed $K{=}50$. Solid lines use the plaintext minimum-uplink denominator and dashed lines use the plaintext maximum-uplink denominator.
 
\section*{Table Legends}

\noindent\textbf{Table 1.} Comparison of representative FHE schemes used in privacy-preserving analytics. ``Threshold'' indicates widely available threshold decryption in mainstream libraries.
 
\noindent\textbf{Table 2.} Representative federated KM systems across three families: integer-HE (BGV/BFV), approximate-HE (CKKS), and differential privacy (DP). Our framework differs by (i) threshold CKKS with \emph{output gating} to the public survival curve $\hat S_{\mathrm{HE}}(t)$ (no per-time-point tables), (ii) KM-specific estimator-level guarantees, and (iii) closed-form \emph{communication} and \emph{computational} scaling laws.
 
\noindent\textbf{Table 3.} Recommended defaults for CKKS federated Kaplan--Meier computation with add-only aggregation.
 
\noindent\textbf{Table 4.} Notation used in the federated CKKS Kaplan--Meier protocol.
 
\noindent\textbf{Table 5.} Protocol keywords grouped by phase: Phase~A performs setup and time-grid construction; Phase~B performs encrypted aggregation; Phase~C performs multiparty decryption and output reveal. The keywords map to the abstract protocol tuple as follows: \textsc{Setup}~$\leftrightarrow$~Phase~A (InitCKKS through Union~\&~Sort); \textsc{Pack}+\textsc{Enc}+\textsc{Agg}~$\leftrightarrow$~Phase~B; \textsc{PartDec}~$\leftrightarrow$~MultipartyDecryptLead/Main; \textsc{Fuse}~$\leftrightarrow$~MultipartyDecryptFusion; \textsc{Reveal}~$\leftrightarrow$~ReconstructCounts+ComputeKM+Reveal+Broadcast($\hat S_{\mathrm{HE}}$).
 
\noindent\textbf{Table 6.} Security-specific notation used in the phase-wise proofs (Theorems~20--22 and Corollary~23). Symbols shared with the main protocol are defined in Table~4; this table contains only notation that is specific to or has a refined meaning within the security analysis.
 
\noindent\textbf{Table 7.} CKKS instantiation used throughout.
 
\noindent\textbf{Table 8.} Equivalence metrics, definitions, and pass thresholds used to judge agreement between HE-federated and pooled (oracle) estimators. Notes: $\tau$ is the horizon; Greenwood SE for bands.
 
\noindent\textbf{Table 9.} Parameter grid used to instantiate the communication formulas.
\end{document}